\documentclass[useAMS,usenatbib]{mn2e}

\pdfoutput=1
\usepackage{url}
\usepackage{lscape}
\usepackage{graphicx}
\usepackage{float}
\usepackage{morefloats}
\usepackage{supertabular}
\usepackage{floatflt}
\usepackage{caption}

\title{Hunting For Eclipses: High Speed Observations of Cataclysmic Variables}
\author{L. K. Hardy,$^{1}$ M. J. McAllister,$^{1}$ V. S. Dhillon,$^{1,2}$ S. P. Littlefair,$^{1}$ M. C. P. Bours,$^{3,4}$ 
\newauthor
E. Breedt,$^{4}$ T. Butterley,$^{5}$ A. Chakpor,$^{6}$ P. Irawati,$^{6}$ P. Kerry,$^{1}$ T. R. Marsh,$^{4}$ 
\newauthor
S. G. Parsons,$^{1}$ C. D. J. Savoury,$^{1}$ R. W. Wilson$^{5}$ and P. A. Woudt$^{7}$\\  
$^{1}$Department of Physics and Astronomy, University of Sheffield, Sheffield, S3 7RH, UK\\
$^{2}$Instituto de Astrof\`{i}sica de Canarias, E-38205 La Laguna, Tenerife, Spain\\
$^{3}$Instituto de F\`{i}sica y Astronom\`{i}a, Universidad de Valpara\`{i}so, Avenida Gran Breta\~{n}a 1111, Valpara\`{i}so, Chile\\
$^{4}$Department of Physics, University of Warwick, Coventry CV4 7AL, UK\\
$^{5}$Center for Advanced Instrumentation, Department of Physics, University of Durham, South Road, Durham, DH1 3LE, UK\\
$^{6}$National Astronomical Research Institute of Thailand, 191 Siriphanich Building, Huay Kaew Road, Chiang Mai 50200, Thailand\\
$^{7}$Department of Astronomy, University of Cape Town, Private Bag X3, Rondebosch 7701, South Africa}

\begin{document}

\date{}

\pagerange{\pageref{firstpage}--\pageref{lastpage}} \pubyear{2016}

\maketitle

\label{firstpage}

\begin{abstract}
We present new time-resolved photometry of 74 cataclysmic variables (CVs), 47 of which are eclipsing. 13 of these eclipsing systems are newly discovered. For all 47 eclipsing systems we show high cadence (1-20 seconds) light curves obtained with the high-speed cameras \textsc{ultracam} and \textsc{ultraspec}. We provide new or refined ephemerides, and supply mid-eclipse times for all observed eclipses. We assess the potential for light curve modelling of all 47 eclipsing systems to determine their system parameters, finding 20 systems which appear to be suitable for future study.

Systems of particular interest include V713 Cep, in which we observed a temporary switching-off of accretion; and ASASSN-14mv and CSS111019:233313-155744, which both have orbital periods well below the CV period minimum. The short orbital periods and light curve shapes suggests that they may be double degenerate (AM CVn) systems or CVs with evolved donor stars. 
\end{abstract}



\section{Introduction}
Cataclysmic variables (CVs) are binary systems in which a white dwarf accretes mass from a (usually) late-type, main sequence companion. The accretion tends to be via an accretion disk, or directly onto the poles of the white dwarf in the case of a strong magnetic field. In all cases, the donor star loses mass via Roche lobe overflow, and steady mass transfer is driven by a loss of angular momentum. For extensive reviews of CVs see \citet{warner95a} and \citet{hellier01}.

If the orientation of the orbital plane of a CV is coincident with our line of sight, eclipses may be observed. CVs often have complex eclipse structures, revealing different components of the system, including the accretion disk, the white dwarf and the bright spot, where the infalling material from the donor star encounters the disk. In certain systems these features are evident and clearly separated from one another, whilst in others they are not present, are very weak, or are blended together. In most CV systems there is the added stochastic variability component known as flickering, which is due to random fluctuations in the accretion flow rate. Some systems show varying rates of accretion and disk radius, which can be observed as changes in the timing and depth of the bright spot eclipse. 

\begin{figure}
 \includegraphics[width=.49\textwidth]{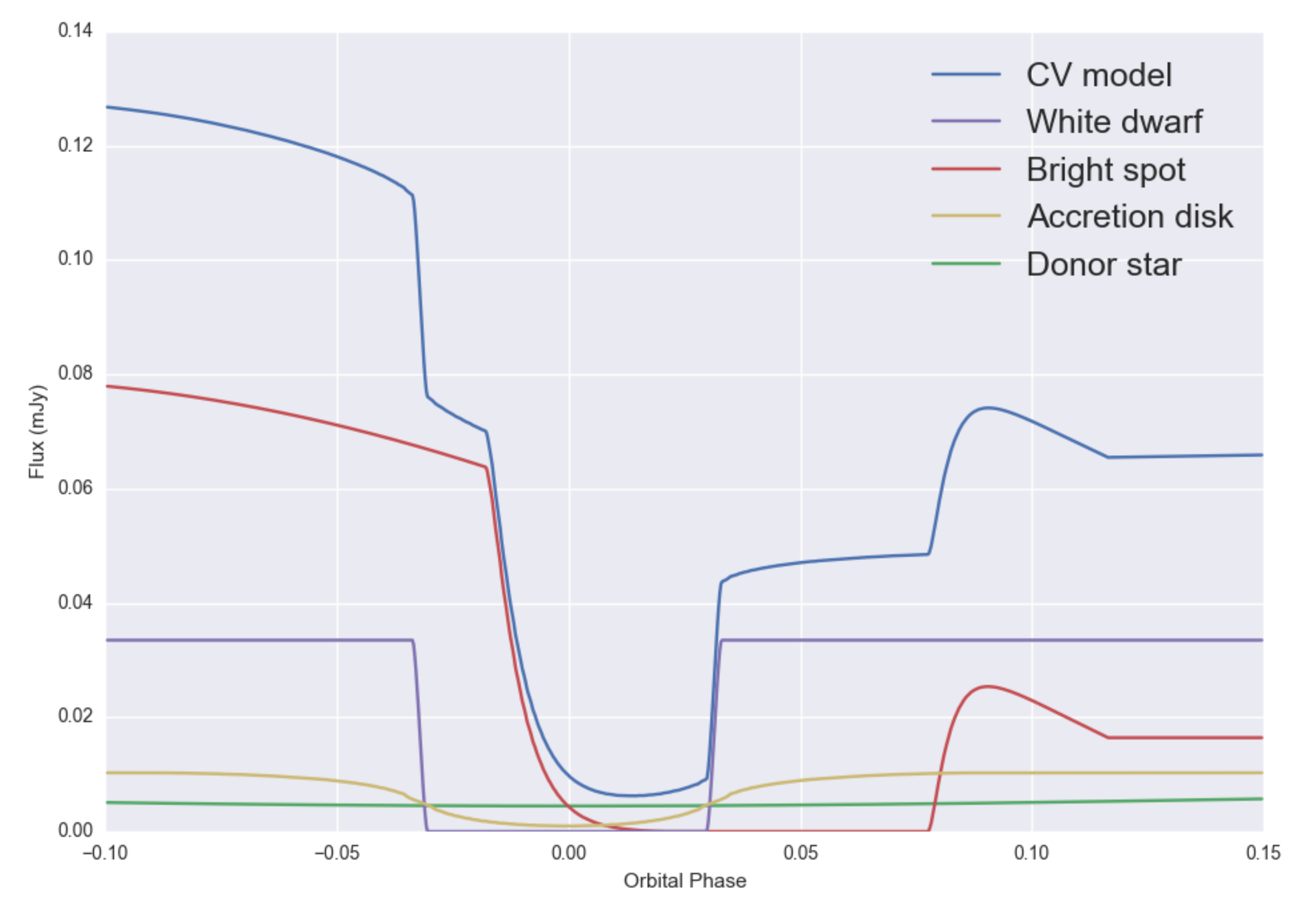}
 \caption{Artificial eclipse light curve, showing the different components of the model, which all contribute to the total light curve. An eclipse light curve showing clearly separated structure like this is ideal for modelling.}
 \label{fig:modeleclipse}
\end{figure}

When dwarf novae CVs display clear and separable white dwarf and bright spot eclipses, their light curves can be modelled allowing precise measurements to be made of the component masses and radii, in addition to other system parameters. In such cases, this information can be extracted without any form of spectroscopic study, e.g. \citet{wood86a,littlefair08,savoury11,mcallister16}, and the validity of this method has been verified using time-resolved spectroscopy \citep{tulloch09,savoury12,copperwheat12}. Figure \ref{fig:modeleclipse} shows an example of an ideal eclipse, with structure that is suitable for modelling. 

Eclipsing CVs can therefore be useful in improving our understanding of CV evolution. Theoretical and empirical studies of the white dwarf mass distribution in CVs, and the properties of CV donor stars, both rely on the limited database of precisely measured CV system parameters. 

For example, the average white dwarf mass is considerably higher in CVs than in pre-CV systems and single white dwarfs. Explanations for this include rapid mass accretion during CV formation \citep{schenker02} or gradual mass growth via long-term stable accretion \citep{savoury11}, however \citet{wijnen15} ruled out both of these scenarios using binary population synthesis models. Recently this mass distribution discrepancy has instead been attributed to enhanced angular momentum loss in low-mass white dwarf CVs during early nova eruptions \citep{schreiber16,nelemans16}.

The study of CV evolution is also directly linked to the properties of the donor star, which can affect the rate of angular momentum loss in the system. Much work has gone in to producing a semi-empirical donor sequence which describes the mass, radius, effective temperature and spectral type of CV donor stars as a function of orbital period \citep{smith98b,patterson05,knigge06,knigge11}. 

However, these models of white dwarf and donor star evolution still require considerable refinement, since precise measurements of CV system parameters exist for fewer than 30 systems \citep{zorotovic11}. More measurements over a range of orbital periods are needed. Finding and characterising new eclipsing CVs will greatly help this process.

In this paper we present a search for new eclipsing CVs, predominantly in the northern hemisphere. Similar studies of CVs in the southern hemisphere have been conducted by \citet{coppejans14}, \citet{woudt12}, and in the preceding papers by that group. In addition, we also present a large collection of high speed observations of existing eclipsing systems, many of which already have published photometry, but of a lower time resolution. For all eclipsing systems we provide an evaluation of their suitability to be used in light curve modelling. These observations are part of an ongoing campaign to derive CV system parameters (\citealt{mcallister16} and references therein), and will form the observational basis of a future paper detailing the modelling of these systems. 


\section{Candidate Selection}\label{sec:selection}
Dwarf novae routinely enter phases of brighter emission known as outbursts, due to thermal instabilities causing changes in viscosity of the accretion disk, dumping large amounts of material onto the white dwarf in a short period of time \citep{osaki96,lasota01}. New dwarf novae are being detected in outburst at a rate of at least one per day by current transient surveys. As of October 2016, the Catalina Real-Time Transient Survey (CRTS, \citealp{drake09}) had detected almost 1400 new CVs, the All-Sky Automated Survey for SuperNovae (ASAS-SN, \citealp{shappee14}) had found over 750, and the Mobile Astronomical System of TElescope-Robots (MASTER, \citealp{lipunov10}) had discovered almost 750. The Gaia astrometric surveying satellite is also now discovering transient events, including CVs, at a rate of several per day \citep{wyrzykowski12,blagorodnova16}. On top of these, many hundreds have been identified in the Sloan Digital Sky Survey (SDSS, \citealp{gansicke09, szkody11}) and by a vast network of amateur astronomers. See for example \textit{vsnet} \citep{kato04} and the Center for Backyard Astrophysics \citep{patterson13}. 

Our aim was to find new eclipsing systems amongst the CVs found in these surveys. Starting in 2014 with a list of nearly 2000, we first rejected those systems which were not visible or too faint in quiescence to be studied with our main facility in La Palma (see Section \ref{sec:obs}). We also rejected any systems which had already been studied at sufficiently high time resolution to rule out eclipses. These include systems which have already been studied with time-resolved photometry (e.g. \citealp{coppejans14}), and systems which have been closely followed by amateur observers. For example, \textit{vsnet} have studied many systems whilst in outburst in search of superhumps\footnote{Superhumps are hump-like variations in the light curve of outbursting dwarf novae, due to a projection effect seen when the tidal torques of the secondary star entice the heated accretion disk into an elliptical geometry \citep{whitehurst88,hellier01}.} sometimes at high enough cadence to determine an orbital period and rule out eclipses. 

At the end of this process we were left with around 80 systems eligible for follow up, and this number has grown gradually since our observing campaign began. We began by observing the brighter targets, and those which were already suspected to be eclipsing according to other studies. This could be either from spectroscopy suggesting that the accretion disk is close to edge-on (e.g. \citealt{breedt14}), or from long-term light curves provided by CRTS, for example, that show regular dips or faint measurements which are often due to eclipses.


\section{Observations}\label{sec:obs}
For newly discovered CVs, initial observations were made with \textit{pt5m}, a 0.5m telescope located at the Roque de los Muchachos Observatory on La Palma \citep{hardy15}. The telescope is robotic and sits on the roof of the control-room building of the 4.2m William Herschel Telescope (WHT). We aimed to obtain light curves of new CVs spanning upwards of 5 hours, with cadences of 1-2 minutes, as this should be sufficient to discover eclipses in most systems (see Section \ref{sec:results2}). 

If eclipses were seen and a period could be measured, we then observed the systems at high time-resolution (few seconds) using either \textsc{ultracam} on the WHT \citep{dhillon07b} or \textsc{ultraspec} on the 2.4m Thai National Telescope (TNT, \citealt{dhillon14}). Both are imagers utilising frame-transfer CCDs for high cadence, low dead-time (of order 10ms) observations. \textsc{ultraspec} has a single channel, whilst \textsc{ultracam} has three channels, and can therefore observe in three filters (e.g. $u'$, $g'$, $r'$) simultaneously. 

In Section \ref{sec:results3} we also present new observations of CVs which were already known to be eclipsing. Along with the WHT and the TNT, some of these observations were conducted at other telescopes, such as the 3.6m New Technology Telescope (NTT) at La Silla or the 8.2m Very Large Telescope (VLT) at Paranal, both in Chile. A select few systems were also observed with the 9.2m South Africa Large Telescope (SALT) and its \textsc{salticam} instrument \citep{odonoghue06}. A log of all observations is presented in Table \ref{tab:obs}, an extract of which is presented here. The full version is available online.

In all observations, we performed differential photometry using bright stars near to the target as comparison stars. The fluxes of the target and comparison stars were extracted using variable aperture photometry with the \textsc{ultracam} pipeline reduction software \citep{dhillon07b}. For particularly faint targets, the optimal photometric extraction technique of \citet{naylor98} was used. We calculated the magnitudes of the comparison stars by extracting their full flux using large, fixed apertures, then flux calibrated these using measured extinction coefficients and instrumental zeropoints. The flux of the target was then calibrated using the calculated magnitude of the comparison star.  

We supply average out-of-eclipse magnitudes in Table \ref{tab:obs} primarily as a guide to show when systems are in outburst, or to demonstrate which systems show strong variability. However, since the zeropoints measurements (based on observations of photometric standard stars) are made less frequently than the science observations, these magnitudes should always be considered estimates. Systematic uncertainties of 0.2 magnitudes or more could be introduced by variable extinction or high altitude clouds. The atmosphere tends to be stable and clear in La Palma and in Chile, but in Thailand there is frequently low altitude haze, as well as thin, high altitude clouds, which strongly affect the extinction coefficients and zeropoints. Therefore all measured magnitudes from the TNT are to be treated with extra caution. 

Most observations made with \textsc{ultraspec} at the TNT were conducted in the standard SDSS filters, but sometimes we used a non-standard \textit{KG5} filter. This filter encompasses the SDSS $u'$, $g'$ and $r'$ pass bands, and can therefore obtain better signal-to-noise ratios on fainter targets than individual SDSS filters. The obvious disadvantage here is that it is not easy to convert the measured counts into a calibrated flux. We describe our solution to this problem in Appendix \ref{sec:kg5calibration}. 

\begin{table*}
\centering\footnotesize
\begin{minipage}{1.\textwidth}
\caption{Journal of Observations. Some object names have been shortened. Start(MJD) is the start time of each observing run, given in MJD(UTC). Mid-eclipse times are given in BMJD(TDB) and the number in parentheses is the uncertainty in the last digit. Note that \textit{pt5m} mid-eclipses times are estimated using a Gaussian fit and will suffer from additional systematic uncertainty. $T_{exp}$ is the exposure time in seconds and is supplied as $T_{blue}/T_{green}/T_{red}$ for the three beams of \textsc{ultracam}. $\Delta T$ is the duration of the observing run in minutes. Mag. is the estimated out-of-eclipse magnitude. This is an extract showing only one object. The full table containing all 74 objects is available online.}
\label{tab:obs}
\begin{tabular}{p{3.7cm} p{1.7cm} p{2.1cm} p{1.3cm} p{0.8cm} p{1.7cm} p{1.1cm} p{1.8cm}}
\textbf{Object} & \textbf{Start (MJD)} & \textbf{Mid-eclipse time (BMJD)} & \textbf{T$_{exp}$ (s)} & \textbf{$\Delta $T (min)} & \textbf{Tel./Inst.}  & \textbf{Filter} & \textbf{Mag.} \\ \hline
  1RXS J180834.7+101041 & 55316.33464 & 55316.3613(3) & 6/2/2 & 141 & \textsc{ntt/ucam} & $u'/g'/r'$ & 16.9/16.9/16.6 \\
       & & 55316.4310(3) & & & & & \\
    & 55334.37156 & 55334.4299(3) & 10/2/2 & 101 & \textsc{ntt/ucam} & $u'/g'/r'$ & 17.0/17.0/16.8 \\
 \hline  
\end{tabular}
\end{minipage}
\end{table*}

\section{Results: New Eclipsing Systems}\label{sec:results1}

We present 13 CV systems which were either unknown to be eclipsing, were suspected eclipsers, or whose eclipsing nature has only recently been published. Most discoveries were made with \textit{pt5m} observations over several nights, in the Johnson \textit{V} filter. When eclipses were found, mid-eclipse times were measured by fitting a Gaussian to the eclipse profile, this being the simplest and fastest method. The quoted uncertainties in these mid-eclipse times are the statistical errors given by the Gaussian fit. We expect the true uncertainty to be up to an order of magnitude larger and dominated by the systematic error inherent in treating these complex eclipse shapes as Gaussians.

Ephemerides were obtained by a linear fit to the eclipse times. When fitting for an ephemeris, we always selected the orbital cycle zero-point ($T_{0}$) which minimised the covariance between $P$ and $T_{0}$. Occasionally we omitted some mid-eclipse times from the ephemeris fit, if for example the system was in outburst, there were orbital cycle ambiguities, or the times had very large errors (often the case with \textit{pt5m} observations). We conducted additional checks on period measurements using phase-dispersion-minimisation (PDM, \citealt{stellingwerf78}) of all the available, normalised light curve data. The resulting PDM periodogram shows troughs at potential orbital periods, and visual inspection of the light curves phase-folded on these potential periods was sufficient to confirm or reject reliable periods. In all cases we had sufficient eclipse and out-of-eclipse data to eliminate period aliases. The derived ephemerides of all systems are presented in Table \ref{tab:ephem}. 

Once a reliable period had been derived, we then attempted to investigate the eclipse structure by observing the system at high time resolution. Details of these observations are included in Table \ref{tab:obs}. In high cadence observations, the Gaussian fit to an eclipse profile is unsuitable, because the different components of the system are clearly visible. In this case, we use the mid-point of the white dwarf eclipse as the mid-eclipse time, determined from measurement of white dwarf mid-ingress and mid-egress. We supply the mid-eclipse times in Table \ref{tab:obs} in case this information is useful to future orbital period variability studies (e.g. \citealt{parsons10,bours16}), but remind the reader that eclipse times measured by \textit{pt5m} suffer from the systematic uncertainty introduced by treating the eclipses as Gaussians. 

\begin{figure*}
 \includegraphics[width=1.29\textwidth,angle=270]{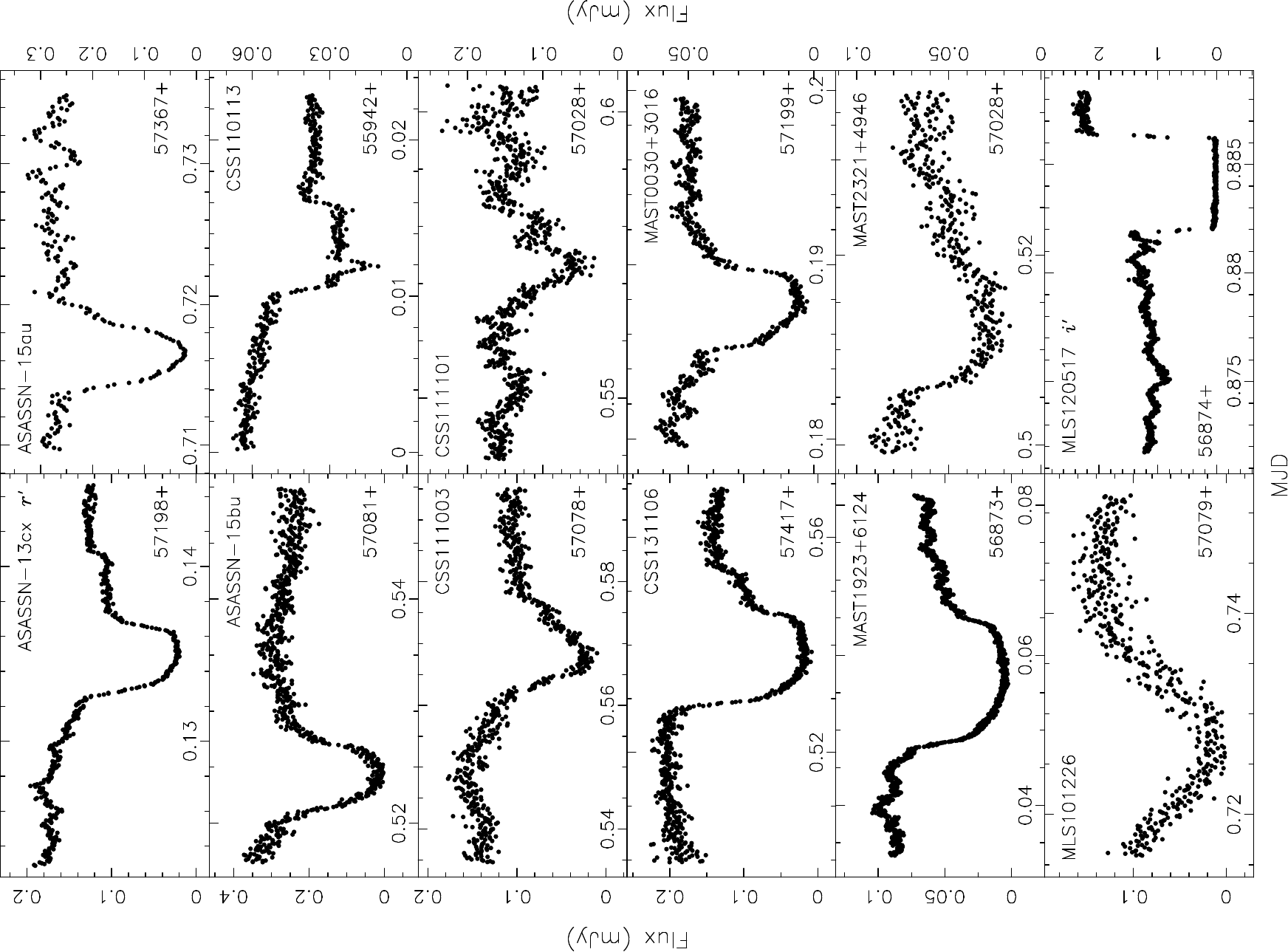}
 \caption{High-speed eclipses of newly discovered eclipsing systems, observed with \textsc{ultracam} and \textsc{ultraspec}. The instrument and filter used in each observation is given in Table \ref{tab:obs}. In the case of \textsc{ultracam} observations, we plot the $g'$-band light curve unless otherwise indicated in the plot. See text for discussion of individual light curve morphology. In these light curves and all subsequent densely-sampled light curves, error bars have been omitted for clarity.}
 \label{fig:comb1}
\end{figure*}

High-speed eclipse light curves for each new object are shown in Figure \ref{fig:comb1}. We briefly discuss each system below, and comment on the light curve morphology. Our judgement on the suitability for light curve modelling of each object is included in Table \ref{tab:ephem}. 

\vspace*{0.1cm}\noindent
\textbf{ASASSN-13cx:} Discovered by ASAS-SN\footnote{References for ASAS-SN transients are not always available. See the ASAS-SN webpage at \url{http://www.astronomy.ohio-state.edu/~assassin/transients.html}} with an outburst to $V=15.5$ magnitudes, from $V=18$ as recorded in the NOMAD catalogue \citep{zacharias04}. ASASSN-13cx also showed previous outbursts in the Catalina Sky Survey (CSS, part of CRTS), as well as fainter measurements indicating eclipses. During an outburst in 2014, \textit{vsnet} observers were able to confirm the eclipsing nature and estimate a period \citep{kato15}. The high-speed eclipse structure shows clearly separated egress for the white dwarf and bright spot. The ingress of the bright spot is less clear. 

\vspace*{0.1cm}\noindent
\textbf{ASASSN-15au:} Discovered by ASAS-SN during an outburst to $V=15.3$ magnitudes, from a quiescent SDSS magnitude of $g=17.8$. It also shows variability of up to 3 magnitudes in archival CSS and Mount Lemmon Survey (MLS, part of CRTS) light curves, including outburst and eclipse-like faint measurements. We found multiple eclipses in \textit{pt5m} light curves, and conducted high speed observations with \textsc{ultraspec}. These observations show weak features that may be the bright spot ingress and egress, but further study is required.

\vspace*{0.1cm}\noindent
\textbf{ASASSN-15bu:} Shows previous outbursts in the CSS light curve, as well as multiple dips signalling eclipses. \textit{vsnet} (alert 18228) folded the CSS data to find a coherent orbital period of 110 minutes. We confirmed the eclipsing nature with 4 \textsc{ultraspec} observations. No clear bright spot features are visible in the light curve, but a higher signal-to-noise light curve is needed to be certain that no features exist.

\vspace*{0.1cm}\noindent
\textbf{CSS110113:043112-031452:} Discovered via a super-outburst in CRTS\footnote{References are often not available for CRTS discoveries, though all transients are available online at \url{http://crts.caltech.edu/} and the discovery of CVs in particular is discussed in \citet{drake14}.}, from a quiescent magnitude of $V=$ 19-20 to $V=15$. The \textit{vsnet} collaboration discovered eclipse signals and estimated an orbital period of 1.58 hours (alerts 12614, 12661), but no detailed studies have yet been published. We confirmed the eclipsing nature with observations of 12 eclipses in varying conditions. Sometimes the bright spot signal is hidden in noisy data or blended with the white dwarf, but throughout 2012 the ingress and egress are clearly separable. In this system it appears that the white dwarf begins exiting eclipse before the bright spot has finished entering eclipse.

\vspace*{0.1cm}\noindent
\textbf{CSS111003:054558+022106:} Initially discovered as a planetary nebula and also known as Te 11 \citep{jacoby10}. It was shown to be eclipsing, with an orbital period of 0.12 days, before its reclassification as a CV \citep{miszalski11}. \citet{drake14} took spectra of the object, which showed very strong double-peaked Balmer emission lines, as well as [OIII] emission presumably associated with the nebula. Recently, \citet{miszalski16} presented a detailed study of the eclipses, using high cadence observations to extract its approximate system parameters using the light curve modelling technique. They also proposed that the nebula surrounding the system may in fact be the remnants of a nova eruption 1500 years ago. Despite being a faint source, a few of our eclipses showed well-separated bright spot and white dwarf features.

\vspace*{0.1cm}\noindent
\textbf{CSS111101:233003+303301:} Discovered via an outburst of around 2 magnitudes from the quiescent level observed in SDSS ($g$=18.9-19.5). The system shows numerous faint dips in its CSS light curve, leading CRTS to suspect this system might be eclipsing. We confirmed the presence of eclipses and conducted high speed observations with \textsc{ultraspec}. The light curve is dominated by strong flickering, and the eclipse is shallow, suggesting a possible grazing eclipse.

\vspace*{0.1cm}\noindent
\textbf{CSS131106:052412+004148:} Has a quiescent magnitude in SDSS of $g'$ = 18.3. The CSS light curve shows multiple outbursts in recent years, as well as occasional faint measurements, prompting us to search for and subsequently discover eclipses with \textit{pt5m}. The high-speed eclipses show separation between the white dwarf egress and bright spot egress, although separation of the ingresses is less clear.

\noindent
\textbf{MASTER OT J003059.39+301634.3:} Discovered by MASTER \citep{shurpakov14a} with an outburst to 16.1 magnitudes (unfiltered). It is listed as $g$ = 18.1 in SDSS \citep{ahn12}, though we suspect this magnitude may have been observed during a brightened state, as the USNO-B1.0 catalogue \citep{monet03} and our own observations suggest it is fainter than 19th magnitude in \textit{B} and \textit{V}. We discovered eclipses and obtained several high cadence light curves, which display a clear white dwarf eclipse. There is also a suspected bright spot component, but one which eclipses slowly and follows directly after the white dwarf eclipse. It is also unclear exactly where the transition from bright spot egress to full out-of-eclipse flux occurs due to flickering.

\vspace*{0.1cm}\noindent
\textbf{MASTER OT J192328.22+612413.5:} Discovered in outburst at 17.5 magnitudes (unfiltered, \citealt{balanutsa14b}). It was suspected to be eclipsing, and this was confirmed by observations at the Vatican Observatory \citep{garnavich14}. We independently confirmed its eclipsing nature with \textit{pt5m}. Very recently, \citet{kennedy16} published further photometric and spectroscopic studies of this system, showing that it shares similarities with the SW Sex class of CVs during outburst. Similar to those of \citet{kennedy16}, our quiescent eclipse observations show clearly the white dwarf ingress and egress, but no bright spot is present.

\vspace*{0.1cm}\noindent
\textbf{MASTER OT J232100.42+494614.0:} Discovered via an outburst to 15.5 magnitudes \citep{shumkov14b} from $V=19$ magnitudes in quiescence. We found MASTER OT J232100.42+494614.0 to have a long period of over 5 hours. The high-speed eclipse is noisy, and although there are signs of bright spot features, better data is needed before we can judge conclusively the potential for modelling this system. This is a priority, as a system with such a long period would be particularly useful to the CV evolution studies, since measurements of system parameters at long periods remain scarce \citep{zorotovic11}.

\vspace*{0.1cm}\noindent
\textbf{MLS101226:072033+172437:} Discovered by CRTS \citep{drake09} with a MLS light curve showing obvious variability, which was interpreted as eclipses \citep{drake14}. However, no detailed studies have, to our knowledge, been published. We confirmed its eclipsing nature with \textit{pt5m} and also observed MLS101226:072033+172437 with \textsc{ultraspec}. Unfortunately observations began a little late, missing the beginning of eclipse ingress. The eclipse is smooth and dominated by the disk feature, showing no strong white dwarf or bright spot components.

\vspace*{0.1cm}\noindent
\textbf{MLS120517:152507-032655:} Flagged as a possible MLS CV candidate in 2012, after also showing substantial variability in the CSS data as far back as 2007. The source has a quiescent magnitude of around \textit{V}=19, but appears to have been brightening for several years and is now seen at \textit{V}$\sim$16. The object, also known as 1RXS J152506.9-032647, is detected in X-rays and the UV \citep{voges99}. It was proposed to show deep eclipses by \citet{drake14}, and although no further studies have been published on MLS120517:152507-032655, it has been observed by amateur astronomers\footnote{\url{http://var2.astro.cz/EN/obslog.php}} and does indeed show eclipses. 

We made several observations of the system with \textit{pt5m} and \textsc{ultracam}, all of which show the system to be brighter after eclipse than before. This behaviour is typical for a polar CV \citep{hellier01}, when the eclipse signal is primarily due to the obscuration of a hot spot on the surface of the white dwarf. Most of the light in the system comes from this hot spot where the accretion stream impacts the surface, and from the stream itself that trails between this spot and the L1 point on the donor star. The projection effect of this stream causes the system to appear bright just after the eclipse, when the stream is being viewed more side-on than before the eclipse. As a polar, the light curve is not suitable for modelling to determine the system parameters.

\begin{figure}
 \includegraphics[width=4.67cm,angle=270]{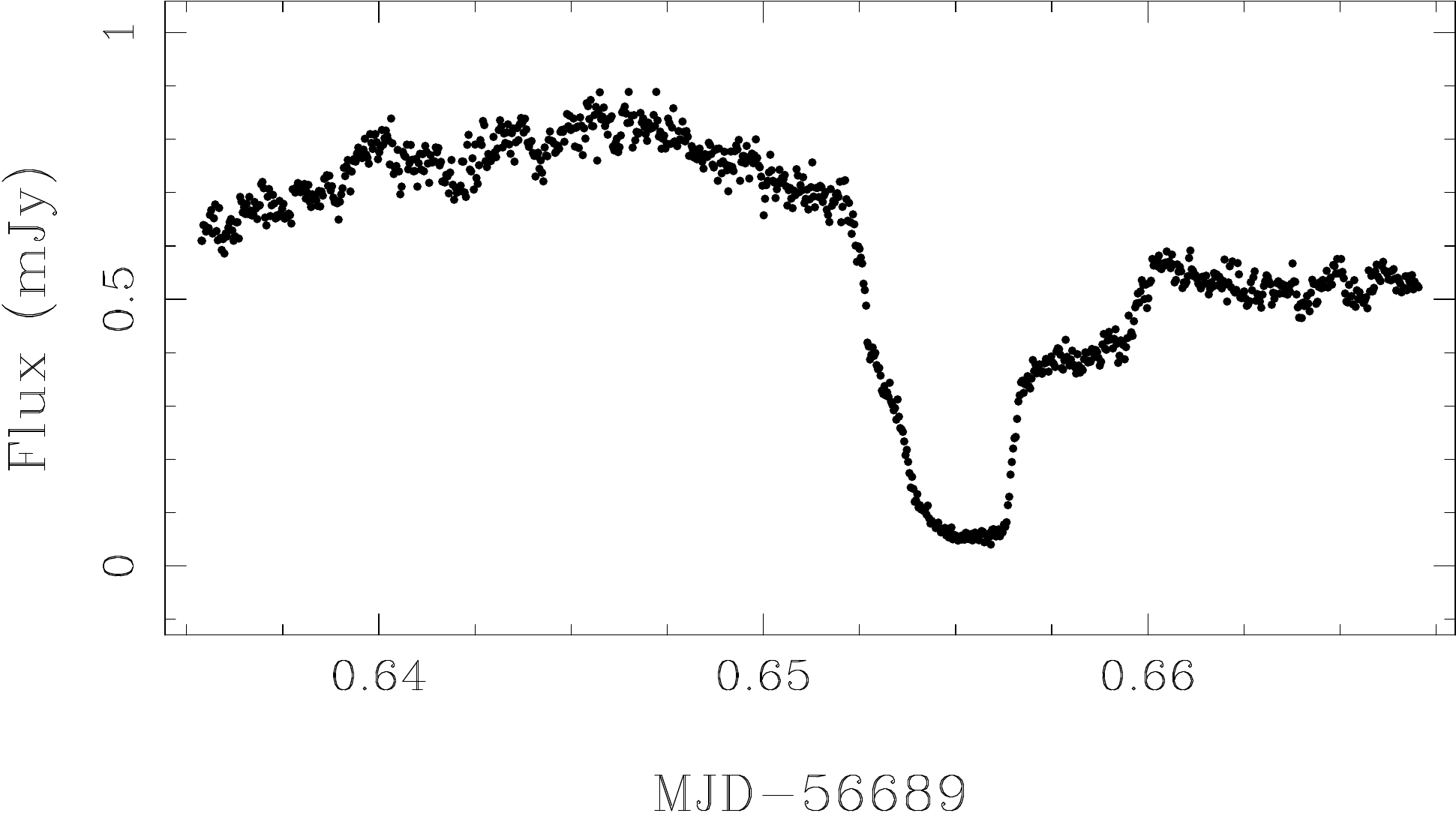}
 \caption{SSS130413:094551-194402 eclipse in the \textit{KG5} filter, showing clearly separated white dwarf and bright spot features.}
 \label{fig:sss130413}
\end{figure}

\vspace*{0.1cm}\noindent
\textbf{SSS130413:094551-194402:} Long suspected of being a variable star, this system was first known as NSV4618 \citep{kukarkin81}. The Siding Springs Survey (SSS, part of CRTS) light curve also shows numerous faint measurements associated with eclipses. The \textit{vsnet} collaboration reported eclipses and estimated an orbital period of 1.6 hours (alert 15615). We confirmed the eclipsing nature and Figure \ref{fig:sss130413} shows the eclipse structure, which has clear white dwarf and bright spot features.

\section{Results: Systems Not Showing Eclipses}\label{sec:results2}
The vast majority ($>$80\%) of CV systems have orbital periods less than 5 hours \citep{ritter03,gansicke09}. In addition, the eclipse duration tends to scale with the orbital period, such that longer period systems have longer eclipses. This means that a 5-hour observation showing no sign of eclipses would suggest the system either does not eclipse, or has a period of at least 5.5 hours. We can therefore say that any systems we have observed continuously for 5 or more hours that show no signs of eclipses are unlikely to be eclipsing. 

Sometimes shorter observations can also help to rule out eclipses, if for example the system shows periodic variability. This could be due to superhumps, the projection effect of the orbiting bright spot, or the oblation of the secondary star as it fills its Roche lobe. These modulations occur on or close to the orbital period, thus any observation covering at least one cycle of such modulations without showing an eclipse is sufficient to rule out the presence of eclipses in the system.  

In Table \ref{tab:noneclipsers} we present 27 systems which did not show eclipses. We discuss and show light curves for only the two most interesting systems. Light curves of the other systems are given in Figures 1-25 in the supplementary online materials. Details of the observations are given in Table \ref{tab:obs}.

\begin{table*}
\centering\footnotesize
\begin{minipage}{1.\textwidth}
\caption{Summary of CVs not showing eclipses. Magnitudes given without filters are MASTER unfiltered magnitudes. SH = superhumps.}
\label{tab:noneclipsers}
\begin{tabular}{p{3.5cm} p{1.1cm} p{1.2cm} p{0.5cm} p{1.0cm} p{3.0cm} p{4.9cm}}
\textbf{Object} & \textbf{Outburst Mag.} & \textbf{Quiescent Mag.} & \textbf{SH?} & \textbf{Approx. Period (mins)} & \textbf{Comments} & \textbf{References} \\ \hline
  ASASSN-14cl & $V$=$10.7$ & $g$=$18.8$ & Y & 85 &  & \citealp{stanek14a,kato15} \\
  ASASSN-14ds & $V$=$15.7$ & $V$=$16.8$ & N &  & Known X-ray source & \citealp{holoien14,masetti14} \\
  ASASSN-14gl & $V$=$15.7$ & $g$=$18.8$ & N &  & Flickering  &  \\
  ASASSN-14gu & $V$=$15.2$ & $V$=$17.4$ & N & 170 & Periodic modulation &  \\
  ASASSN-14hk & $V$=$14.7$ & $B$=$20.8$ & Y & 90 &  & \citealp{stanek14b,kato15} \\
  ASASSN-14mv & $V$=$12$ & $B$=$17.3$ & Y & 40 & Candidate AM CVn & \citealp{denisenko14c} \\
  ASASSN-15ni & $V$=$12.9$ & Unknown & N &  & Flickering & \citealp{dong15,berardi15} \\
  CSS090219:044027+023301 & $V$=$16.5$ & $V$=$18.5$ & N & & Flickering & \citealp{thorstensen12} \\
  CSS091116:232551-014024 & $V$=$15.9$ & $V$=$18.5$ & N & & Flickering & \citealp{djorgovski11} \\
  CSS100508:085604+322109 & $V$=$16.5$ & $g$=$19.6$ & N & & Flickering & \citealp{kato12} \\
  CSS100520:214426+222024 & $V$=$14.7$ & $g$=$17.6$ & N & &  &  \\
  CSS110114:091246-034916 & $V$=$16.0$ & $R$=$18$ & N & & Flickering & \citealp{thorstensen12} \\
  CSS110226:112510+231036 & $V$=$16.0$ & $R$=$19$ & N & & Double-peaked lines & \citealp{kato12,breedt14} \\
  CSS130906:064726+491542 & $V$=$14$ & $V$=$17$ & N & & Also seen by MASTER & \citealp{tiurina13,thorstensen16b} \\
  CSS140402:173048+554518 & $V$=$14.5$ & $g$=$20.1$ & Y & 39 & Known AM CVn type & \citealp{carter14,kato15} \\
  CSS140901:013309+133234 & $V$=$15.3$ & $g$=$21.3$ & Y & 150 & A.K.A ASASSN-14gk & \citealp{kaur14} \\
  CSS141005:023428-045431 & $V$=$14.5$ & $g$=$16.3$ & Y & 83 & A.K.A ASASSN-14dx & \citealp{kaur14,thorstensen16b} \\
  CSS141117:030930+263804 & $V$=$12$ & $g$=1$8.9$ & N &  &  & \citealp{kato15} \\
  Gaia15aan & $G$=$13$ & $g$=$19.7$ & N &  & A.K.A ASASSN-14mo  & \citealp{carter13,campbell15a} \\
  MAST034045.31+471632.2 & $15.6$ & $g$=$18$ & N &  &  & \citealp{denisenko13c} \\
  MAST041923.57+653004.3 & $13.9$ & $V$=$17.9$ & N &  & Known X-ray source & \citealp{balanutsa13} \\
  MAST171921.40+640309.8 & $15.6$ & $g$=$21.3$ & Y & 80 &  & \citealp{balanutsa14a} \\
  MAST194955.17+455349.6 & $16.4$ & $g$=$19.7$ & N &  & A.K.A KIC 9358280 & \citealp{greiss12,denisenko14a} \\
  MAST201121.95+565531.1 & $16.5$ & $R>20$ & N &  &  & \citealp{shurpakov14b} \\
  MAST202157.69+212919.4 & $16.3$ & $V$=$20$ & N &  & Known X-ray source & \citealp{voges00,denisenko14b} \\
  MAST203421.90+120656.9 & $17.4$ & $B$=$20.4$ & N &  &  & \citealp{yecheistov14} \\
  MAST210316.39+314913.6 & $13.9$ & $B$=$19.4$ & N &  &  & \citealp{shumkov14a} \\
 \hline  
\end{tabular}
\end{minipage}
\end{table*}

\begin{figure}
 \includegraphics[width=2.5cm,angle=270]{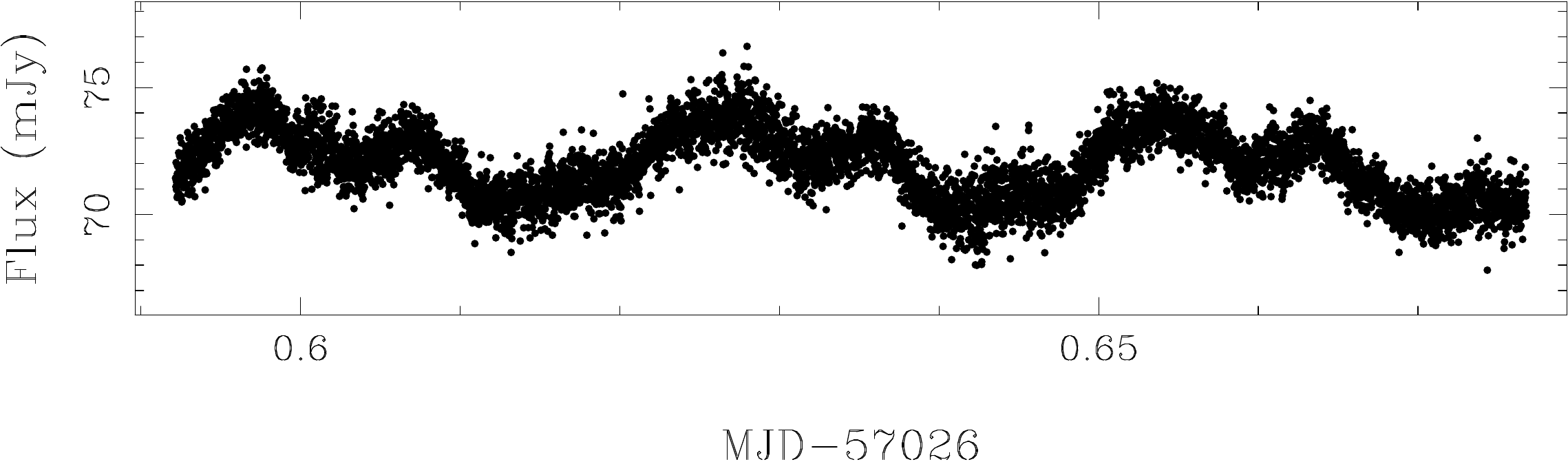}
 \caption{ASASSN-14mv light curve in the $g'$ filter. The periodic modulation is associated with superhumps, and has a period of approximately 40 minutes. Error bars have been omitted for clarity.}
 \label{fig:asassn14mv}
\end{figure}

\vspace*{0.1cm}\noindent
\textbf{ASASSN-14mv:} We observed ASASSN-14mv for two hours during outburst with \textsc{ultraspec}. The light curve is shown in Figure \ref{fig:asassn14mv}, showing strong periodic variability associated with superhumps, suggesting an orbital period of the order of 40 minutes. This, along with other observations including \textit{vsnet} alerts 18124 and 18160 (which links to an amateur spectrum taken by M. Potter showing helium absorption lines during outburst), suggests that ASASSN-14mv is likely to be an AM CVn system \citep{nelemans05}. 

\begin{figure}
 \includegraphics[width=2.5cm,angle=270]{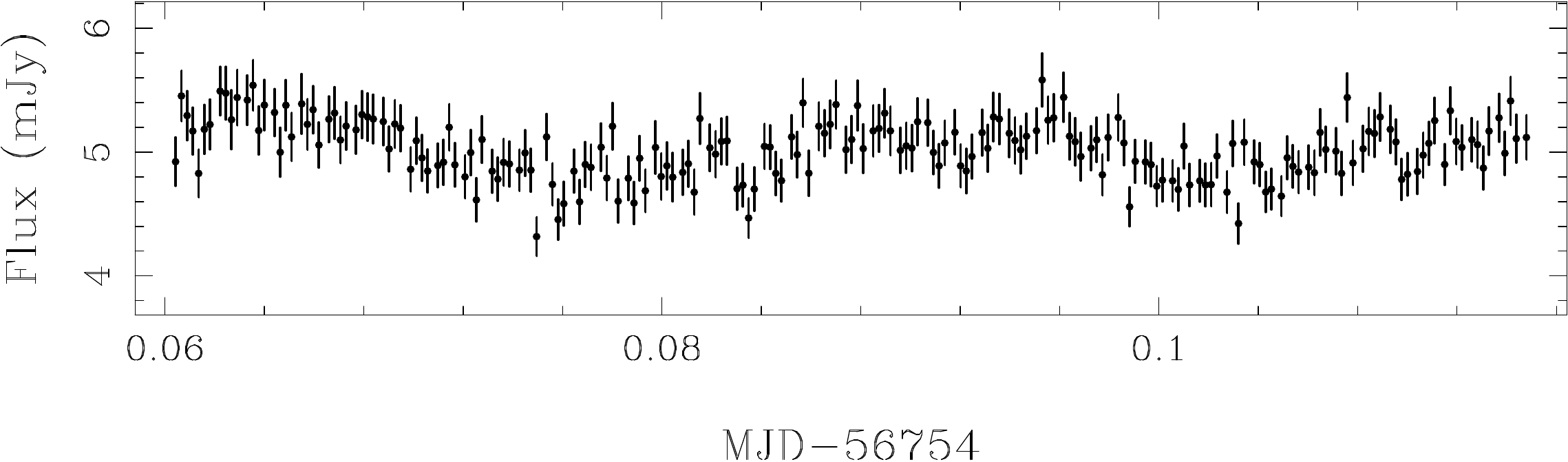}
 \caption{CSS140402:173048+554518 $V$-band light curve showing superhumps, observed during outburst.}
 \label{fig:sdss1730}
\end{figure}

\vspace*{0.1cm}\noindent
\textbf{CSS140402:173048+554518:} This object, also known as SDSS J173047.59+554518.5 is a confirmed AM CVn system, with an orbital period of 35 minutes \citep{carter14}. Although not likely to be eclipsing due to it being a relatively face-on system, we observed CSS140402:173048+554518 during outburst with automated follow-up from the initial CRTS transient trigger \citep{hardy15}. A 1.3 hour light curve is shown in Figure \ref{fig:sdss1730}, showing superhump behaviour with a period of 39 minutes, close to the orbital period. This is also similar to the 35 minutes measured by \citet{kato15}. Unfortunately, because of the large uncertainty in the spectroscopic orbital period, and since the accretion disk is made predominantly of helium and not hydrogen, the superhump excess cannot be reliably used to measure a mass ratio \citep{pearson07,kato14b}, as can be done for classical SU UMa systems \citep{patterson05}. 


\section{Results: Previously Known Eclipsing Systems}\label{sec:results3}
We present high cadence light curves of CV systems which are already known to be eclipsing. Studies at such high time resolution do not appear elsewhere in the literature, and therefore many of these systems have never been studied before in such detail. Some systems have already been studied in a time-resolved manner, but our additional observations should enable further refinement of existing measurements of system parameters. Details of the observations are given in Table \ref{tab:obs}, and mid-eclipse times are provided whenever an eclipse was observed. In those systems with resolved eclipse structure, we measure the mid-eclipse time as the mid point between the white dwarf mid-ingress and mid-egress, when these features are visible, rather than using a Gaussian fitting procedure. If the white dwarf ingress or egress is not clear but there is visible structure, we use the ingress and egress of the most consistent features, and account for this in the uncertainties. We derived updated ephemerides for most systems, and these are given in Table \ref{tab:ephem}.

\begin{figure*}
 \includegraphics[width=1.29\textwidth,angle=270]{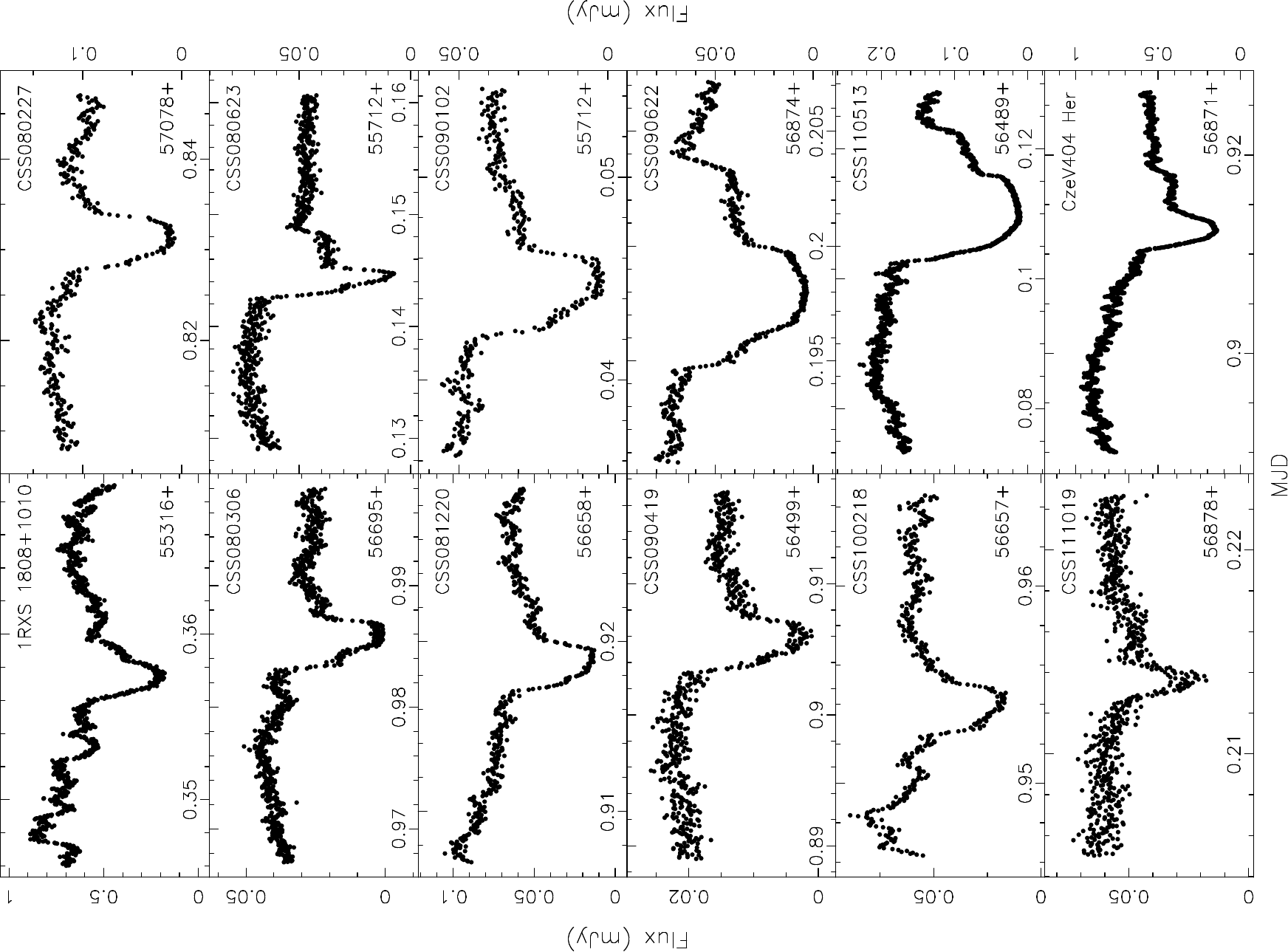}
 \caption{High-speed eclipses of known eclipsing systems, observed with \textsc{ultracam}, \textsc{ultraspec} and \textsc{salticam}. The instrument and filter used in each observation is given in Table \ref{tab:obs}. The $g'$-band light curve is shown for all \textsc{ultracam} observations. See text for discussion of individual light curve morphology.}
 \label{fig:comb2}
\end{figure*}

\begin{figure*}
 \includegraphics[width=1.29\textwidth,angle=270]{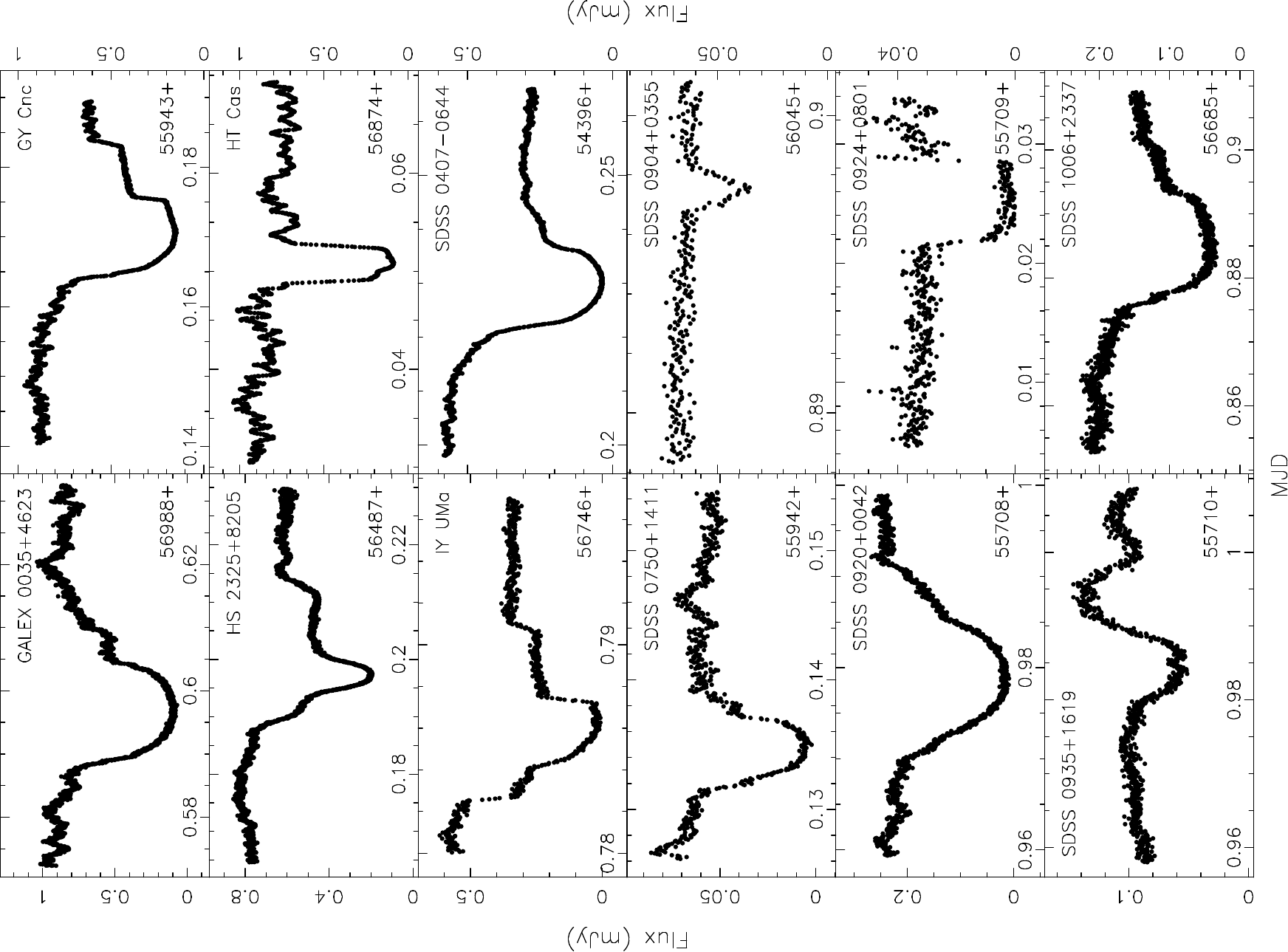}
 \caption{High-speed eclipses of known eclipsing systems, observed with \textsc{ultracam} and \textsc{ultraspec}. The instrument and filter used in each observation is given in Table \ref{tab:obs}. The $g'$-band light curve is shown for all \textsc{ultracam} observations. See text for discussion of individual light curve morphology.}
 \label{fig:comb3}
\end{figure*}

\begin{figure}
 \includegraphics[width=1.27\textwidth,angle=270]{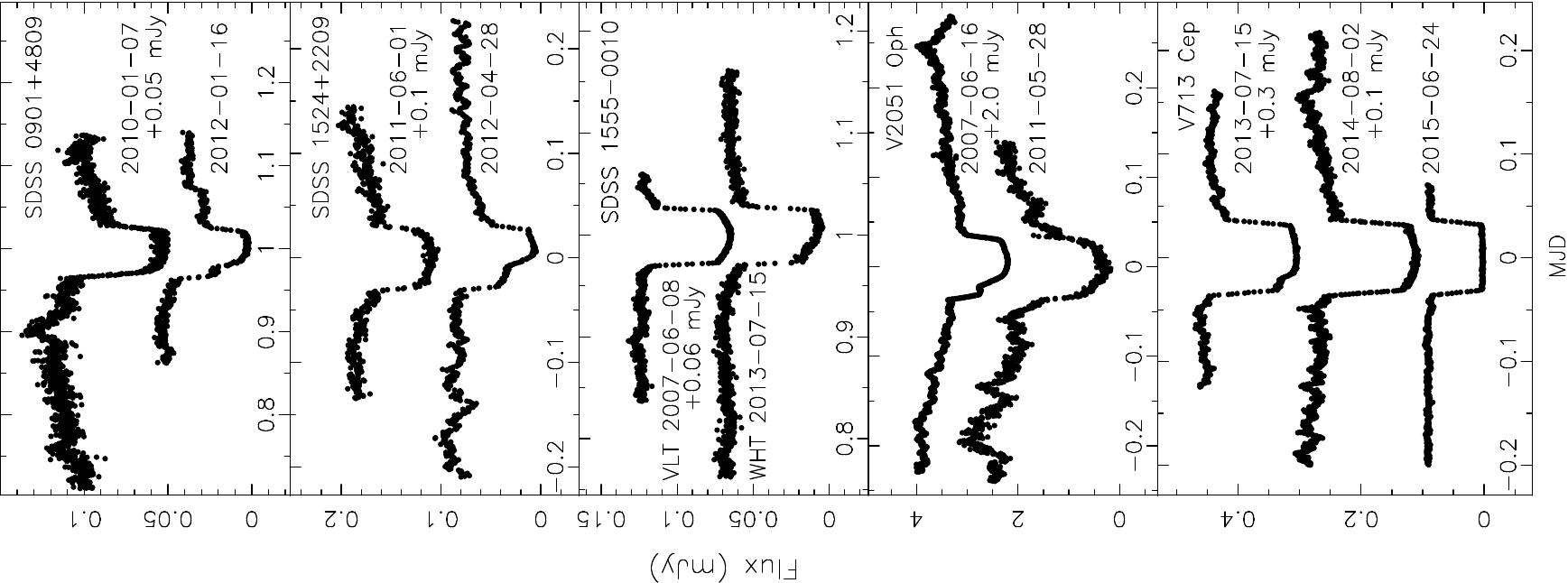}
 \caption{Phase-folded high-speed eclipses of known eclipsing systems showing variable bright spot features. Observations were conducted in $g'$ with \textsc{ultracam}. Individual eclipses have been offset vertically in each panel by the specified amount. See text for discussion of individual light curve morphology.}
 \label{fig:phaseFolds}
\end{figure}

\begin{figure}
 \includegraphics[width=1.07\textwidth,angle=270]{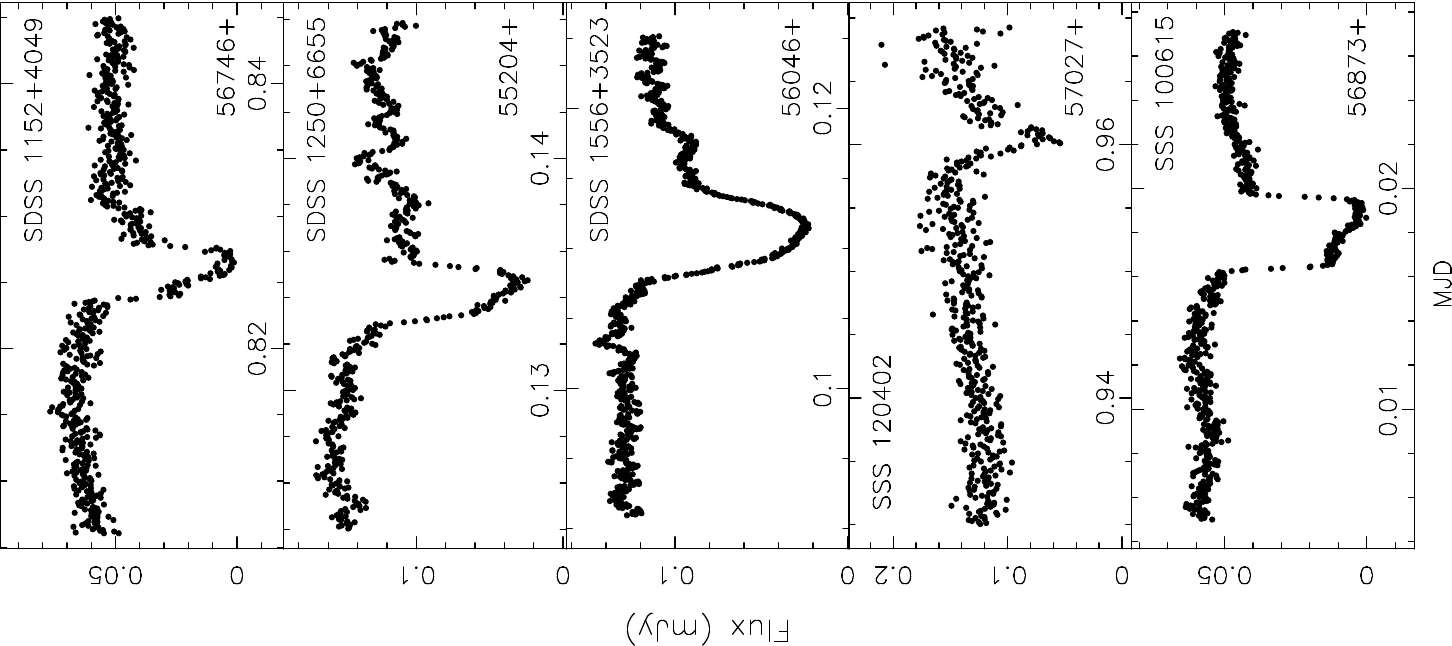}
 \caption{High-speed eclipses of known eclipsing systems, observed with \textsc{ultracam} and \textsc{ultraspec}. The instrument and filter used in each observation is given in Table \ref{tab:obs}. The $g'$-band light curve is shown for all \textsc{ultracam} observations. See text for discussion of individual light curve morphology.}
 \label{fig:comb4}
\end{figure}

We provide discovery references and discuss the eclipse structure of each system below. Whether or not each system is suitable for light curve modelling to determine the system parameters is included in Table \ref{tab:ephem}. Light curves for each system are shown in Figures \ref{fig:comb2}, \ref{fig:comb3}, \ref{fig:phaseFolds} and \ref{fig:comb4}.

\vspace*{0.1cm}\noindent
\textbf{1RXS J180834.7+101041:} Originally classified as a polar \citep{denisenko08}, spectra later showed that it has an accretion disk \citep{bikmaev08}. Eclipses have been studied previously \citep{yakin10,southworth11}, but attempts to measure the system parameters have had limited success because the white dwarf itself might not be eclipsed. We found strong flickering and only weak eclipse structure, possibly associated soley with the accretion disk and bright spot but not the white dwarf. 

\vspace*{0.1cm}\noindent
\textbf{CSS080227:112634-100210:} Discovered by CRTS \citep{drake08,djorgovski08,drake09}. The CSS light curve shows numerous dips due to eclipses, and these were later observed with time-resolved photometry \citep{woudt10}. The eclipse displays a separated bright spot ingress and egress, although the egress is less clear due to flickering. Other eclipses show similar features, and with additional observations this system may be suitable for light curve modelling.

\vspace*{0.1cm}\noindent
\textbf{CSS080306:082655-000733:} Discovered by CRTS. Photometry by \citet{woudt12} found eclipses and persistent negative superhumps, even in quiescence. Clear bright spot features are present.

\vspace*{0.1cm}\noindent
\textbf{CSS080623:140454-102702:} Time-resolved photometry revealed eclipses \citep{woudt12} and spectroscopy found both the white dwarf and the secondary star to be present in the spectrum \citep{breedt14}. This system shows classic CV eclipse structure with resolved white dwarf and bright spot features.

\vspace*{0.1cm}\noindent
\textbf{CSS081220:011614+092216:} CSS found faint measurements signalling eclipses. Time-resolved photometry confirmed deep eclipses \citep{coppejans14} and found an orbital period of around 1.6 hours. Most of our observations found either the bright spot ingress blended with that of the white dwarf, or the bright spot egress lost in flickering. Occasionally the system was found in a low-flickering state, with the bright spot features visible.

\vspace*{0.1cm}\noindent
\textbf{CSS090102:132536+210037:} Numerous eclipse-like dips are visible in the CSS light curve. Also known as SDSS J132536.05+210036.7, it was flagged as a dwarf nova in retrospective data mining of SDSS objects by \citet{wils10} because of its strong blue colour ($u'$-$g'$ of -1 magnitudes) an odd $g'$-$r'$ colour (almost 3 magnitudes), which must have been affected by an eclipse. \citet{southworth15} studied the system in detail and confirmed its eclipsing nature. Most of our observations show resolved white dwarf and bright spot features.

\vspace*{0.1cm}\noindent
\textbf{CSS090419:162620-125557:} This system is relatively faint in quiescence ($r'$=20.4) and all observations are noisy. Follow-up photometry identified eclipses \citep{woudt12}, and we have observed these at high speed. The light curve displays clear white dwarf and bright spot eclipses.

\vspace*{0.1cm}\noindent
\textbf{CSS090622:215636+193242:} CSS light curve showed many dips suggestive of eclipses. The eclipsing nature was confirmed by \citet{drake14} and studied later by \citet{thorstensen16b}. The high speed eclipse shows clear white dwarf and bright spot features.

\vspace*{0.1cm}\noindent
\textbf{CSS100218:043829+004016:} Found to be eclipsing by \citet{coppejans14}, who derived a period of 1.5 hours. The system is faint, with strong flickering and a weak bright spot.

\vspace*{0.1cm}\noindent
\textbf{CSS110513:210846-035031:} Shows faint measurements in the long-term light curve, and \citet{coppejans14} confirmed eclipses with an orbital period of 3.8 hours. Unfortunately most of our observations found the system in outburst, showing an eclipse dominated by the disk. However, one high quality quiescent eclipse was observed, which showed separated white dwarf and bright spot features.

\vspace*{0.1cm}\noindent
\textbf{CSS111019:233313-155744:} Found with a short period of around 62 minutes \citep{woudt11}. The eclipses are shallow, and the bright spot eclipse appears to arrive very late - almost after the white dwarf has come out of eclipse. This behaviour is similar to that seen in the AM CVn system SDSS J092638.71+362402.4 \citep{marsh07,copperwheat11}, where the secondary star is so bright and small in size (e.g. a white dwarf), that the eclipses are shallow and the bright spot ingress occurs after the primary white dwarf egress. The short period also fits with a possible AM CVn classification. CSS111019:233313-155744 would be an interesting object for further study.

\vspace*{0.1cm}\noindent
\textbf{CzeV404 Her:} Discovered in outburst by \citet{cagas14}, and further studied by \citet{bakowska14}, who observed multiple super-outbursts and estimated an orbital period within the period gap. The bright spot and white dwarf ingresses appear to be blended, although there may be a hint of separation.

\vspace*{0.1cm}\noindent
\textbf{GALEX J003535.7+462353:} Long-term variability in SuperWASP data \citep{butters10} prompted time-resolved photometry which discovered dwarf novae outbursts and eclipses \citep{wils11}. Most of our observations showed little or no sign of bright spot features. Some distinct structure is visible in the eclipse light curve, but more observations are needed. 

\vspace*{0.1cm}\noindent
\textbf{GY Cnc:} Already has spectroscopically derived masses \citep{thorstensen00} and published \textsc{ultracam} data \citep{feline05}. We found signs of a varying orbital period for this system, although since the system appears to have a variable disk radius, we cannot be certain we are always measuring the exact mid-eclipse time of the same features. The eclipse sometimes has clear bright spot features, though often we found these components blended with the white dwarf.

\vspace*{0.1cm}\noindent
\textbf{HS 2325+8205:} Already studied extensively \citep{pyrzas12}, and found to be a candidate eclipsing Z Cam type CV, with a long orbital period. The eclipse appears to be well suited for modelling with clearly visible white dwarf and bright spot features.

\vspace*{0.1cm}\noindent
\textbf{HT Cas:} Well-studied \citep{horne91b} and observed with \textsc{ultracam} before \citep{feline05}. There are signs of an evolving orbital period for this system according to the linear fit to the mid-eclipse times. This is discussed in detail in \citet{bours16}. Normally HT Cas shows lots of flickering, and eclipses with no visible bright spot. For a short time in 2014 the bright spot ingress was visible, but with the egress masked by flickering.

\vspace*{0.1cm}\noindent
\textbf{IY UMa:} Known to be eclipsing for many years \citep{uemura00}, IY UMa has been studied extensively \citep{steeghs03}. System parameter determinations already exist, but have large uncertainties.

\vspace*{0.1cm}\noindent
\textbf{SDSS J040714.78-064425.1:} Also known as LT Eri, this system was discovered in SDSS follow up \citep{szkody03}. Further studies confirmed a period of around 4 hours \citep{ak05}, placing it well above the period gap. The CSS light curve shows many rapid $\sim$2 magnitude outbursts. We often found the system in outburst, but even in quiescence it usually shows blended white dwarf and bright spot ingresses.

\vspace*{0.1cm}\noindent
\textbf{SDSS J075059.97+141150.1:} Discovered in SDSS \citep{szkody07} and shows frequent outbursts in its CSS light curve. \citet{southworth10} presented time-resolved photometry and were able to measure the system parameters through light curve modelling. However, these measurements suffered from high uncertainties due to the low cadence of the observations.

\vspace*{0.1cm}\noindent
\textbf{SDSS J090103.93+480911.1:} Discovered in SDSS \citep{szkody03}, SDSS J090103.93+480911.1 has been studied photometrically already \citep{dillon08,shears12,kato13}. Our observations spanning 6 years found an evolution of bright spot features. At first the bright spot was clearly visible, then it disappeared and the flickering intensified. Later the flickering decreased and the bright spot returned. Two phase-folded light curves are shown in Figure \ref{fig:phaseFolds} to demonstrate this behaviour.

\vspace*{0.1cm}\noindent
\textbf{SDSS J090403.49+035501.2:} Discovered in SDSS \citep{szkody04}, and already studied photometrically \citep{woudt12}. It shows no outbursts in the CSS archival light curve. We observed a single eclipse of this system confirming the grazing nature of its eclipses, but the large orbital hump seen in other observations \citep{woudt12} is not present. We cannot improve on the precision of the ephemeris in \citet{woudt12}.

\vspace*{0.1cm}\noindent
\textbf{SDSS J092009.54+004245.0:} Discovered in SDSS \citep{szkody03}, further studies revealed a period of 3.5 hours \citep{gansicke09,camurdan10}. It has since been suggested to be a SW Sex system \citep{schmidtobreick12}, a nova-like with a strong, flared disk. We found that the eclipses have some structure, with a probable strong disk component, but without any clear sign of a white dwarf. This supports the SW Sex classification as a nova-like variable.

\vspace*{0.1cm}\noindent
\textbf{SDSS J092444.48+080150.9:} Also known as HU Leo and discovered in SDSS spectra \citep{szkody05}. Numerous eclipses have been observed \citep{southworth10,southworth15}, and we present one additional eclipse. The system shows no signs of an accretion disk, and has the possible characteristics of a polar light curve, showing a bright post-eclipse hump due to the projection effect of the accretion stream. We are unable to refine the ephemeris further.

\vspace*{0.1cm}\noindent
\textbf{SDSS J093537.46+161950.8:} Discovered in SDSS \citep{szkody09}, with a spectrum that could have been classified as a polar or an old nova. It has also been studied photometrically \citep{southworth15}. We observed a single eclipse with \textsc{ultracam}, finding a typical polar light curve similar to that seen in HU Aqr \citep{harropallin99}.

\vspace*{0.1cm}\noindent
\textbf{SDSS J100658.40+233724.4:} Discovered in SDSS \citep{szkody07}, this system has already been studied in detail. \citet{southworth09} published full system parameters based on time-resolved photometry and spectroscopy. We continued to observe SDSS J100658.40+233724.4 in an attempt to improve on these measurements. The eclipse light curve displays clear white dwarf and bright spot features, but some observations suffer from strong flickering.

\vspace*{0.1cm}\noindent
\textbf{SDSS J115207.00+404947.8:} Time-resolved \textsc{ultracam} observations already exist, with parameters determined through light curve modelling \citep{savoury11}. We later discovered that the ephemeris presented in \citet{savoury11} suffered from a cycle count ambiguity. We have since observed 4 more eclipses of this CV, and a new ephemeris is given in Table \ref{tab:ephem}. The new eclipse observations have stronger bright spot features than previously seen. 

\vspace*{0.1cm}\noindent
\textbf{SDSS J125023.84+665525.4:} Also known as OV Dra, discovered with a high inclination in SDSS spectra \citep{szkody03}. The archival CSS light curve shows long-term variability, eclipse-like dips, and several outbursts. Time-resolved photometry confirmed eclipses \citep{dillon08,kato14a}, but no high-speed studies have been published so far. Our observations found the bright spot eclipse to be visible, but unfortunately it is often interrupted by flickering.

\vspace*{0.1cm}\noindent
\textbf{SDSS J152419.33+220920.1:} Discovered in SDSS spectra \citep{szkody09}, which revealed double-peaked emission lines suggesting a high inclination. It was also independently discovered as a dwarf nova by CRTS and is therefore also known as CSS090329:152419+220920. Time-resolved photometry later revealed eclipses \citep{southworth10}, which were modelled to estimate the inclination and mass ratio of the system, but the data were not of sufficient quality to measure the individual component masses. Further photometry of 27 eclipses improved the ephemeris greatly \citep{michel13}. In our first observations SDSS J152419.33+220920.1 showed minimal bright spot features, and often its eclipse was completely blended with that of the white dwarf. However, the disk radius and/or bright spot flux changed in later observations, showing a clear bright spot component which is separate from the white dwarf. Figure \ref{fig:phaseFolds} shows two phase-folded eclipses which demonstrate this change in eclipse structure.

\vspace*{0.1cm}\noindent
\textbf{SDSS J155531.99-001055.0:} Discovered in SDSS \citep{szkody02}, with double-peaked emission lines indicating a high inclination. The CSS light curve showed no outbursts, but does hint at eclipses with occasional faint measurements. Photometric follow-up confirmed the eclipses and an orbital period of 113 minutes \citep{southworth07}. The eclipse structure clearly changes over time, as shown in Figure \ref{fig:phaseFolds}. In 2007 there were no bright spot features, but in 2013 the bright spot ingress, at least, is clear.

\vspace*{0.1cm}\noindent
\textbf{SDSS J155656.92+352336.6:} Also known as BT CrB, discovered with an orbital period of around 2 hours in SDSS \citep{szkody06}. High cadence observations of the eclipse often show the white dwarf and bright spot egresses well separated, but with blended ingresses. On one occasion the ingresses are marginally separated. Clearly this system has a variable disk radius, and at times could present an eclipse structure that can be modelled.

\vspace*{0.1cm}\noindent
\textbf{SSS100615:200331-284941:} Discovered via an outburst in CRTS, and later observed by \citet{coppejans14}. They found deep eclipses and an orbital period of 1.4 hours. Our observations show clearly visible bright spot features.

\vspace*{0.1cm}\noindent
\textbf{SSS120402:134015-350512:} Discovered by CRTS showing multiple dwarf nova outbursts in its archival light curve, as well as fainter observations associated with eclipses. \citet{coppejans14} observed two eclipses, and derived a period of 1.4 hours. We observed a single eclipse with \textsc{ultraspec} and confirm the eclipse is only partial in nature, with the eclipse of the white dwarf suspected to be grazing. We cannot improve on the ephemeris due to cycle count ambiguities.

\vspace*{0.1cm}\noindent
\textbf{V2051 Oph:} Originally discovered on objective prism plates \citep{sanduleak72}. It was studied photometrically by \citet{warner83}, who found bright spot features, making this system a good target for eclipse modelling. V2051 Oph has been the subject of numerous studies, via eclipse mapping and Doppler tomography \citep{vrielmann02,baptista04,saito06,longa15}, as well as eclipse-timing observations which show it has cyclical period changes and a larger than expected orbital decay rate \citep{baptista03,qian15} - this may be due to the presence of a third body and/or magnetic braking. In the hope of being able to model the eclipse structure of V2051 Oph, we have observed a total of 26 high-speed eclipses over 8 years (Table \ref{tab:obs}). On many occasions the light curves are not useful, due to either cloudy conditions, the system being in outburst, the presence of strong flickering, or the bright spot eclipse not being visible. However, several observations appear to be useful, with visible bright spot features. Figure \ref{fig:phaseFolds} shows two phase-folded eclipses showing the differences in eclipse structure at different epochs. We confirm the period changes discussed by \citet{baptista03} and \citet{qian15}, though we cannot confirm that these changes are cyclical in nature due to the relatively short baseline of our observations.

\begin{table*}
\centering\footnotesize
\begin{minipage}{1.\textwidth}
\caption{Ephemerides and suitability for modelling of eclipsing systems. Ephemerides are measured and quoted in BMJD(TDB). Some ephemerides have been calculated using additional mid-eclipse times from the literature. In these cases the relevant references are supplied. Where we have been unable to refine the ephemeris, we refer the reader to previous papers rather than repeating it here. Evaluation of the potential for light curve modelling is given as `Y' for yes, `N' for no, and `M' for maybe/more data are needed.}
\label{tab:ephem}
\begin{tabular}{p{3.8cm} p{2.3cm} p{2.4cm} p{5.5cm} p{1.5cm}}
\textbf{Object} & \textbf{$T_{0}$} & \textbf{$P$} & \textbf{Additional Eclipse Times} & \textbf{Modelling Potential} \\ \hline
  1RXS J180834.7+101041 & 54926.3108(1) & 0.06949063(5) & \citealp{southworth11} & N \\
  ASASSN-13cx & 57104.30636(1) & 0.07965006(1) & - & M \\
  ASASSN-15au & 57148.04781(4) & 0.06894973(2) & - & M \\
  ASASSN-15bu & 57080.98512(1) & 0.0768262(9) & - & M \\
  CSS080227:112634-100210 & 56294.94350(3) & 0.077421572(3) & \citealp{southworth15} & M \\
  CSS080306:082655-000733 & 56444.203767(8) & 0.059764424(1) & - & Y \\
  CSS080623:140454-102702 & 55488.728318(8) & 0.059578972(2) & - & Y \\
  CSS081220:011614+092216 & 56713.570050(8) & 0.065843019(3) & Coppejans (priv.comm.) & M \\
  CSS090102:132536+210037 & 56207.945385(5) & 0.0623849134(7) & \citealp{southworth15} & Y \\
  CSS090419:162620-125557 & 56551.813968(9) & 0.075442719(5) & - & Y \\
  CSS090622:215636+193242 & 56877.82014(2) & 0.0709293(2) & - & Y \\
  CSS100218:043829+004016 & 56658.417503(1) & 0.0654948(1) & - & M \\
  CSS110113:043112-031452 & 56396.181193(3) & 0.0660508721(6) & - & Y \\
  CSS110513:210846-035031 & 56767.65993(3) & 0.15692657(3) & - & Y \\
  CSS111003:054558+022106 & 56852.23197(9) & 0.12097138(4) & \citealp{miszalski16} & Y \\
  CSS111019:233313-155744 & 56866.13556(2) & 0.04285020(1) & - & M \\
  CSS111101:233003+303301 & 57015.4708(2) & 0.1559784(6) & - & N \\
  CSS131106:052412+004148 & 57126.37256(2) & 0.17466647(2) & - & M \\
  CzeV404 Her & 56871.91730(4) & 0.09802125(2) & - & M \\
  GALEX J003535.7+462353 & 56489.69294(4) & 0.17227491(1) & \citealp{wils11} & M \\
  GY Cnc & 55938.26369(1) & 0.175442402(1) & \citealp{kato02} & Y \\
  HS 2325+8205 & 55661.080122(7) & 0.194334533(1) & \citealp{pyrzas12} & Y \\
  HT Cas & 55550.16974(3) & 0.073647175(1) & \citealp{feline05} & N \\
  IY UMa & 54679.998816(9) & 0.0739089285(3) & \citealp{steeghs03} & Y \\
  MAST003059.39+301634.3 & 57159.48849(5) & 0.07026247(3) & - & M \\
  MAST192328.22+612413.5 & 56852.10211(9) & 0.1676465(3) & \citealp{kennedy16} & N \\
  MAST232100.42+494614.0 & 57010.6753(1) & 0.2123715(9) & - & M \\
  MLS101226:072033+172437 & 57023.1777(2) & 0.1504066(3) & - & N \\
  MLS120517:152507-032655 & 56821.89967(5) & 0.06438163(7) & - & N \\
  SDSS J040714.78-064425.1 & 56332.1340(1) & 0.17020393(2) & - & N \\
  SDSS J075059.97+141150.1 & 56306.51099(1) & 0.093165496(2) & \citealp{southworth10,southworth15} & Y \\
  SDSS J090103.93+480911.1 & 55229.666015(7) & 0.0778805326(6) & - & Y \\
  SDSS J090403.49+035501.2 & - & - & - & N \\
  SDSS J092009.54+004245.0 & 56145.35967(3) & 0.147875676(7) & - & N \\
  SDSS J092444.48+080150.9 & - & - & - & N \\
  SDSS J093537.46+161950.8 & 55670.30839(6) & 0.06405514(3) & \citealp{southworth15} & N \\
  SDSS J100658.40+233724.4 & 56396.98203(6) & 0.18591303(2) & \citealp{southworth09,southworth15} & Y \\
  SDSS J115207.00+404947.8 & 56082.544698(2) & 0.0677497017(2) & \citealp{southworth10,savoury11} & Y \\
  SDSS J125023.84+665525.4 & 54614.43088(1) & 0.0587356818(6) & \citealp{dillon08} & M \\
  SDSS J152419.33+220920.1 & 56287.823054(3) & 0.0653187310(8) & \citealp{southworth10} & Y \\
  SDSS J155531.99-001055.0 & 55267.794254(4) & 0.0788455518(3) & \citealp{southworth07} & M \\
  SDSS J155656.92+352336.6 & 55633.75890(3) & 0.088091490(4) & - & M \\
  SSS100615:200331-284941 & 56874.0223924(8) & 0.05870445(6) & - & Y \\
  SSS120402:134015−350512 & - & - & - & N \\
  SSS130413:094551-194402 & 56809.688700(9) & 0.065769292(3) & - & Y \\
  V2051 Oph & 55314.156237(4) & 0.06242785751(8) & \citealp{vrielmann02,baptista03,qian15} & Y \\
  V713 Cep & 56532.192754(5) & 0.085418509(1) & - & Y \\
 \hline  
\end{tabular}
\end{minipage}
\end{table*}

\vspace*{0.1cm}\noindent
\textbf{V713 Cep:} Originally discovered as variable by inspection of archival photometric plates \citep{antipin03}, eclipses were soon discovered in this CV (\textit{vsnet} alert 9516). More detailed photometry showed that it had the classic CV eclipse structure that lends itself to light curve modelling \citep{boyd11}. There are signs of a varying orbital period for this system according to our linear fit to the mid-eclipse times. This is discussed in detail in \citet{bours16}. We observed the bright spot in V713 Cep changing significantly over time. In some observations it is clearly seen, and in others there is no bright spot feature at all. The top two light curves in the bottom panel of Figure \ref{fig:phaseFolds} demonstrate this. 

In addition to this, we also observed an apparent complete switch-off of accretion on 2015-06-24. This eclipse, the bottom light curve shown in the bottom panel of Figure \ref{fig:phaseFolds}, is typical of detached binaries in which the white dwarf primary is not accreting gas from the secondary star. Three months later, on 2015-09-17, the system appeared to be accreting again. This low-state behaviour is unexpected and very rare among non-magnetic dwarf novae. The only other example of this was seen in IR Com, with an extended low state in which accretion was confirmed to have switched off \citep{manser14}. Low states are however found in the VY Scl class of nova-like variables, in both magnetic and non-magnetic systems \citep{hellier01}. 

One theory for temporary interruptions to mass transfer is that they occur when a starspot crosses the L1 point on the donor star, lowering the stellar surface and disconnecting it from the L1 point. In CVs with strong magnetic fields (i.e. polars), accretion onto the white dwarf ceases immediately, and the system enters a low state until mass transfer resumes. However, in non-magnetic dwarf novae the system remains bright, as the accretion disk is thought to outlive the short break in mass transfer, only draining slowly onto the white dwarf \citep{hessman00,manser14}. 

Our observations of V713 Cep show that this break in mass transfer is not constrained to nova-likes, polars and the single instance of IR Com, but may be more common than first thought. Unfortunately V713 Cep is not covered in the CSS footprint, and we cannot say exactly how long the low state lasted. Our observations of eclipses before and after the non-accreting observation on 2015-06-24 both show normal eclipse structure including an accretion disk. These are separated by 403 days and thus we constrain the low state of V713 Cep to a maximum duration of 403 days. 


\section{Conclusions}\label{sec:conclusion}
We have presented time-resolved photometry of 74 CVs, most of which have never before been observed in detail. We have discovered or confirmed 13 new eclipsing systems with orbital periods ranging from 1.5 to 5.1 hours. We studied 27 new systems which did not show any eclipses. The remaining 34 systems presented here were already known to be eclipsing, but we provide the highest cadence observations of these objects to date. For all eclipsing systems we have discussed the feasibility of modelling their eclipse structure to determine their system parameters. We found that approximately 20 of the objects discussed above should be suitable for system parameter studies using light curve modelling, while a further 15 systems may be feasibly modelled if additional observations are acquired. 


\section*{Acknowledgements}
We would like to thank the editor and anonymous referee for constructive feedback which has helped improve this paper. We thank Deanne Coppejans for providing additional mid-eclipse times of CSS081220:011614+092216. 

LKH acknowledges support via a Harry Worthington Scholarship at the University of Sheffield. MJM acknowledges the support of a UK Science and Technology Facilities Council (STFC) funded PhD. VSD, SPL and
\textsc{ultracam} are supported by STFC grant ST/J001589/1. TB and RWW are grateful to the STFC for financial support (grant reference ST/J001236/1). EB and TRM acknowledge support from the STFC in the form of a Consolidated Grant ST/L000733. PI acknowledges the support of NRC-Thailand. SGP acknowledges financial support from FONDECYT in the form of grant number 3140585. PAW acknowledges UCT and the NRF for their financial support. 

This work was based in part on observations made with \textit{pt5m}, a collaborative effort between the Universities of Durham and Sheffield. Other observations have been made with telescopes operated by the Isaac Newton Group in La Palma, the National Astronomical Research Institute of Thailand, the European Southern Observatory in Chile, and the South African Astronomical Observatory. This work has made use of NASA's Astrophysics Data System Bibliographic Services, the Vizier database operated at CDS, Strasbourg, France, and the International Variable Star Index (VSX) database operated at AAVSO, Cambridge, Massachusetts, USA.

\bibliographystyle{mn2e}
\bibliography{abbrev.bib,refs.bib}


\appendix

\section{Calibrating the \textit{KG5} filter}\label{sec:kg5calibration}

\begin{figure}
 \includegraphics[width=.42\textwidth,angle=270]{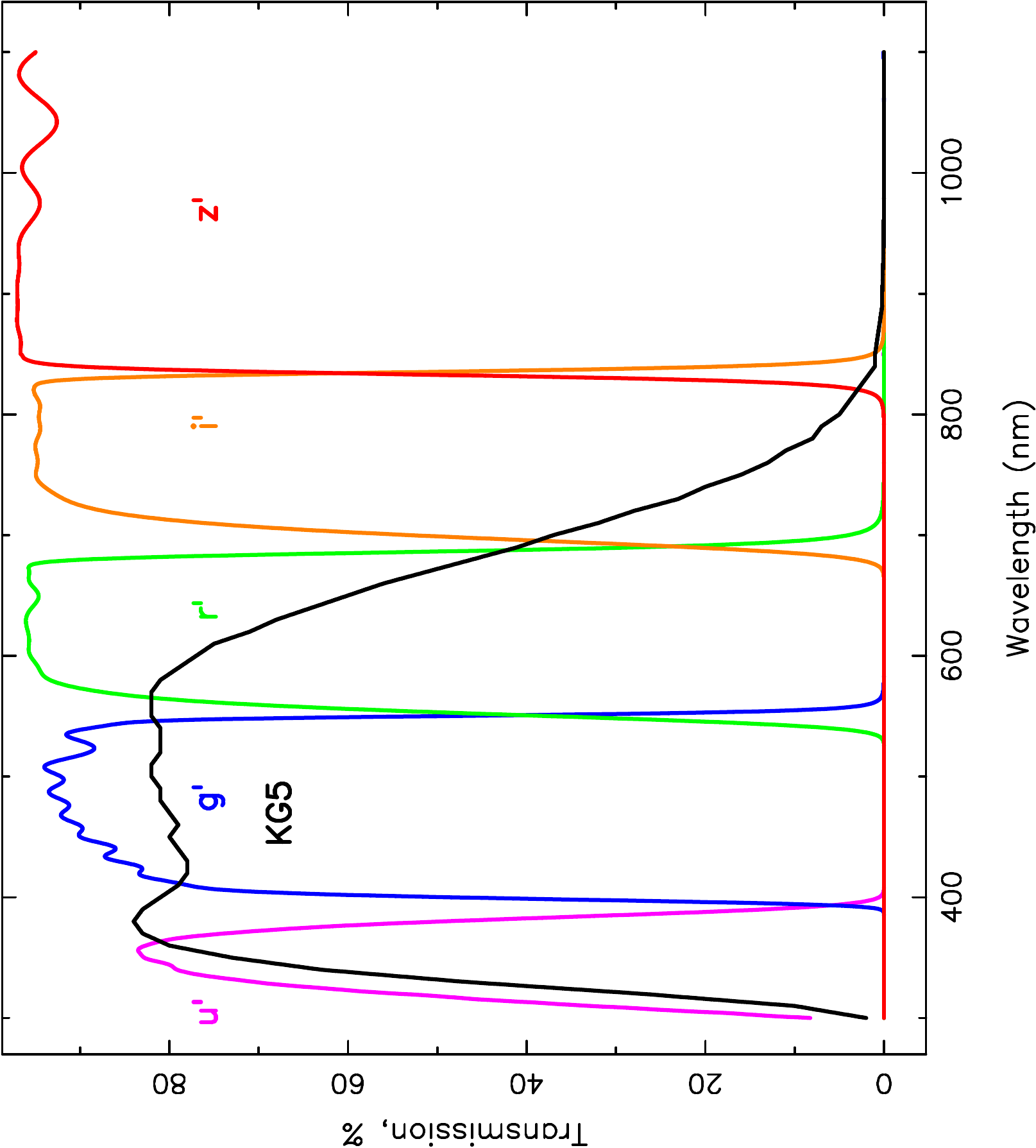}
 \caption{Transmission as a function of wavelength for the five SDSS filters and the \textit{KG5} filter used in \textsc{ultraspec}.}
 \label{fig:kg5throughput}
\end{figure}

In Section \ref{sec:obs} we explained that many observations made with \textsc{ultraspec} at the TNT were conducted with the wide-throughput \textit{KG5} filter. The transmission of this filter and of the standard SDSS filters are shown in Figure \ref{fig:kg5throughput}. 

To calibrate the fluxes of observations in the standard SDSS filters, we simply compared the observed counts of a stable well-studied photometric standard star with its catalogued magnitude. However, this approach breaks down for the \textit{KG5} filter, because no catalogued \textit{KG5} magnitudes exist for standard stars. Instead, we needed to produce theoretical \textit{KG5} magnitudes for these stars. 

As discussed by \citet{bell12}, one method for calculating theoretical magnitudes in any filter is to use bolometric corrections (BCs). BCs are unique to every star, and are the difference between the magnitude of the star in a particular filter, and the bolometric (all wavelengths) magnitude of that star, such that $M_{bol} = m_{g} + BC_{g}$ for the SDSS $g$-filter for example. If we know the BC for a photometric standard star in both the SDSS $g$-filter and the \textsc{ultraspec} \textit{KG5}-filter, those corrections can be used along with the SDSS catalogue $g$-band magnitude to predict the \textit{KG5} magnitude, as shown in Equation \ref{eq:bolconversion}: 

\begin{equation}\label{eq:bolconversion}
m_{KG5} + BC_{KG5} = m_{g} + BC_{g}
\end{equation}

To calculate a BC for any given star, we need the star's spectrum and a model of our telescope, instrument and detector throughput as a function of wavelength. Example stellar spectra are available from spectral atlases (e.g. \citealt{gunn83,pickles98,castelli04}), and we had already produced a throughput model for \textsc{ultraspec} in order to quantify the performance of the instrument during commissioning \citep{dhillon14}. Our model for the system includes the throughput of the atmosphere, telescope mirrors, instrument lenses and adhesives, the filters themselves, and the quantum efficiency of the detector. We used the most accurate available transmission data, with the largest uncertainties expected to come from the atmospheric extinction and the mirror reflectivity. We have no wavelength dependent information for the mirrors, or in fact any measurement of reflectivity at all, so we assumed a uniform reflectivity of 85\%, repeated across all four mirrors of the TNT. We discuss the accuracy of this throughput model below in Section \ref{sec:throughputaccuracy}. 

The formula for calculating a BC for the $g$-filter in the AB magnitude system is shown in Equation \ref{eq:bolcor} and follows from the derivations in \citet{bell12}.

\begin{eqnarray}\label{eq:bolcor}
BC_{g} = M_{bol\odot} - 2.5\log_{10}\left(\frac{4\pi(10pc)^{2}F_{bol}}{L_{\odot}}\right) \nonumber \\
+ 2.5\log_{10}\left(\frac{\int_{g}\lambda F_{\lambda}10^{-0.4A_{\lambda}}R_{\lambda}d\lambda}{\int_{g}\frac{c}{\lambda}\,f^{0}_{\nu}R_{\lambda}d\lambda}\right),
\end{eqnarray}
where $F_{bol} = \sigma T^{4}_{e\!f\!f}$ by the Stefan-Boltzmann law, $F_{\lambda}$ is the flux at the stellar surface according to the model atmosphere spectrum, $A_{\lambda}$ is the appropriate galactic extinction curve, and $R_{\lambda}$ is the instrument throughput model as a function of wavelength. All functions of wavelength are handled in units of \AA. We used the following values for the Sun's bolometric magnitude and the AB magnitude zero flux: $M_{bol\odot}=4.74$  and $f^{0}_{\nu}=3631\!\times\!10^{-23}$erg s$^{-1}$cm$^{-2}$Hz$^{-1}$.

Since the BCs are dependent on the spectrum of the star in question, they will vary as a function of effective temperature ($T_{e\!f\!f}$) and surface gravity (log $g$). Thus our final aim is to create a table of \textit{KG5} magnitude transformations from a given $g$-band magnitude, as a function of effective temperature. This could then be used to estimate a \textit{KG5} magnitude for any photometric standard star.

We began by creating a list of $T_{e\!f\!f}$/log $g$ pairs from the Dartmouth isochrone database \citep{dotter08}, to sample a range of effective temperatures. Since most standard stars are bright and nearby, we assumed no reddening ($E(B-V) = 0$), and solar metallicity (Fe/H = 0). Incorporating a value for the reddening into the production of the isochrone allows us to ignore the $A_{\lambda}$ term in Equation \ref{eq:bolcor}. 

\begin{figure}
 \includegraphics[width=.41\textwidth,angle=270]{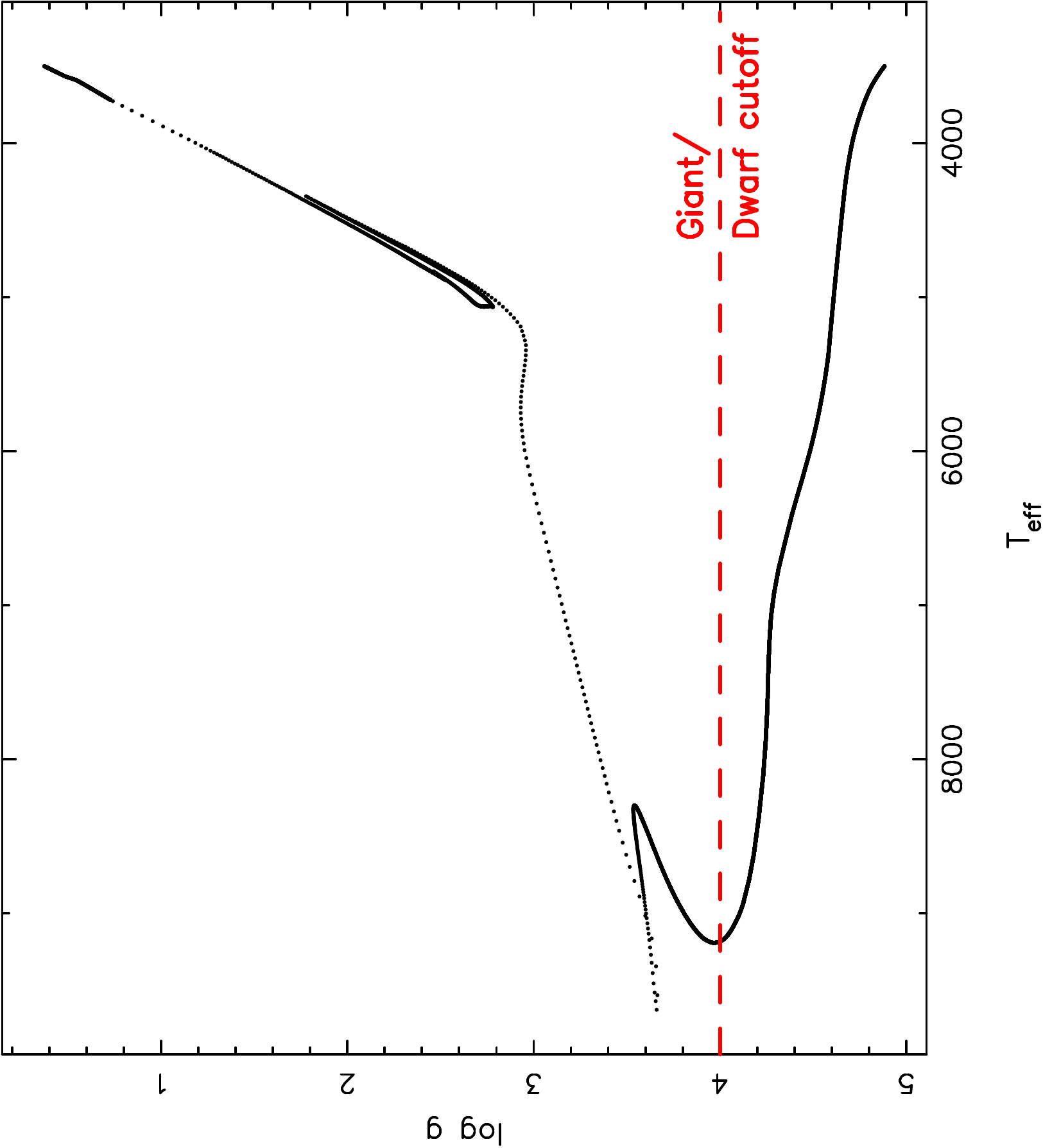}
 \caption{Theoretical $T_{e\!f\!f}$/log $g$ pairs for stars with no reddening, solar metallicity and an age of 0.5 Gyrs, generated from the Dartmouth isochrone database. Axes have been inverted to simulate a classical colour-magnitude diagram. We select only the main sequence stars by making a cut at log $g$ = 4.}
 \label{fig:isochrone}
\end{figure}

To investigate a wider ranger of available temperatures, we sampled two age scenarios; 0.5 Gyrs (log age = 8.699) and 2 Gyrs (log age = 9.301). Since the model spectra available are generally constrained to main sequence stars, we rejected those pairs with log $g$ $<4$, as these values are usually only found in evolved stars. This means that our final theoretical \textit{KG5} magnitudes are only applicable to main sequence standard stars. An example isochrone of $T_{e\!f\!f}$/log $g$ pairs for the 0.5 Gyr scenario is shown in Figure \ref{fig:isochrone}.

For every theoretical star described by the Dartmouth isochrone, we generated a model spectrum using the atlas of \citet{castelli04} and the Python module \textsc{pysynphot} \citep{pysynphot}. We then calculated a \textit{KG5} BC for every star using Equation \ref{eq:bolcor}, along with these model spectra and our instrument throughput model. The BCs apply to given effective temperatures, sampling the range 3500K to 9200K, in steps of a few Kelvin. 

Now, supplied with the catalogue $g$-band magnitude and the effective temperature (estimated from the spectral type, within the range 3500-9200K) of any given main sequence standard star, we are able to calculate a theoretical \textit{KG5} magnitude for that standard star using Equation \ref{eq:bolconversion} and linear interpolation of the existing sample. This enabled us to flux calibrate observations taken in the \textit{KG5} filter.

\subsection{Accuracy of the instrumental throughput model}\label{sec:throughputaccuracy}
If we are to trust our \textit{KG5} flux calibration procedure outlined above, we need to be sure that our throughput model for \textsc{ultraspec} is accurate and reliable. When testing the throughput model by comparison with observations, we found that the model matched observations better in some filters than in others. This suggested that the model may be non-uniformly accurate as a function of wavelength. To investigate this, we followed the procedure of \citet{bell12} to calculate values of d$m$, the difference between theoretical and observed magnitudes in a standard filter $m$, as a function of SDSS colour ($g$-$r$), for a large selection of stars. 

We obtained observations of the open cluster NGC 6940, and used these to test the throughput model accuracy. We measured observational magnitudes for stars in the cluster, and also computed theoretical magnitudes for these stars using BCs and their SDSS catalogue magnitudes. The model spectra used in the BC calculations were tailored to the properties of the cluster, being derived from a suitable isochrone; log age = 8.858, metallicity Fe/H = 0.01, E(B-V) = 2.14 \citep{wu09}. We used the SDSS colour versus effective temperature relation given in Equation \ref{eq:colourtemp} and taken from \citet{fukugita11} to estimate temperatures for each star in the cluster. 

\begin{equation}\label{eq:colourtemp}
T_{e\!f\!f} = \frac{1.09\times 10^{4}}{1.47+g-r}
\end{equation}

We also rejected anomalous stars based on a $\chi^{2}$ analysis, in which we compared the actual measured magnitudes of stars with the predicted magnitudes in a given filter, calculated using BCs and the catalogue magnitudes in the other filters. Stars with measured magnitudes more than 3-$\sigma$ away from their predicted magnitudes in any filter were rejected. 

\begin{figure}
 \includegraphics[width=.37\textwidth,angle=270]{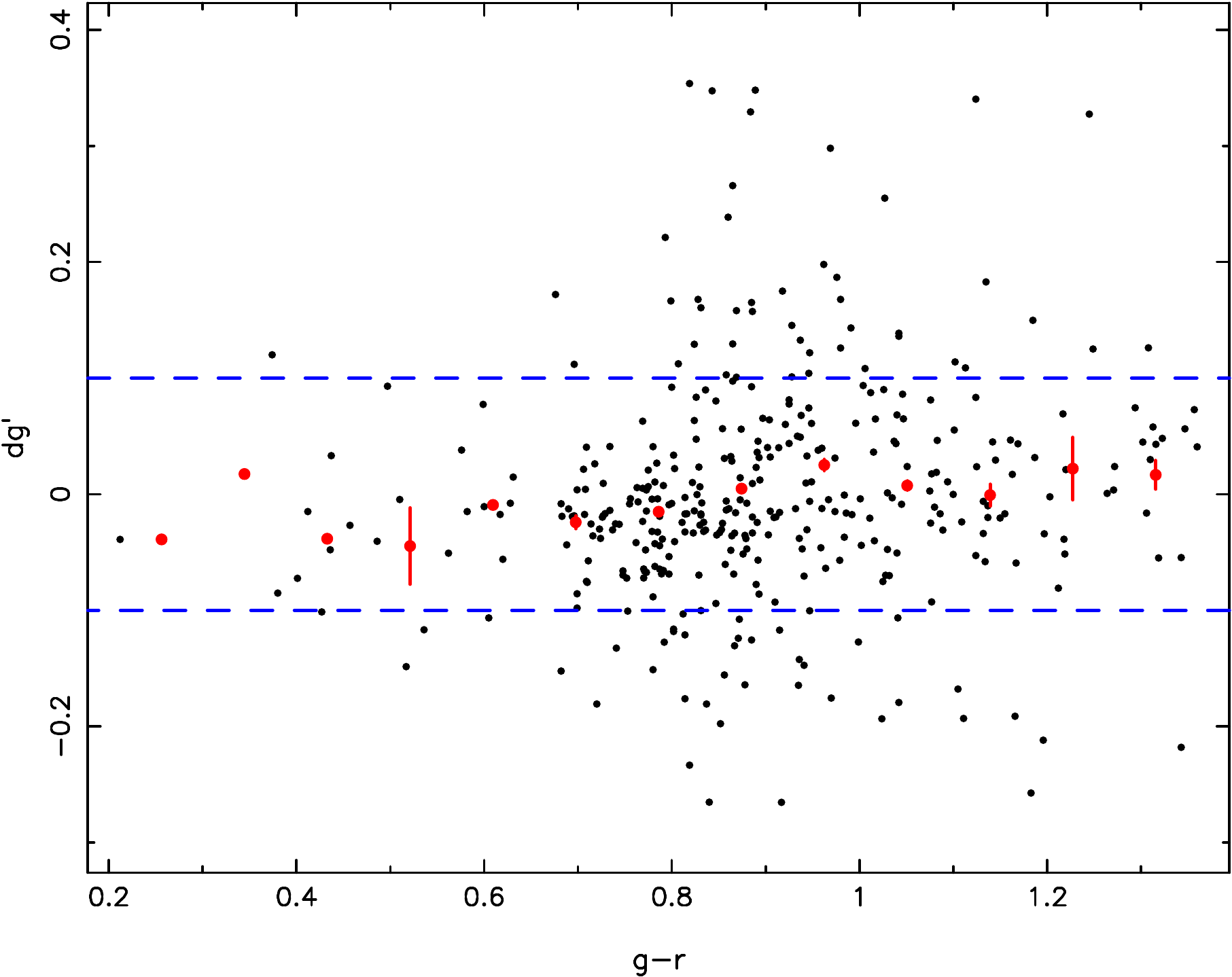}
 \caption{The difference between the theoretical and measured \textsc{ultraspec} $g'$-band magnitudes of stars in NGC 6940, as a function of their colour. Individual stars are represented by black points, whilst red circles show the mean d$g'$ of 13 equally spaced bins in colour. The error bars on the binned points are the maximum photometric uncertainty within each bin, divided by the number of points in the bin. The blue dashed lines show the $\pm0.1$ magnitude level with respect to zero.}
 \label{fig:dgVcolour}
\end{figure}

We compared the predicted magnitudes produced by the BC analysis with the SDSS catalogue magnitude for each star, calculating d$m$ as the difference between the two. Any non-uniform inaccuracy in the throughput model should be apparent as a trend of d$m$ against the colour of the star. An example plot of d$m$ against $g$-$r$ is shown in Figure \ref{fig:dgVcolour} for the \textsc{ultraspec} $g'$ filter. The same behaviour was seen in other filters, except in $u'$ where the data is so sparse and scattered that we cannot conclude anything with certainty. 

We found that in all filters the theoretical magnitudes matched the observed magnitudes (except for a uniform efficiency factor, likely due to the actual mirror reflectivity being different to our assumed value), with no significant trend as a function of colour. However, the available data were not sufficient to constrain this behaviour strongly, with a scatter of up to 0.2 magnitudes. 

These investigations have confirmed that our throughput model is correct as a function of wavelength, and would give accurate BCs to within a systematic error of 0.2 magnitudes. 

\section{Supplementary Online Materials}

\onecolumn
\centering
\topcaption{Journal of Observations. The systems are ordered alphabetically, and some system names have been shortened. Start(MJD) is the start time of each observing run, given in MJD(UTC). Mid-eclipse times are given in BMJD(TDB) and the number in parentheses is the uncertainty in the last digit. $T_{exp}$ is the exposure time in seconds and is supplied as $T_{blue}/T_{green}/T_{red}$ for the three beams of \textsc{ultracam}. $\Delta T$ is the duration of the observing run in minutes. Mag. is the estimated out-of-eclipse magnitude.}

\tablefirsthead{ 
\hline
\textbf{Object} & \textbf{Start (MJD)} & \textbf{Mid-eclipse time (BMJD)} & \textbf{T$_{exp}$ (s)} & \textbf{$\Delta $T (min)} & \textbf{Tel./Inst.}  & \textbf{Filter} & \textbf{Mag.} \\ \hline}
\tablehead{
\multicolumn{8}{|l|}{\small\sl continued from previous page.}\\
\hline
\textbf{Object} & \textbf{Start (MJD)} & \textbf{Mid-eclipse time (BMJD)} & \textbf{T$_{exp}$ (s)} & \textbf{$\Delta $T (min)} & \textbf{Tel./Inst.}  & \textbf{Filter} & \textbf{Mag.} \\ \hline}
\tabletail{
\hline
\multicolumn{8}{|r|}{\small\sl continued on next page}\\
}
\tablelasttail{}
\label{tab:obs}
\small
\begin{supertabular}{p{3.7cm} p{1.7cm} p{2.1cm} p{1.3cm} p{0.7cm} p{1.8cm} p{1.1cm} p{1.8cm}}

  1RXS J180834.7+101041 & 55316.33464 & 55316.3613(3) & 6/2/2 & 141 & \textsc{ntt/ucam} & $u'/g'/r'$ & 16.9/16.9/16.6 \\
 & & 55316.4310(3)  \\
    & 55334.37156 & 55334.4299(3) & 10/2/2 & 101 & \textsc{ntt/ucam} & $u'/g'/r'$ & 17.0/17.0/16.8 \\
\\
  ASASSN-13cx & 56943.91536 & 56943.9710(3) & 150 & 137 & \textit{pt5m} & \textit{V} & 16.1 \\
              & 56968.81315 & 56968.8218(3) & 150 & 206 & \textit{pt5m} & \textit{V} & 18.0 \\
                          & & 56968.9015(3) \\
              & 56973.81141 & 56973.8401(7) & 150 & 406 & \textit{pt5m} & \textit{V} & 18.3 \\
                          & & 56973.9186(7) \\
              & 56989.65725 & 56989.68988(3) & 3.9 & 60 & \textsc{tnt/uspec} & $g'$ & 18.9 \\
              & 56993.59267 & - & 3.9 & 53 & \textsc{tnt/uspec} & \textit{KG5} & 17.9 \\
              & 57014.83182 & 57014.8596(2) & 120 & 69  & \textit{pt5m} & \textit{V} & 16.2 \\
              & 57015.86791 & 57015.89471(7) & 120 & 69  & \textit{pt5m} & \textit{V} & 16.1 \\
              & 57023.51317 & 57023.54121(3) & 3.9 & 57 & \textsc{tnt/uspec} & \textit{KG5} & 18.5 \\
              & 57024.56187 & 57024.57665(3) & 4.9 & 46 & \textsc{tnt/uspec} & $g'$ & 18.3 \\
              & 57027.56687 & 57027.60339(3) & 5.9 & 69 & \textsc{tnt/uspec} & \textit{KG5} & 18.9 \\
              & 57198.12288 & 57198.13413(15) & 12/4/4 & 31 & \textsc{wht/ucam} & $u'/g'/r'$ & 18.4/18.4/18.3 \\
\\
  ASASSN-14cl & 56827.07177 & - & 10 & 188 & \textit{pt5m} & \textit{V} & 11.3 \\
    & 56829.08098 & - & 10 & 180 & \textit{pt5m} & \textit{V} & 11.6 \\
    & 56852.08564 & - & 10 & 184 & \textit{pt5m} & \textit{V} & 16.7 \\
\\
  ASASSN-14ds & 56853.00227 & - & 60 & 304 & \textit{pt5m} & \textit{V} & 16.7 \\
    & 56854.99663 & - & 60 & 314 & \textit{pt5m} & \textit{V} & 16.7 \\
    & 56855.99399 & - & 60 & 318 & \textit{pt5m} & \textit{V} & 16.9 \\
    & 56856.99123 & - & 60 & 183 & \textit{pt5m} & \textit{V} & 16.8 \\
    & 56858.98604 & - & 60 & 310 & \textit{pt5m} & \textit{V} & 17.1 \\
    & 56859.98344 & - & 60 & 170 & \textit{pt5m} & \textit{V} & 17.0 \\
\\
  ASASSN-14gl & 56995.80900 & - & 60 & 124 & \textit{pt5m} & \textit{V} & 18.2 \\
    & 56998.80903 & - & 60 & 147 & \textit{pt5m} & \textit{V} & 18.6 \\
    & 57016.88881 & - & 120 & 129 & \textit{pt5m} & \textit{V} & 19.1 \\
    & 57018.81524 & - & 120 & 348 & \textit{pt5m} & \textit{V} & 19.3 \\
    & 57016.88881 & - & 120 & 129 & \textit{pt5m} & \textit{V} & 18.2 \\
\\
  ASASSN-14gu & 57057.83442 & - & 120 & 141 & \textit{pt5m} & \textit{V} & 18.0 \\
    & 57058.83810 & - & 120 & 105 & \textit{pt5m} & \textit{V} & 17.3 \\
    & 57084.89539 & - & 120 & 253 & \textit{pt5m} & \textit{V} & 16.4 \\
\\
  ASASSN-14hk & 56922.12637 & - & 60 & 172 & \textit{pt5m} & \textit{V} & 15.8 \\
    & 56932.99821 & - & 60 & 128 & \textit{pt5m} & \textit{V} & 17.1 \\
\\
  ASASSN-14mv & 57026.59217 & - & 1.0 & 122 & \textsc{tnt/uspec} & $g'$ & 11.8 \\
\\
  ASASSN-15au & 57059.83530 & 57059.8614(1) & 120 & 226 & \textit{pt5m} & \textit{V} & 18.1 \\
 & & 57059.9302(3) \\
    & 57067.00211 & 57067.0322(2) & 120 & 148 & \textit{pt5m} & \textit{V} & 18.0 \\
 & & 57067.1011(2) \\
    & 57067.89695 & 57067.9283(2) & 120 & 69  & \textit{pt5m} & \textit{V} & 18.4 \\
    & 57067.97227 & 57067.9974(3) & 120 & 69  & \textit{pt5m} & \textit{V} & 18.1 \\
    & 57068.03489 & 57068.0665(2) & 120 & 69  & \textit{pt5m} & \textit{V} & 18.1 \\
    & 57076.52507 & 57076.54688(5) & 7.0 & 40  & \textsc{tnt/uspec} & $g'$ & 18.3 \\
    & 57077.62875 & 57077.65003(5) & 7.0 & 39  & \textsc{tnt/uspec} & $r'$ & 18.5 \\
    & 57106.52037 & 57106.54011(5) & 4.0 & 37  & \textsc{tnt/uspec} & $g'$ & 18.2 \\
    & 57367.70969 & 57367.72165(5) & 10  & 36  & \textsc{tnt/uspec} & $g'$ & 17.8 \\
\\
  ASASSN-15bu & 57079.51212 & 57079.52541(5) & 3.9 & 38 & \textsc{tnt/uspec} & $g'$ & 17.6 \\
    & 57080.51411 & 57080.52418(5) & 3.8 & 51 & \textsc{tnt/uspec} & $g'$ & 17.8 \\
    & 57081.51652 & 57081.52293(5) & 3.9 & 48 & \textsc{tnt/uspec} & $g'$ & 17.9 \\
   & 57082.51514 & 57082.52163(5) & 4.9 & 35 & \textsc{tnt/uspec} & $g'$ & 18.0 \\
\\
  ASASSN-15ni & 57231.92628 & - & 10 & 114 & \textit{pt5m} & \textit{V} & 13.2 \\
 & 57232.90219 & - & 20 & 311 & \textit{pt5m} & \textit{V} & 13.6 \\
 & 57235.90511 & - & 30 & 287 & \textit{pt5m} & \textit{V} & 14.1 \\
 & 57236.89403 & - & 30 & 138 & \textit{pt5m} & \textit{V} & 14.2 \\
\\
  CSS080227:112634-100210 & 55333.97234 & 55333.98693(3) & 9/3/3 & 154 & \textsc{ntt/ucam} & $u'/g'/r'$ & 18.6/18.7/18.4 \\
 & & 55334.06433(3)  \\
    & 55711.01379 & 55711.03000(3) & 12/4/4  & 41 & \textsc{ntt/ucam} & $u'/g'/r'$ & 18.3/18.5/18.1 \\
    & 56046.90666 & - & 8.5/4.2/4.2 & 25 & \textsc{wht/ucam} & $u'/g'/r'$ & 18.4/18.5/18.3 \\
    & 56685.91260 & 56685.92246(3) & 4.9 & 31 & \textsc{tnt/uspec} & \textit{KG5} & 18.3 \\
    & 57076.88173 & 57076.90134(3) & 9.9 & 41 & \textsc{tnt/uspec} & \textit{KG5} & 19.1 \\
    & 57077.81348 & 57077.83043(3) & 9.9 & 39 & \textsc{tnt/uspec} & \textit{KG5} & 19.0 \\
    & 57078.80800 & 57078.83694(3) & 9.9 & 56 & \textsc{tnt/uspec} & \textit{KG5} & 19.0 \\
\\
  CSS080306:082655-000733 & 56045.86872 & 56045.87387(3) & 6/2/2 & 14 & \textsc{wht/ucam} & $u'/g'/r'$ & 20.3/19.9/20.0 \\
    & 56045.90974 & 56045.93365(3) & 12/4/4  & 42 & \textsc{wht/ucam} & $u'/g'/r'$ & 20.2/20.0/19.9 \\
    & 56046.86831 & 56046.88989(3) & 12/4/4  & 36 & \textsc{wht/ucam} & $u'/g'/r'$ & 19.8/19.8/19.6 \\
    & 56689.94364 & 56689.95507(3) & 3.0 & 26 & \textsc{salt/salticam} & $r'$ & 19.7 \\
    & 56693.87936 & 56693.89955(3) & 3.0 & 39 & \textsc{salt/salticam} & $r'$ & 20.1 \\
    & 56694.95652 & 56694.97525(3) & 3.0 & 33 & \textsc{salt/salticam} & $g'$ & 20.1 \\
    & 56695.96725 & 56695.99128(3) & 3.0 & 44 & \textsc{salt/salticam} & $g'$ & 20.3 \\
    & 56988.76551 & 56988.77724(5) & 5.9 & 31 & \textsc{tnt/uspec} & \textit{KG5} & 19.7 \\
\\
  CSS080623:140454-102702 & 55329.21114 & 55329.23537(3) & 13/3.3/3.3  & 41 & \textsc{ntt/ucam} & $u'/g'/r'$ & 19.8/19.8/19.6 \\
    & 55334.09572 & 55334.12090(3) & 15/5/5  & 59 & \textsc{ntt/ucam} & $u'/g'/r'$ & 20.1/19.8/19.8 \\
    & 55334.15978 & 55334.18049(3) & 16/4/4  & 60 & \textsc{ntt/ucam} & $u'/g'/r'$ & 20.1/19.9/19.8 \\
    & 55334.20345 & 55334.24003(3) & 12/4/4  & 67 & \textsc{ntt/ucam} & $u'/g'/r'$ & 20.1/19.9/19.9 \\
    & 55355.01296 & 55355.03309(3) & 7.7/3.8/3.8 & 41 & \textsc{ntt/ucam} & $u'/g'/r'$ & 19.7/19.7/19.5 \\
    & 55355.12826 & 55355.15230(3) & 7.7/3.8/3.8 & 49 & \textsc{ntt/ucam} & $u'/g'/r'$ & 20.0/19.8/19.7 \\
    & 55709.03970 & 55709.05135(3) & 6/3/3 & 21 & \textsc{ntt/ucam} & $u'/g'/r'$ & 19.9/19.8/19.9 \\
    & 55711.95380 & 55711.97074(3) & 12/4/4  & 29 & \textsc{ntt/ucam} & $u'/g'/r'$ & 19.8/19.7/19.9 \\
    & 55711.99633 & 55712.03031(3) & 12/4/4  & 53 & \textsc{ntt/ucam} & $u'/g'/r'$ & 20.1/19.8/19.9 \\
    & 55712.11378 & 55712.14946(3) & 12/4/4  & 67 & \textsc{ntt/ucam} & $u'/g'/r'$ & 20.0/19.7/19.9 \\
\\
  CSS081220:011614+092216 & 56489.16741 & 56489.17702(3) & 9.8/3.2/3.2 & 28 & \textsc{wht/ucam} & $u'/g'/i'$ & 19.5/19.6/19.3 \\
    & 56500.15832 & 56500.17282(3) & 9.8/3.2/3.2 & 35 & \textsc{wht/ucam} & $u'/g'/r'$ & 19.2/19.4/19.3 \\
    & 56504.17318 & 56504.18927(3) & 9.8/3.2/3.2 & 33 & \textsc{wht/ucam} & $u'/g'/i'$ & 19.1/19.2/18.9 \\
    & 56510.10544 & 56510.11505(3) & 6.5/3.2/3.2 & 30 & \textsc{wht/ucam} & $u'/g'/r'$ & 17.7/17.8/17.8 \\
    & 56657.84860 & 56657.86688(3) & 9.8/3.2/3.2 & 48 & \textsc{wht/ucam} & $u'/g'/r'$ & 18.8/18.9/18.7 \\
    & 56657.92579 & 56657.93272(3) & 9.8/3.2/3.2 & 24 & \textsc{wht/ucam} & $u'/g'/r'$ & 19.1/19.3/19.1 \\
    & 56658.89473 & 56658.92033(3) & 12/4/4  & 50 & \textsc{wht/ucam} & $u'/g'/r'$ & 19.2/19.4/19.2 \\
    & 56683.54083 & 56683.54564(3) & 5.9 & 21 & \textsc{tnt/uspec} & \textit{KG5} & 19.5 \\
    & 56685.57737 & 56685.58670(8) & 9.4 & 33 & \textsc{tnt/uspec} & \textit{KG5} & 19.6 \\
    & 56689.53020 & 56689.53730(4) & 9.4 & 22 & \textsc{tnt/uspec} & \textit{KG5} & 19.2 \\
    & 56873.08525 & 56873.10772(3) & 9/3/3 & 57 & \textsc{wht/ucam} & $u'/g'/r'$ & 19.0/19.0/18.8 \\
    & 56874.07296 & 56874.09532(3) & 9/3/3 & 41 & \textsc{wht/ucam} & $u'/g'/r'$ & 18.8/18.9/18.8 \\
    & 56988.71680 & 56988.72799(3) & 4.1 & 24 & \textsc{tnt/uspec} & \textit{KG5} & 19.0 \\
    & 56989.50680 & 56989.51817(3) & 4.1 & 20 & \textsc{tnt/uspec} & \textit{KG5} & 19.2 \\
    & 57366.64759 & 57366.66722(3) & 7.9 & 48 & \textsc{tnt/uspec} & \textit{KG5} & 16.7 \\
\\
  CSS090102:132536+210037 & 55711.97933 & 55711.98534(2) & 15/5/5  & 19 & \textsc{ntt/ucam} & $u'/g'/r'$ & 19.9/19.9/19.7 \\
    & 55712.03626 & 55712.04774(2) & 15/5/5  & 26 & \textsc{ntt/ucam} & $u'/g'/r'$ & 19.8/19.9/19.6 \\
    & 55714.01658 & 55714.04404(2) & 15/5/5  & 100& \textsc{ntt/ucam} & $u'/g'/r'$ & 19.8/19.9/19.6 \\
 & & 55714.10641(2) \\
    & 55943.10286 & 55943.12141(1) & 18/6/6  & 50 & \textsc{wht/ucam} & $u'/g'/r'$ & 20.4/20.2/20.1 \\
    & 55943.21976 & 55943.24618(1) & 18/6/6  & 55 & \textsc{wht/ucam} & $u'/g'/r'$ & 19.9/19.8/19.8 \\
    & 56873.87607 & 56873.90434(1) & 9/3/3 & 57 & \textsc{wht/ucam} & $u'/g'/r'$ & 20.2/20.1/20.2 \\
    & 56874.89075 & 56874.90250(1) & 9/3/3 & 31 & \textsc{wht/ucam} & $u'/g'/i'$ & 20.1/20.1/20.2 \\
\\
  CSS090219:044027+023301 & 56960.08801 & - & 150 & 50 & \textit{pt5m} & \textit{V} & 18.1 \\
    & 56961.02405 & - & 150 & 340 & \textit{pt5m} & \textit{V} & 18.5 \\
    & 57003.94554 & - & 150 & 54 & \textit{pt5m} & \textit{V} & 18.6 \\
    & 57015.91831 & - & 150 & 295 & \textit{pt5m} & \textit{V} & 18.6 \\
\\
  CSS090419:162620-125557 & 56486.92629 & 56486.93325(5) & 12/4/4  & 18 & \textsc{wht/ucam} & $u'/g'/i'$ & 21.3/21.1/20.6 \\
    & 56498.90349 & 56498.92860(5) & 12/4/4  & 58 & \textsc{wht/ucam} & $u'/g'/i'$ & 20.7/20.6/20.0 \\
    & 56499.87896 & 56499.90938(5) & 12/4/4  & 55 & \textsc{wht/ucam} & $u'/g'/r'$ & 20.7/20.8/20.4 \\
    & 56501.92501 & 56501.94635(5) & 15/3.9/3.9  & 50 & \textsc{wht/ucam} & $u'/g'/i'$ & 20.9/20.7/20.1 \\
    & 56508.87512 & 56508.88707(5) & 12/4/4  & 30 & \textsc{wht/ucam} & $u'/g'/r'$ & 20.5/20.4/20.2 \\
    & 56508.95486 & 56508.96249(5) & 12/4/4  & 26 & \textsc{wht/ucam} & $u'/g'/r'$ & 21.1/21.0/20.5 \\
    & 56872.87452 & 56872.89818(5) & 4.2/4.2 & 51 & \textsc{wht/ucam} & $g'/r'$ & 21.8/20.4 \\
\\
  CSS090622:215636+193242 & 56873.91771 & 56873.9890(5) & 120 & 243 & \textit{pt5m} & \textit{V} & 19.2 \\
 & & 56874.0621(5) \\
    & 56874.10512 & 56874.13184(5) & 9/3/3 & 49 & \textsc{wht/ucam} & $u'/g'/r'$ & 19.6/19.5/19.1 \\
    & 56874.19057 & 56874.20268(5) & 9/3/3 & 24 & \textsc{wht/ucam} & $u'/g'/r'$ & 19.5/19.6/19.2 \\
    & 56874.96361 & 56874.98303(5) & 9/3/3 & 36 & \textsc{wht/ucam} & $u'/g'/i'$ & 19.2/19.2/18.0 \\
    & 56877.93877 & 56877.96200(5) & 9/3/3 & 39 & \textsc{wht/ucam} & $u'/g'/i'$ & 20.0/19.9/18.7 \\
    & 56878.15537 & 56878.17474(5) & 9/3/3 & 39 & \textsc{wht/ucam} & $u'/g'/i'$ & 19.7/19.6/18.3 \\
    & 56880.20188 & 56880.23174(5) & 9/3/3 & 47 & \textsc{wht/ucam} & $u'/g'/i'$ & 19.7/19.8/18.3 \\
    & 56992.49766 & 56992.5128(2) & 5.9 & 37 & \textsc{tnt/uspec} & \textit{KG5} & 19.5 \\
\\
  CSS091116:232551-014024 & 56924.94657 & - & 100 & 289 & \textit{pt5m} & \textit{V} & 18.9 \\
    & 56939.84366 & - & 150 & 369 & \textit{pt5m} & \textit{V} & 18.4 \\
\\
  CSS100218:043829+004016 & 56657.88694 & 56657.89355(2) & 12/4/4 & 11 & \textsc{wht/ucam} & $u'/g'/r'$ & 19.3/19.5/19.3 \\
    & 56657.94634 & 56657.95904(1) & 12/4/4 & 26 & \textsc{wht/ucam} & $u'/g'/r'$ & 19.3/19.5/19.3 \\
    & 56658.99048 & 56659.00696(1) & 12/4/4 & 20 & \textsc{wht/ucam} & $u'/g'/r'$ & 20.2/19.9/19.7 \\
    & 56991.77750 & - & 6.0 & 53 & \textsc{tnt/uspec} & \textit{KG5} & 19.9 \\
    & 57081.57619 & - & 30  & 50 & \textsc{tnt/uspec} & \textit{KG5} & 20.2 \\
\\
  CSS100508:085604+322109 & 57091.85566 & - & 150 & 141 & \textit{pt5m} & \textit{V} & 19.5 \\
                          & 57097.91622 & - & 150 & 288 & \textit{pt5m} & \textit{V} & 19.8 \\
\\
  CSS100520:214426+222024 & 56938.83073 & - & 120 & 353 & \textit{pt5m} & \textit{V} & 18.0 \\
\\
  CSS110113:043112-031452 & 55580.09688 & 55580.12266(5) & 7.2/2.4/2.4 & 47 & \textsc{ntt/ucam} & $u'/g'/i'$ & 16.1/15.8/15.9 \\
    & 55940.94016 & 55940.95858(3) & 12/4/4  & 37 & \textsc{wht/ucam} & $u'/g'/r'$ & 19.9/19.9/19.5 \\
    & 55942.00021 & 55942.01540(3) & 20/5/5  & 32 & \textsc{wht/ucam} & $u'/g'/r'$ & 20.1/20.0/19.7 \\
    & 55942.85431 & 55942.87405(3) & 12/4/4  & 34 & \textsc{wht/ucam} & $u'/g'/r'$ & 20.0/20.0/19.8 \\
    & 56180.11393 & 56180.12877(3) & 12/4/4  & 30 & \textsc{wht/ucam} & $u'/g'/r'$ & 19.7/19.8/19.6 \\
    & 56214.05905 & 56214.07895(3) & 9/3/3 & 35 & \textsc{wht/ucam} & $u'/g'/r'$ & 19.8/19.8/19.6 \\
    & 56659.90460 & 56659.92234(3) & 12/4/  & 27 & \textsc{wht/ucam} & $u'/g'/r'$ & 19.1/19.0/18.7 \\
    & 56685.54492 & 56685.55006(3) & 9.4 & 19 & \textsc{tnt/uspec} & \textit{KG5} & 20.0 \\
    & 56685.60074 & 56685.61612(3) & 9.4 & 35 & \textsc{tnt/uspec} & \textit{KG5} & 20.1 \\
    & 56686.50253 & 56686.54082(3) & 9.4 & 81 & \textsc{tnt/uspec} & $g'$ & 19.8 \\
    & 56689.56937 & 56689.57917(3) & 9.0 & 44 & \textsc{tnt/uspec} & $g'$ & 19.8 \\
    & 56730.84840 & 56730.86095(3) & 8.0/2.6/2.6 & 42 & \textsc{wht/ucam} & $u'/g'/r'$ & 19.7/19.7/19.4 \\
\\
  CSS110114:091246-034916 & 57046.98180 & - & 120 & 105 & \textit{pt5m} & \textit{V} & 18.0 \\
    & 57051.94906 & - & 120 & 105 & \textit{pt5m} & \textit{V} & 16.4 \\
    & 57053.99243 & - & 120 & 295 & \textit{pt5m} & \textit{V} & 17.8 \\
    & 57054.99178 & - & 120 & 291 & \textit{pt5m} & \textit{V} & 18.4 \\
\\
  CSS110226:112510+231036 & 57049.06558 & - & 150 & 318 & \textit{pt5m} & \textit{V} & 19.5 \\
    & 57068.08501 & - & 120 & 121 & \textit{pt5m} & \textit{V} & 19.5 \\
    & 57070.05512 & - & 120 & 304 & \textit{pt5m} & \textit{V} & 19.7 \\
\\
  CSS110513:210846-035031 & 56176.01207 & 56176.0486(4) & 120 & 126 & \textit{pt5m} & \textit{V} & 17.7 \\
    & 56176.91862 & 56176.9894(3) & 120 & 239 & \textit{pt5m} & \textit{V} & 17.9 \\
    & 56177.85543 & 56177.9306(3) & 120 & 157 & \textit{pt5m} & \textit{V} & 17.9 \\
    & 56486.97477 & 56487.0769(3) & 60  & 209 & \textit{pt5m} & \textit{V} & 18.3 \\
    & 56489.07325 & 56489.11534(8) & 11/3.8/3.8  & 80 & \textsc{wht/ucam} & $u'/g'/i'$ & 18.6/18.3/17.4 \\
    & 56508.97803 & 56509.04487(8) & 11/3.8/3.8  & 115 & \textsc{wht/ucam} & $u'/g'/r'$ & 18.5/18.3/17.7 \\
    & 56871.98925 & 56872.01611(5) & 8/2/2 & 64 & \textsc{wht/ucam} & $u'/g'/r'$ & 18.0/17.9/17.5 \\
    & 56872.94317 & 56872.95764(5) & 9.1/2.2/2.2 & 40 & \textsc{wht/ucam} & $u'/g'/r'$ & 18.0/17.9/17.5 \\
\\
  CSS111003:054558+022106 & 56976.09353 & 56976.1086(6) & 150 & 61  & \textit{pt5m} & \textit{V} & 18.6 \\
    & 56987.07470 & - & 150 & 135 & \textit{pt5m} & \textit{V} & 18.2 \\
    & 57017.93566 & 57017.9642(5) & 150 & 76  & \textit{pt5m} & \textit{V} & 18.8 \\
    & 57020.91573 & 57020.9908(8) & 150 & 342 & \textit{pt5m} & \textit{V} & 19.1 \\
 & & 57021.1107(7) \\
    & 57024.71894 & 57024.7369(2) & 3.9 & 78  & \textsc{tnt/uspec} & \textit{KG5} & 18.7 \\
    & 57027.61787 & 57027.6402(2) & 9.7 & 53  & \textsc{tnt/uspec} & \textit{KG5} & 18.7 \\
    & 57028.93920 & 57028.9727(5) & 120 & 69  & \textit{pt5m} & \textit{V} & 19.2 \\
    & 57077.58668 & 57077.6014(2) & 8.9 & 57  & \textsc{tnt/uspec} & $g'$ & 19.1 \\
    & 57078.53462 & 57078.5690(2) & 8.9 & 87  & \textsc{tnt/uspec} & $r'$ & 18.5 \\
    & 57365.74728 & - & 15  & 38  & \textsc{tnt/uspec} & $g'$ & 19.1 \\
    & 57367.68063 & 57367.6917(2) & 16  & 36  & \textsc{tnt/uspec} & $g'$ & 19.0 \\
\\
  CSS111019:233313-155744 & 55867.80455 & 55867.8116(6) & 20/10/10 & 53 & \textsc{wht/ucam} & $u'/g'/r'$ & 20.1/19.7/19.4 \\
    & 56874.14164 & 56874.1486(1) & 9/3/3 & 130 & \textsc{wht/ucam} & $u'/g'/r'$ & 18.4/18.2/18.2 \\
 & & 56874.1914(1) \\
 & & 56874.2342(1) \\
    & 56878.18496 & 56878.2193(2) & 7/2.3/2.3 & 54 & \textsc{wht/ucam} & $u'/g'/i'$ & 20.0/19.5/19.3 \\
    & 56879.18534 & 56879.2049(5) & 9/3/3 & 80 & \textsc{wht/ucam} & $u'/g'/i'$ & 20.1/19.7/19.4 \\
\\
  CSS111101:233003+303301 & 56874.09009 & 56874.155(1) & 120 & 191 & \textit{pt5m} & \textit{V} & 18.1 \\
    & 56874.95930 & 56875.092(1) & 120 & 256 & \textit{pt5m} & \textit{V} & 18.2 \\
    & 56876.95942 & 56877.118(1) & 120 & 259 & \textit{pt5m} & \textit{V} & 18.1 \\
    & 57016.81470 & 57016.8744(1) & 150 & 104 & \textit{pt5m} & \textit{V} & 17.6 \\
    & 57017.83588 & - & 120 & 140 & \textit{pt5m} & \textit{V} & 18.4 \\
    & 57020.81609 & - & 120 & 140 & \textit{pt5m} & \textit{V} & 18.4 \\
    & 57026.53984 & 57026.5459(3) & 7.9 & 71 & \textsc{tnt/uspec} & \textit{KG5} & 18.5 \\
    & 57026.83120 & 57026.8580(5) & 120 & 69  & \textit{pt5m} & \textit{V} & 18.0 \\
    & 57028.53937 & 57028.5731(3) & 7.9 & 94 & \textsc{tnt/uspec} & \textit{KG5} & 18.4 \\
    & 57028.85964 & 57028.8857(6) & 120 & 69  & \textit{pt5m} & \textit{V} & 18.5 \\
\\
  CSS130906:064726+491542 & 56542.20638 & - & 2   & 49  & \textit{pt5m} & \textit{V} & 13.8 \\
    & 57043.96248 & - & 120 & 321 & \textit{pt5m} & \textit{V} & 17.5 \\
\\
  CSS131106:052412+004148 & 56604.20805 & - & 35 & 99  & \textit{pt5m} & \textit{V} & 15.6 \\
    & 56940.10939 & 56940.181(9) & 120 & 204 & \textit{pt5m} & \textit{V} & 17.8 \\
    & 56943.14634 & 56943.150(9) & 150 & 153 & \textit{pt5m} & \textit{V} & 18.2 \\
    & 56946.10810 & 56946.117(1) & 150 & 210 & \textit{pt5m} & \textit{V} & 15.8 \\
    & 56956.18535 & 56956.248(9) & 150 & 105 & \textit{pt5m} & \textit{V} & 18.1 \\
    & 56970.03219 & 56970.0459(2) & 150 & 142 & \textit{pt5m} & \textit{V} & 16.2 \\
    & 56976.15126 & 56976.160(3) & 150 & 152 & \textit{pt5m} & \textit{V} & 18.1 \\
    & 56987.00005 & - & 150 & 104 & \textit{pt5m} & \textit{V} & 18.2 \\
    & 56990.78805 & 56990.83142(5) & 4.9 & 96 & \textsc{tnt/uspec} & $g'$ & 18.6 \\
    & 56992.71556 & 56992.75265(5) & 4.9 & 86 & \textsc{tnt/uspec} & \textit{KG5} & 17.9 \\
    & 57023.63566 & 57023.66863(5) & 7.9 & 68 & \textsc{tnt/uspec} & \textit{KG5} & 18.1 \\
    & 57027.65700 & 57027.68606(5) & 8.0 & 73 & \textsc{tnt/uspec} & \textit{KG5} & 18.3 \\
    & 57080.59617 & 57080.60994(5) & 15  & 65 & \textsc{tnt/uspec} & $g'$ & 18.0 \\
    & 57361.78732 & 57361.82302(5) & 5.0 & 95 & \textsc{tnt/uspec} & $g'$ & 18.0 \\
    & 57365.64517 & 57365.666(2) & 15  & 55 & \textsc{tnt/uspec} & $u'$ & 18.3 \\
    & 57417.49956 & 57417.54151(5) & 8.0 & 100 & \textsc{tnt/uspec} & $g'$ & 18.2 \\
\\
  CSS140402:173048+554518 & 56754.06043 & - & 15 & 78 & \textit{pt5m} & \textit{V} & 14.6 \\
\\
  CSS140901:013309+133234 & 56904.17086 & - & 170 & 97 & \textit{pt5m} & \textit{V} & 16.7 \\
    & 56908.07756 & - & 150 & 233 & \textit{pt5m} & \textit{V} & 17.1 \\
    & 56913.97688 & - & 150 & 177 & \textit{pt5m} & \textit{V} & 17.8 \\
\\
  CSS141005:023428-045431 & 56969.93998 & - & 60 & 120 & \textit{pt5m} & \textit{V} & 14.6 \\
    & 56995.95685 & - & 60 & 174 & \textit{pt5m} & \textit{V} & 14.7 \\
    & 56997.88016 & - & 60 & 276 & \textit{pt5m} & \textit{V} & 14.9 \\
    & 56998.91431 & - & 60 & 107 & \textit{pt5m} & \textit{V} & 14.7 \\
    & 57027.88719 & - & 60 & 148 & \textit{pt5m} & \textit{V} & 14.8 \\
\\
  CSS141117:030930+263804 & 57038.82444 & - & 120 & 51  & \textit{pt5m} & \textit{V} & 18.2 \\
    & 57041.82943 & - & 120 & 158 & \textit{pt5m} & \textit{V} & 18.5 \\
    & 57052.85159 & - & 120 & 48  & \textit{pt5m} & \textit{V} & 18.3 \\
    & 57053.83222 & - & 120 & 160 & \textit{pt5m} & \textit{V} & 18.4 \\
    & 57054.83256 & - & 120 & 225 & \textit{pt5m} & \textit{V} & 18.3 \\
\\
  CzeV404 Her & 56740.83024 & 56740.86289(8) & 4.4 & 77 & \textsc{tnt/uspec} & $g'$ & 17.1 \\
    & 56742.88593 & 56742.9211(2) & 3.4 & 82 & \textsc{tnt/uspec} & $g'$ & 15.1 \\
    & 56854.02607 & 56854.0773(2) & 60  & 98 & \textit{pt5m} & \textit{V} & 14.8 \\
    & 56856.89675 & 56856.9201(2) & 60  & 69 & \textit{pt5m} & \textit{V} & 14.1 \\
    & 56871.87078 & 56871.91729(5) & 3.6/1.8/1.8 & 80 & \textsc{wht/ucam} & $u'/g'/r'$ & 17.1/17.0/16.6 \\
    & 57079.90232 & 57079.91840(5) & 3.4 & 58 & \textsc{tnt/uspec} & \textit{KG5} & 16.1 \\
    & 57083.92911 & 57083.9373(2) & 3.4 & 33 & \textsc{tnt/uspec} & \textit{KG5} & 14.9 \\
\\
  Gaia15aan & 57054.25900 & - & 30 & 40  & \textit{pt5m} & \textit{V} & 18.3 \\
    & 57055.25668 & - & 30 & 42  & \textit{pt5m} & \textit{V} & 18.6 \\
    & 57057.21084 & - & 60 & 107 & \textit{pt5m} & \textit{V} & 18.6 \\
    & 57059.17340 & - & 60 & 40  & \textit{pt5m} & \textit{V} & 18.8 \\
\\
  GALEX J003535.7+462353 & 55940.81525 & 55940.8250(1) & 5.6/1.8/1.8 & 50 & \textsc{wht/ucam} & $u'/g'/r'$ & 17.1/17.0/16.6 \\
    & 55941.83001 & 55941.8592(5) & 5.6/1.8/1.8 & 65 & \textsc{wht/ucam} & $u'/g'/r'$ & 16.5/15.9/15.7 \\
    & 56179.91441 & 56179.9424(1) & 4.6/1.5/1.5 & 65 & \textsc{wht/ucam} & $u'/g'/r'$ & 16.7/16.4/16.1 \\
    & 56209.90088 & 56209.9185(1) & 6/2/2 & 55 & \textsc{wht/ucam} & $u'/g'/r'$ & 16.4/16.3/15.9 \\
    & 56210.93470 & 56210.9523(1) & 5/1/1 & 39 & \textsc{wht/ucam} & $u'/g'/r'$ & 16.6/16.3/16.1 \\
    & 56500.18602 & 56500.2015(1) & 6/2/2 & 78 & \textsc{wht/ucam} & $u'/g'/r'$ & 16.4/16.3/16.0 \\
    & 56505.17727 & 56505.1977(1) & 6/2/2 & 53 & \textsc{wht/ucam} & $u'/g'/i'$ & 17.0/16.9/16.3 \\
    & 56509.99699 & 56510.0215(1) & 6/2/2 & 75 & \textsc{wht/ucam} & $u'/g'/r'$ & 16.9/16.8/16.4 \\
    & 56510.16914 & 56510.1936(1) & 6/2/2 & 65 & \textsc{wht/ucam} & $u'/g'/r'$ & 16.7/16.5/16.2 \\
    & 56657.79859 & 56657.8333(1) & 6/2/2 & 68 & \textsc{wht/ucam} & $u'/g'/r'$ & 16.6/16.3/16.0 \\
    & 56658.83889 & 56658.8669(1) & 6/2/2 & 75 & \textsc{wht/ucam} & $u'/g'/r'$ & 16.4/16.4/15.9 \\
    & 56872.09341 & 56872.1433(1) & 6/2/2 & 100 & \textsc{wht/ucam} & $u'/g'/r'$ & 16.6/16.4/16.1 \\
    & 56873.12930 & 56873.1770(1) & 6/2/2 & 117 & \textsc{wht/ucam} & $u'/g'/r'$ & 16.6/16.4/16.0 \\
    & 56988.57231 & 56988.6012(1) & 1.9 & 87 & \textsc{tnt/uspec} & \textit{KG5} & 16.4 \\
    & 57023.55486 & 57023.5727(1) & 1.9 & 45 & \textsc{tnt/uspec} & \textit{KG5} & 16.4 \\
    & 57365.69822 & 57365.7111(1) & 10  & 56 & \textsc{tnt/uspec} & $g'$ & 17.2 \\
\\
  GY Cnc & 55938.24332 & 55938.26366(4) & 12/4/4 & 56 & \textsc{wht/ucam} & $u'/g'/r'$ & 16.7/16.6/16.2 \\
    & 55941.22656 & 55941.24626(4) & 6.2/3.1/3.1 & 52 & \textsc{wht/ucam} & $u'/g'/r'$ & 16.7/16.8/16.2 \\
    & 55942.98285 & 55943.00068(4) & 7.5/2.5/2.5 & 46 & \textsc{wht/ucam} & $u'/g'/r'$ & 17.0/16.9/16.3 \\
    & 55943.14044 & 55943.17605(4) & 7.5/2.5/2.5 & 71 & \textsc{wht/ucam} & $u'/g'/r'$ & 16.7/16.5/16.1 \\
    & 55947.18590 & 55947.21130(4) & 12/4/4  & 51 & \textsc{wht/ucam} & $u'/g'/r'$ & 16.6/16.6/16.1 \\
    & 56657.18967 & 56657.22682(5) & 12/4/4  & 74 & \textsc{wht/ucam} & $u'/g'/r'$ & 16.3/16.4/16.0 \\
    & 56683.69249 & 56683.71865(5) & 1.3 & 53 & \textsc{tnt/uspec} & \textit{KG5} & 16.5 \\
    & 57080.55294 & 57080.56902(5) & 2.5 & 50 & \textsc{tnt/uspec} & \textit{KG5} & 16.7 \\
    & 57367.73864 & 57367.76812(5) & 4.0 & 91 & \textsc{tnt/uspec} & \textit{KG5} & 16.6 \\
\\
  HS 2325+8205 & 55941.87952 & 55941.89303(5) & 5.6/1.8/1.8 & 48 & \textsc{wht/ucam} & $u'/g'/r'$ & 14.7/14.4/14.4 \\
    & 56487.16448 & 56487.19622(5) & 5.6/1.8/1.8 & 94 & \textsc{wht/ucam} & $u'/g'/i'$ & 17.1/16.7/16.1 \\
    & 56509.13073 & 56509.15629(5) & 5.6/1.8/1.8 & 76 & \textsc{wht/ucam} & $u'/g'/r'$ & 15.2/14.9/14.8 \\
\\
  HT Cas & 54400.23091 & 54400.24274(5) & 9.4/2.3/2.3 & 23 & \textsc{wht/ucam} & $u'/g'/r'$ & 16.4/16.6/16.5 \\
    & 56872.19923 & 56872.21011(5) & 6.6/2.2/2.2 & 46 & \textsc{wht/ucam} & $u'/g'/r'$ & 16.5/16.6/16.4 \\
    & 56874.03047 & 56874.05133(5) & 6.3/2.1/2.1 & 56 & \textsc{wht/ucam} & $u'/g'/r'$ & 16.2/16.5/16.3 \\
    & 56988.48215 & 56988.49911(5) & 1.0 & 36 & \textsc{tnt/uspec} & $g'$ & 16.8 \\
    & 56993.48629 & 56993.50711(5) & 2.0 & 32 & \textsc{tnt/uspec} & $g'$ & 16.5 \\
    & 57023.61271 & 57023.62882(5) & 2.0 & 29 & \textsc{tnt/uspec} & \textit{KG5} & 16.8 \\
    & 57024.65162 & 57024.65988(5) & 2.0 & 22 & \textsc{tnt/uspec} & $g'$ & 16.7 \\
    & 57285.06842 & 57285.07646(5) & 6.6/2.2/2.2 & 15 & \textsc{wht/ucam} & $u'/g'/r'$ & 16.0/16.3/16.1 \\
\\
  IY UMa & 56746.62944 & 56746.64028(2) & 2.2 & 46 & \textsc{tnt/uspec} & \textit{KG5} & 17.6 \\
    & 56746.70652 & 56746.71419(2) & 2.2 & 28 & \textsc{tnt/uspec} & \textit{KG5} & 17.7 \\
    & 56746.78005 & 56746.78810(2) & 2.2 & 33 & \textsc{tnt/uspec} & \textit{KG5} & 17.6 \\
    & 56991.93503 & 56991.94411(5) & 3.4 & 28 & \textsc{tnt/uspec} & \textit{KG5} & 14.5 \\
    & 57025.92282 & 57025.94209(3) & 3.4 & 40 & \textsc{tnt/uspec} & $g'$ & 17.4 \\
    & 57028.88243 & 57028.89840(5) & 4.0 & 30 & \textsc{tnt/uspec} & $r'$ & 17.3 \\
    & 57076.85098 & 57076.86533(3) & 3.9 & 41 & \textsc{tnt/uspec} & $r'$ & 17.2 \\
\\
  MAST003059.39+301634.3 & 56904.95602 & 56904.9976(3) & 90 & 407 & \textit{pt5m} & \textit{V} & 19.4 \\
 & & 56905.0685(3) \\
 & & 56905.1396(5) \\
 & & 56905.2084(5) \\
    & 56907.94797 & - & 120 & 117 & \textit{pt5m} & \textit{V} & 19.7 \\
    & 56921.95125 & 56922.002(1) & 120 & 148 & \textit{pt5m} & \textit{V} & 19.7 \\
    & 56922.95876 & 56922.984(1) & 120 & 343 & \textit{pt5m} & \textit{V} & 19.8 \\
 & & 56923.0559(9) \\
 & & 56923.1263(3) \\
 & & 56923.1965(8) \\
    & 56992.64865 & 56992.6851(1) & 5.9 & 61 & \textsc{tnt/uspec} & \textit{KG5} & 19.9 \\
    & 57023.58867 & 57023.6008(1) & 6.9 & 33 & \textsc{tnt/uspec} & \textit{KG5} & 19.2 \\
    & 57199.17958 & 57199.18681(4) & 10.5/3.5/3.5 & 29 & \textsc{wht/ucam} & $u'/g'/r'$ & 19.6/19.7/19.5 \\
    & 57284.89172 & 57284.90695(8) & 20/6.7/6.7 & 28 & \textsc{wht/ucam} & $u'/g'/r'$ & 19.5/19.5/19.3 \\
\\
  MAST034045.31+471632.2 & 56936.97202 & - & 120 & 402 & \textit{pt5m} & \textit{V} & 18.7 \\
\\
  MAST041923.57+653004.3 & 56944.02661 & - & 120 & 326 & \textit{pt5m} & \textit{V} & 17.2 \\
    & 56960.12942 & - & 120 & 132 & \textit{pt5m} & \textit{V} & 17.3 \\
    & 56967.11017 & - & 120 & 129 & \textit{pt5m} & \textit{V} & 17.4 \\
\\
  MAST171921.40+640309.8 & 56732.06257 & - & 30 & 100 & \textit{pt5m} & \textit{V} & 16.2 \\
    & 56735.02752 & - & 40 & 155 & \textit{pt5m} & \textit{V} & 16.6 \\
\\
  MAST192328.22+612413.5 & 56774.04124 & 56774.148(5) & 120 & 263 & \textit{pt5m} & \textit{V} & 19.3 \\
    & 56779.11794 & 56779.1753(2) & 35/7/7  & 164 & \textsc{wht/ucam} & $u'/g'/r'$ & 18.2/17.9/17.7 \\
    & 56799.97068 & - & 180 & 176 & \textit{pt5m} & \textit{V} & 17.4 \\
    & 56825.92616 & 56825.9488(3) & 180 & 80  & \textit{pt5m} & \textit{V} & 18.8 \\
    & 56826.92700 & 56826.9551(3) & 180 & 119 & \textit{pt5m} & \textit{V} & 18.9 \\
    & 56834.92790 & 56835.0026(3) & 180 & 398 & \textit{pt5m} & \textit{V} & 17.2 \\
 & & 56835.1704(3) \\
    & 56873.03328 & 56873.05815(5) & 13/3.3/3.3  & 69 & \textsc{wht/ucam} & $u'/g'/r'$ & 19.3/19.1/18.8 \\
\\
  MAST194955.17+455349.6 & 56874.88619 & - & 90 & 106 & \textit{pt5m} & \textit{V} & 18.6 \\
    & 56878.95944 & - & 90 & 292 & \textit{pt5m} & \textit{V} & 18.7 \\
\\
  MAST201121.95+565531.1 & 56836.00727 & - & 60 & 203 & \textit{pt5m} & \textit{V} & 17.2 \\
\\
  MAST202157.69+212919.4 & 56894.91143 & - & 120 & 304 & \textit{pt5m} & \textit{V} & 20.3 \\
\\
  MAST203421.90+120656.9 & 56849.02716 & - & 100 & 264 & \textit{pt5m} & \textit{V} & 17.9 \\
    & 56849.96328 & - & 100 & 262 & \textit{pt5m} & \textit{V} & 17.8 \\
\\
  MAST210316.39+314913.6 & 56858.03866 & - & 80 & 255 & \textit{pt5m} & \textit{V} & 18.0 \\
    & 56879.88002 & - & 80 & 425 & \textit{pt5m} & \textit{V} & 18.3 \\
\\
  MAST232100.42+494614.0 & 56799.13236 & 56799.152(9) & 60  & 111 & \textit{pt5m} & \textit{V} & 18.9 \\
    & 56854.16419 & - & 90  & 61  & \textit{pt5m} & \textit{V} & 18.7 \\
    & 56878.02050 & 56878.1546(4) & 100 & 295 & \textit{pt5m} & \textit{V} & 16.9 \\
    & 56880.96494 & 56881.128(9) & 100 & 377 & \textit{pt5m} & \textit{V} & 18.5 \\
    & 56881.99945 & 56882.191(6) & 100 & 329 & \textit{pt5m} & \textit{V} & 18.7 \\
    & 56884.95665 & 56885.160(9) & 150 & 392 & \textit{pt5m} & \textit{V} & 18.8 \\
    & 56895.12463 & 56895.143(4) & 150 & 157 & \textit{pt5m} & \textit{V} & 19.0 \\
    & 56921.88054 & 56921.902(9) & 150 & 99  & \textit{pt5m} & \textit{V} & 19.0 \\
    & 56922.05619 & 56922.124(9) & 150 & 98  & \textit{pt5m} & \textit{V} & 18.7 \\
    & 56943.07065 & 56943.136(3) & 150 & 105 & \textit{pt5m} & \textit{V} & 18.9 \\
    & 56956.03000 & 56956.0949(3) & 150 & 105 & \textit{pt5m} & \textit{V} & 16.8 \\
    & 56988.63813 & - & 4.0 & 87 & \textsc{tnt/uspec} & \textit{KG5} & 17.9 \\
    & 56992.57897 & 56992.6242(1) & 5.0 & 97 & \textsc{tnt/uspec} & $g'$ & 19.3 \\
    & 57025.50124 & 57025.5411(1) & 6.0 & 96 & \textsc{tnt/uspec} & \textit{KG5} & 19.0 \\
    & 57028.49920 & 57028.5144(1) & 8.0 & 55 & \textsc{tnt/uspec} & \textit{KG5} & 19.1 \\
    & 57366.60377 & - & 14  & 56 & \textsc{tnt/uspec} & $g'$ & 18.5 \\
\\
  MLS101226:072033+172437 & 56770.86964 & 56770.9453(8) & 120 & 132 & \textit{pt5m} & \textit{V} & 18.0 \\
    & 56772.87068 & 56772.9016(8) & 120 & 156 & \textit{pt5m} & \textit{V} & 18.0 \\
    & 57045.01865 & 57045.1377(6) & 120 & 232 & \textit{pt5m} & \textit{V} & 18.3 \\
    & 57048.84401 & 57048.897(5) & 120 & 69  & \textit{pt5m} & \textit{V} & 18.3 \\
    & 57049.01683 & 57049.0479(6) & 120 & 67  & \textit{pt5m} & \textit{V} & 18.4 \\
    & 57051.87424 & 57051.9049(4) & 120 & 69  & \textit{pt5m} & \textit{V} & 18.6 \\
    & 57056.83720 & 57056.8694(7) & 120 & 69  & \textit{pt5m} & \textit{V} & 18.4 \\
    & 57056.98798 & 57057.0204(4) & 120 & 69  & \textit{pt5m} & \textit{V} & 18.4 \\
    & 57079.71591 & 57079.7303(2) & 6.9 & 52  & \textsc{tnt/uspec} & $g'$ & 18.7 \\
\\
  MLS120517:152507-032655 & 56772.03312 & - & 180 & 92  & \textit{pt5m} & \textit{V} & 16.3 \\
    & 56772.98238 & 56773.0340(5) & 120 & 343 & \textit{pt5m} & \textit{V} & 16.5 \\
 & & 56773.0981(5) \\
 & & 56773.1630(5) \\
    & 56775.95890 & 56775.9956(5) & 120 & 308 & \textit{pt5m} & \textit{V} & 16.5 \\
 & & 56776.0600(5) \\
 & & 56776.1244(4) \\
    & 56864.90455 & 56864.9069(5) & 120 & 104 & \textit{pt5m} & \textit{V} & 17.0 \\
 & & 56864.9710(3) \\
    & 56874.87061 & 56874.8857(3) & 3.6/1.7/1.7 & 25 & \textsc{wht/ucam} & $u'/g'/i'$ & 18.0/17.4/16.3 \\
\\
  SDSS J040714.78-064425.1 & 54392.19660 & - & 10/2.5/2.5 & 109 & \textsc{wht/ucam} & $u'/g'/r'$ & 17.8/17.7/17.3 \\
    & 54396.19804 & 54396.23455(8) & 20/5/5 & 98 & \textsc{wht/ucam} & $u'/g'/r'$ & 17.3/17.0/16.7 \\
    & 55939.82002 & 55939.9837(4) & 10 & 281 & \textit{pt5m} & \textit{R} & 15.7 \\
    & 56215.00769 & 56215.03344(8) & 2.8/1.4/1.4 & 78 & \textsc{wht/ucam} & $u'/g'/r'$ & 16.0/15.8/15.7 \\
    & 57026.70536 & 57026.7366(2) & 6.9 & 75 & \textsc{tnt/uspec} & \textit{KG5} & 15.6 \\
    & 57082.54167 & 57082.5629(1) & 6.9 & 74 & \textsc{tnt/uspec} & \textit{KG5} & 18.4 \\
    & 57083.56625 & 57083.5841(1) & 6.9 & 64 & \textsc{tnt/uspec} & \textit{KG5} & 17.8 \\
    & 57367.63138 & 57367.6551(2) & 5.1 & 64 & \textsc{tnt/uspec} & $g'$ & 15.6 \\
\\
  SDSS J075059.97+141150.1 & 55940.89611 & 55940.92964(4) & 12/4/4  & 58 & \textsc{wht/ucam} & $u'/g'/r'$ & 18.9/19.1/18.9 \\
    & 55940.99589 & 55941.02271(4) & 12/4/4  & 66 & \textsc{wht/ucam} & $u'/g'/r'$ & 19.2/19.3/19.1 \\
    & 55942.12618 & 55942.14077(4) & 12/4/4  & 41 & \textsc{wht/ucam} & $u'/g'/r'$ & 19.3/19.6/16.4 \\
    & 55943.05921 & 55943.07241(4) & 12/4/4  & 30 & \textsc{wht/ucam} & $u'/g'/r'$ & 19.3/19.7/19.5 \\
    & 56211.19435 & 56211.20263(5) & 9/3/3 & 29 & \textsc{wht/ucam} & $u'/g'/r'$ & 19.1/19.4/19.3 \\
    & 56657.98984 & 56658.02444(4) & 12/4/4  & 69 & \textsc{wht/ucam} & $u'/g'/r'$ & 18.9/19.2/19.0 \\
    & 56658.93310 & 56658.95609(4) & 12/4/4  & 39 & \textsc{wht/ucam} & $u'/g'/r'$ & 18.8/19.0/18.8 \\
    & 56659.02043 & 56659.04924(4) & 12/4/4  & 47 & \textsc{wht/ucam} & $u'/g'/r'$ & 18.8/19.0/18.8 \\
    & 56683.62169 & 56683.64491(4) & 10  & 45 & \textsc{tnt/uspec} & \textit{KG5} & 18.7 \\
    & 56684.64142 & 56684.66967(5) & 7.0 & 53 & \textsc{tnt/uspec} & \textit{KG5} & 18.4 \\
\\
  SDSS J090103.93+480911.1 & 53803.84353 & 53803.90711(2) & 5/5/5 & 114 & \textsc{wht/ucam} & $u'$/$g'$/$r'$ & 19.5/19.7/19.5 \\
    & 53804.98276 & 53804.99742(2) & 5/5/5 & 25 & \textsc{wht/ucam} & $u'$/$g'$/$r'$ & 19.2/19.4/19.2 \\
    & 53805.12890 & 53805.15321(2) & 5/5/5 & 49 & \textsc{wht/ucam} & $u'$/$g'$/$r'$ & 19.1/19.3/19.1 \\
    & 55203.93743 & 55203.96531(5) & 6.8/1.7/1.7 & 48 & \textsc{wht/ucam} & $u'$/$g'$/$r'$ & 19.4/19.6/19.5 \\
    & 55942.08419 & 55942.11712(2) & 13/4.5/4.5  & 56 & \textsc{wht/ucam} & $u'$/$g'$/$r'$ & 20.0/20.0/19.8 \\
    & 55942.15686 & 55942.19500(2) & 13/4.5/4.5  & 74 & \textsc{wht/ucam} & $u'$/$g'$/$r'$ & 20.2/20.2/20.0 \\
    & 55942.95787 & 55942.97383(2) & 13/4.5/4.5  & 31 & \textsc{wht/ucam} & $u'$/$g'$/$r'$ & 19.5/19.6/19.5 \\
    & 55943.01684 & 55943.05169(2) & 13/4.5/4.5  & 56 & \textsc{wht/ucam} & $u'$/$g'$/$r'$ & 19.7/19.8/19.6 \\
    & 55943.19203 & 55943.20746(2) & 13/4.5/4.5  & 34 & \textsc{wht/ucam} & $u'$/$g'$/$r'$ & 19.5/19.6/19.5 \\
    & 55943.26137 & 55943.28535(2) & 13/4.5/4.5  & 51 & \textsc{wht/ucam} & $u'/g'/r'$ & 19.7/19.7/19.5 \\
\\
  SDSS J090403.49+035501.2 & 56045.88320 & 56045.89912(2) & 12/12/12  & 26 & \textsc{wht/ucam} & $u'/g'/r'$ & 19.6/19.3/19.3 \\
\\
  SDSS J092009.54+004245.0 & 55700.02192 & 55700.10597(5) & 8.7/2.9/2.9 & 154 & \textsc{ntt/ucam} & $u'/g'/r'$ & 17.7/18.0/17.7 \\
    & 55708.95938 & 55708.97858(5) & 5.6/2.8/2.8 & 57 & \textsc{ntt/ucam} & $u'/g'/r'$ & 18.2/18.0/17.7 \\
    & 57026.77655 & 57026.84657(5) & 5.9 & 118 & \textsc{tnt/uspec} & \textit{KG5} & 17.9 \\
\\
  SDSS J092444.48+080150.9 & 55709.00465 & 55709.02440(7) & 15/5/5  & 42 & \textsc{ntt/ucam} & $u'/g'/r'$ & 20.0/20.1/19.2 \\
\\
  SDSS J093537.46+161950.8 & 55710.94728 & 55710.98340(6) & 15/5/5 & 88 & \textsc{ntt/ucam} & $u'/g'/r'$ & 19.5/19.2/18.7 \\
\\
  SDSS J100658.40+233724.4 & 55942.02575 & 55942.0530(1) & 12/4/4  & 79 & \textsc{wht/ucam} & $u'/g'/r'$ & 18.4/18.5/17.9 \\
    & 56682.70289 & 56682.7302(2) & 3.4 & 65 & \textsc{tnt/uspec} & \textit{KG5} & 18.2 \\
    & 56682.87478 & 56682.9163(2) & 3.4 & 85 & \textsc{tnt/uspec} & \textit{KG5} & 18.2 \\
    & 56683.81609 & 56683.8457(2) & 3.4 & 75 & \textsc{tnt/uspec} & \textit{KG5} & 18.2 \\
    & 56685.85274 & 56685.8907(2) & 3.4 & 81 & \textsc{tnt/uspec} & \textit{KG5} & 18.1 \\
    & 56690.66858 & 56690.7244(2) & 5.9 & 118 & \textsc{tnt/uspec} & $g'$ & 18.5 \\
    & 56690.87758 & 56690.9103(2) & 3.4 & 70 & \textsc{tnt/uspec} & $r'$ & 18.6 \\
    & 57362.77084 & 57362.8006(2) & 4.9 & 75 & \textsc{tnt/uspec} & $g'$ & 18.6 \\
\\
  SDSS J115207.00+404947.8 & 56730.88610 & 56730.90933(1) & 12/4/4  & 45 & \textsc{wht/ucam} & $u'/g'/r'$ & 19.6/19.5/19.5 \\
    & 56746.66410 & 56746.69503(1) & 5.4 & 57 & \textsc{tnt/uspec} & \textit{KG5} & 19.7 \\
    & 56746.73621 & 56746.76277(1) & 5.4 & 57 & \textsc{tnt/uspec} & \textit{KG5} & 19.4 \\
    & 56746.80606 & 56746.83053(1) & 5.4 & 56 & \textsc{tnt/uspec} & \textit{KG5} & 19.5 \\
\\
  SDSS J125023.84+665525.4 & 53798.98823 & 53799.00334(4) & 4/4/4 & 32 & \textsc{wht/ucam} & $u'$/$g'$/$r'$ & 18.2/18.3/18.1 \\
    & 53799.22336 & 53799.23838(4) & 4/4/4 & 23 & \textsc{wht/ucam} & $u'$/$g'$/$r'$ & 18.3/18.4/18.2 \\
    & 53799.93218 & 53799.94313(4) & 4/4/4 & 23 & \textsc{wht/ucam} & $u'$/$g'$/$r'$ & 18.6/18.6/18.5 \\
    & 53802.92144 & 53802.93866(4) & 4/4/4 & 37 & \textsc{wht/ucam} & $u'$/$g'$/$r'$ & 18.6/18.7/18.6 \\
    & 53802.97429 & 53802.99743(4) & 4/4/4 & 40 & \textsc{wht/ucam} & $u'$/$g'$/$r'$ & 18.5/18.7/18.5 \\
    & 53803.10330 & 53803.11493(4) & 4/4/4 & 23 & \textsc{wht/ucam} & $u'$/$g'$/$r'$ & 18.6/18.8/18.7 \\
    & 53803.15397 & 53803.17368(4) & 3/3/3 & 41 & \textsc{wht/ucam} & $u'$/$g'$/$r'$ & 18.6/18.7/18.6 \\
    & 53803.21193 & 53803.23237(4) & 4/4/4 & 37 & \textsc{wht/ucam} & $u'$/$g'$/$r'$ & 18.6/18.9/18.6 \\
    & 55202.24092 & 55202.25763(4) & 9/2.2/2.2 & 61 & \textsc{wht/ucam} & $u'$/$g'$/$z'$ & 18.6/19.0/18.7 \\
    & 55204.10992 & 55204.13721(4) & 11/3.8/3.8 & 52 & \textsc{wht/ucam} & $u'$/$g'$/$r'$ & 18.4/18.7/18.5 \\
    & 55204.17561 & 55204.19589(4) & 11/3.8/3.8 & 33 & \textsc{wht/ucam} & $u'$/$g'$/$r'$ & 18.5/18.8/18.6 \\
    & 55204.24567 & 55204.25463(4) & 11/3.8/3.8 & 22 & \textsc{wht/ucam} & $u'$/$g'$/$r'$ & 18.6/18.9/18.8 \\
    & 57076.91313 & 57076.92429(4) & 9.9 & 32 & \textsc{tnt/uspec} & \textit{KG5} & 17.9 \\
    & 57077.78863 & 57077.80532(4) & 9.9 & 32 & \textsc{tnt/uspec} & \textit{KG5} & 18.0 \\
\\
  SDSS J152419.33+220920.1 & 55709.06004 & 55709.09910(3) & 20/5/5 & 80 & \textsc{ntt/ucam} & $u'/g'/r'$ & 19.3/19.3/18.9 \\
                           & 55709.14436 & 55709.16439(3) & 20/5/5 & 35 & \textsc{ntt/ucam} & $u'/g'/r'$ & 19.3/19.2/18.9 \\
                           & 55712.08247 & 55712.10372(3) & 12/3/3 & 39 & \textsc{ntt/ucam} & $u'/g'/r'$ & 19.2/19.2/18.9 \\
                           & 55714.11519 & 55714.12865(3) & 12/3/3 & 26 & \textsc{ntt/ucam} & $u'/g'/r'$ & 19.0/19.1/18.8 \\
                           & 56046.05227 & 56046.07845(1) & 12/4/4 & 51 & \textsc{wht/ucam} & $u'/g'/r'$ & 18.8/19.1/19.0 \\
                           & 56046.12728 & 56046.14374(1) & 12/4/4 & 28 & \textsc{wht/ucam} & $u'/g'/r'$ & 18.6/18.9/18.7 \\
                           & 56046.95909 & 56046.99289(1) & 12/4/4 & 58 & \textsc{wht/ucam} & $u'/g'/r'$ & 18.5/18.9/18.6 \\
                           & 56486.89370 & 56486.91454(1) & 10.6/3.5/3.5 & 39 & \textsc{wht/ucam} & $u'/g'/i'$ & 19.1/19.3/18.8 \\
                           & 56494.86606 & 56494.88342(1) & 10.6/3.5/3.5 & 40 & \textsc{wht/ucam} & $u'/g'/i'$ & 19.3/19.3/19.0 \\
                           & 56503.87498 & 56503.89742(1) & 13.6/4.5/4.5 & 49 & \textsc{wht/ucam} & $u'/g'/i'$ & 19.3/19.4/19.1 \\
                           & 56509.95159 & 56509.97206(1) & 10.6/3.5/3.5 & 60 & \textsc{wht/ucam} & $u'/g'/r'$ & 19.2/19.5/19.1 \\
                           & 56719.23748 & 56719.25331(4) & 12/4/4 & 37 & \textsc{wht/ucam} & $u'/g'/r'$ & 19.5/19.7/19.6 \\
                           & 56746.87519 & 56746.88311(3) & 3.4 & 23 & \textsc{tnt/uspec} & \textit{KG5} & 19.4 \\
                           & 56871.95572 & 56871.96849(4) & 8/2/2 & 43 & \textsc{wht/ucam} & $u'/g'/r'$ & 19.0/19.2/18.9 \\
                           & 56873.91985 & 56873.92807(4) & 9/2.2/2.2 & 24 & \textsc{wht/ucam} & $u'/g'/r'$ & 19.0/19.2/18.8 \\
\\
  SDSS J155531.99-001055.0 & 54260.20835 & 54260.22695(1) & 6/2/2 & 25 & \textsc{vlt/ucam} & $u'/g'/r'$ & 19.1/19.4/19.2 \\
    & 54260.27914 & 54260.30580(1) & 6/2/2 & 54 & \textsc{vlt/ucam} & $u'/g'/r'$ & 19.0/19.4/19.2 \\
    & 54272.98205 & 54272.99993(1) & 6/2/2 & 27 & \textsc{vlt/ucam} & $u'/g'/r'$ & 19.0/19.2/19.0 \\
    & 55334.26865 & 55334.33987(2) & 15/5/5 & 123 & \textsc{ntt/ucam} & $u'/g'/r'$ & 19.0/19.3/19.0 \\
    & 55712.20386 & 55712.24662(2) & 11/3.5/3.5 & 71 & \textsc{ntt/ucam} & $u'/g'/r'$ & 19.0/19.4/19.2 \\
    & 56046.21549 & 56046.23642(2) & 12/4/4 & 27 & \textsc{wht/ucam} & $u'/g'/r'$ & 19.0/19.3/19.0 \\
    & 56176.86930 & 56176.88349(2) & 14/3.5/3.5 & 40 & \textsc{wht/ucam} & $u'/g'/r'$ & 19.0/19.4/19.2 \\
    & 56489.01055 & 56489.03300(2) & 9/3/3 & 45 & \textsc{wht/ucam} & $u'/g'/i'$ & 19.0/19.3/19.0 \\
    & 56498.88028 & 56498.88873(2) & 9.8/3.2/3.2 & 23 & \textsc{wht/ucam} & $u'/g'/i'$ & 19.1/19.4/19.1 \\
    & 56501.87310 & 56501.88484(2) & 9/3/3 & 24 & \textsc{wht/ucam} & $u'/g'/i'$ & 18.9/19.3/18.9 \\
    & 56872.91245 & 56872.93200(2) & 9.8/3.2/3.2 & 35 & \textsc{wht/ucam} & $u'/g'/r'$ & 18.8/19.2/18.9 \\
    & 56873.93949 & 56873.95698(2) & 9.8/3.2/3.2 & 48 & \textsc{wht/ucam} & $u'/g'/r'$ & 19.0/19.2/18.9 \\
    & 57081.86284 & 57081.87272(2) & 15 & 32 & \textsc{tnt/uspec} & \textit{KG5} & 19.6 \\
    & 57198.07630 & 57198.09102(2) & 9/3/3 & 24 & \textsc{wht/ucam} & $u'/g'/r'$ & 18.8/19.1/18.8 \\
\\
  SDSS J155656.92+352336.6 & 54683.88334 & 54683.95644(4) & 3.4/3.4/3.4 & 121 & \textsc{wht/ucam} & $u'/g'/r'$ & 17.5/17.2/17.2 \\
    & 55942.22648 & 55942.25538(4) & 12/4/4  & 63 & \textsc{wht/ucam} & $u'/g'/r'$ & 17.7/17.4/17.4 \\
    & 56046.01331 & 56046.02708(8) & 12/4/4  & 34 & \textsc{wht/ucam} & $u'/g'/r'$ & 18.9/18.8/18.7 \\
    & 56046.09093 & 56046.11505(8) & 12/4/4  & 49 & \textsc{wht/ucam} & $u'/g'/r'$ & 18.9/18.7/18.6 \\
    & 56046.18489 & 56046.20324(8) & 12/4/4  & 41 & \textsc{wht/ucam} & $u'/g'/r'$ & 18.9/18.8/18.8 \\
    & 56499.92019 & - & 12/4/4  & 51 & \textsc{wht/ucam} & $u'/g'/r'$ & 19.1/19.1/19.0 \\
    & 56501.89201 & 56501.9004(3) & 120 & 156 & \textit{pt5m} & \textit{V} & 18.9 \\
 & & 56501.9885(3)  \\
    & 56873.97532 & 56873.9990(1) & 6.5/2.1/2.1 & 46 & \textsc{wht/ucam} & $u'/g'/r'$ & 19.0/18.8/18.8 \\
    & 57079.85619 & 57079.8685(1) & 8.9 & 39 & \textsc{tnt/uspec} & \textit{KG5} & 18.8 \\
\\
  SSS100615:200331-284941 & 56872.97323 & 56873.024417(5) & 12/3/3  & 78 & \textsc{wht/ucam} & $u'/g'/r'$ & 19.9/19.7/19.6 \\
    & 56874.01104 & 56874.022391(5) & 9/3/3 & 21 & \textsc{wht/ucam} & $u'/g'/r'$ & 19.7/19.6/19.4 \\
    & 56874.94861 & 56874.961664(5) & 15/5/5  & 18 & \textsc{wht/ucam} & $u'/g'/i'$ & 19.4/19.5/19.8 \\
\\
  SSS120402:134015−350512 & 57027.91986 & 57027.95944(5) & 6.9 & 71 & \textsc{tnt/uspec} & \textit{KG5} & 18.6 \\
\\
  SSS130413:094551-194402 & 56683.65823 & 56683.67469(5) & 5.8 & 35 & \textsc{tnt/uspec} & \textit{KG5} & 17.0 \\
    & 56683.72993 & 56683.74054(5) & 5.8 & 21 & \textsc{tnt/uspec} & \textit{KG5} & 17.1 \\
    & 56683.78617 & 56683.80628(5) & 5.8 & 40 & \textsc{tnt/uspec} & \textit{KG5} & 17.0 \\
    & 56684.68236 & 56684.72712(5) & 5.8 & 90 & \textsc{tnt/uspec} & \textit{KG5} & 17.1 \\
    & 56685.69475 & 56685.71357(5) & 2.9 & 37 & \textsc{tnt/uspec} & \textit{KG5} & 17.1 \\
    & 56689.63539 & 56689.65982(5) & 2.9 & 46 & \textsc{tnt/uspec} & \textit{KG5} & 17.1 \\
    & 56689.84202 & 56689.85708(5) & 2.9 & 31 & \textsc{tnt/uspec} & \textit{KG5} & 17.1 \\
    & 56690.75257 & 56690.77779(5) & 2.9 & 49 & \textsc{tnt/uspec} & $g'$ & 17.1 \\
    & 56690.82365 & 56690.84355(5) & 2.9 & 33 & \textsc{tnt/uspec} & $r'$ & 17.0 \\
    & 56699.05164 & 56699.06472(3) & 1.0 & 25 & \textsc{salt/salticam} & $g'$ & 17.0 \\
    & 56739.62758 & 56739.6449(1) & 2.9 & 31 & \textsc{tnt/uspec} & \textit{KG5} & 17.2 \\
    & 56777.84234 & 56777.85636(3) & 1.0 & 35 & \textsc{salt/salticam} & $g'$ & 17.0 \\
    & 57023.75635 & 57023.76773(3) & 2.9 & 29 & \textsc{tnt/uspec} & $i'$ & 16.8 \\
    & 57416.70864 & 57416.73923(5) & 9.3 & 52 & \textsc{tnt/uspec} & $u'$ & 16.9 \\
    & 57418.69218 & 57418.71238(5) & 8.1 & 29 & \textsc{tnt/uspec} & $u'$ & 16.6 \\
\\
  V2051 Oph & 54261.28607 & 54261.31011(3) & 1/1/1 & 45 & \textsc{vlt/ucam} & $u'/g'/r'$ & 15.5/15.6/15.4 \\
    & 54264.16307 & 54264.18181(3) & 1/1/1 & 30 & \textsc{vlt/ucam} & $u'/g'/r'$ & 15.6/15.6/15.5 \\
    & 54268.27875 & 54268.30206(3) & 3/1/1 & 41 & \textsc{vlt/ucam} & $u'/g'/r'$ & 15.8/15.9/15.7 \\
    & 54268.34368 & 54268.36449(3) & 3/1/1 & 46 & \textsc{vlt/ucam} & $u'/g'/r'$ & 16.0/16.0/15.9 \\
    & 55314.14160 & 55314.15612(3) & 6/2/2 & 128 & \textsc{ntt/ucam} & $u'/g'/r'$ & 15.5/15.4/15.1 \\
   & & 55314.21854(3) &  &  &  &  &  \\
    & 55314.30287 & 55314.34340(3) & 6/2/2 & 86 & \textsc{ntt/ucam} & $u'/g'/r'$ & 15.7/15.5/15.4 \\
    & 55315.38811 & 55315.40460(3) & 6/2/2 & 59 & \textsc{ntt/ucam} & $u'/g'/r'$ & 15.7/15.6/15.4 \\
    & 55329.30344 & 55329.32629(3) & 3/1.5/1.5 & 55 & \textsc{ntt/ucam} & $u'/g'/r'$ & 15.2/15.0/14.8 \\
    & 55353.32613 & 55353.36077(3) & 4/1.3/1.3 & 69 & \textsc{ntt/ucam} & $u'/g'/r'$ & 15.6/15.5/15.3 \\
    & 55354.01710 & 55354.04752(3) & 4/1.3/1.3 & 57 & \textsc{ntt/ucam} & $u'/g'/r'$ & 15.8/15.7/15.5 \\
    & 55700.37290 & 55700.39738(3) & 9/3/3 & 44 & \textsc{ntt/ucam} & $u'/g'/r'$ & 16.0/15.7/15.5 \\
    & 55701.32116 & 55701.33373(3) & 9/3/3 & 26 & \textsc{ntt/ucam} & $u'/g'/r'$ & 15.8/15.7/15.5 \\
    & 55701.38384 & 55701.39615(3) & 9/3/3 & 57 & \textsc{ntt/ucam} & $u'/g'/r'$ & 15.5/15.5/15.3 \\
    & 55705.28872 & 55705.32906(3) & 9/3/3 & 77 & \textsc{ntt/ucam} & $u'/g'/r'$ & 15.7/15.6/15.4 \\
    & 55706.29035 & 55706.32798(3) & 10/3.3/3.3 & 91 & \textsc{ntt/ucam} & $u'/g'/r'$ & 16.0/15.9/15.6 \\
    & 55709.31110 & 55709.32452(3) & 6/2/2 & 95 & \textsc{ntt/ucam} & $u'/g'/r'$ & 16.0/15.8/15.6 \\
   & & 55709.38697(3) &  &  &  &  &  \\
    & 55710.23956 & 55710.26091(3) & 10/2/2 & 34 & \textsc{ntt/ucam} & $u'/g'/r'$ & 15.7/15.5/15.4 \\
    & 55713.23903 & 55713.25747(3) & 9/3/3 & 34 & \textsc{ntt/ucam} & $u'/g'/r'$ & 15.8/15.8/15.6 \\
    & 55713.30486 & 55713.31989(3) & 6/2/2 & 31 & \textsc{ntt/ucam} & $u'/g'/r'$ & 15.8/15.7/15.5 \\
    & 55713.36174 & 55713.38235(3) & 7.4/2.4/2.4 & 40 & \textsc{ntt/ucam} & $u'/g'/r'$ & 15.7/15.7/15.5 \\
    & 55714.24128 & 55714.25632(3) & 8.4/2.8/2.8 & 64 & \textsc{ntt/ucam} & $u'/g'/r'$ & 15.8/15.6/15.5 \\
    & 57080.91855 & 57080.92712(3) & 1.9 & 29 & \textsc{tnt/uspec} & \textit{KG5} & 15.6 \\
    & 57082.91711 & 57082.92489(3) & 2.0 & 37 & \textsc{tnt/uspec} & \textit{KG5} & 15.6 \\
    & 57084.91585 & 57084.92251(3) & 2.0 & 22 & \textsc{tnt/uspec} & \textit{KG5} & 15.6 \\
\\
  V713 Cep & 55801.14595 & 55801.18114(2) & 4.4/2.2/2.2 & 64 & \textsc{wht/ucam} & $u'/g'/i'$ & 18.9/18.8/18.7 \\
    & 56046.15044 & - & 6.3/2.1/2.1 & 46 & \textsc{wht/ucam} & $u'/g'/r'$ & 18.7/18.6/18.5 \\
    & 56176.90266 & 56176.93715(2) & 10/3.4/3.4  & 61 & \textsc{wht/ucam} & $u'/g'/r'$ & 18.4/18.3/18.2 \\
    & 56177.84463 & 56177.87676(2) & 10/3.4/3.4  & 65 & \textsc{wht/ucam} & $u'/g'/r'$ & 18.3/18.2/18.1 \\
    & 56179.99153 & 56180.01222(2) & 10/3.4/3.4  & 47 & \textsc{wht/ucam} & $u'/g'/r'$ & 18.3/18.3/18.2 \\
    & 56488.17430 & 56488.20223(2) & 10/3.4/3.4  & 74 & \textsc{wht/ucam} & $u'/g'/z'$ & 18.6/18.5/17.6 \\
    & 56489.04428 & 56489.05642(2) & 10/3.4/3.4  & 35 & \textsc{wht/ucam} & $u'/g'/i'$ & 18.6/18.4/18.2 \\
    & 56489.13138 & 56489.14184(2) & 10/3.4/3.4  & 26 & \textsc{wht/ucam} & $u'/g'/i'$ & 18.6/18.5/18.3 \\
    & 56489.18909 & 56489.22726(2) & 10/3.4/3.4  & 66 & \textsc{wht/ucam} & $u'/g'/i'$ & 18.7/18.5/18.3 \\
    & 56499.19598 & 56499.22121(2) & 10/3.4/3.4  & 59 & \textsc{wht/ucam} & $u'/g'/z'$ & 18.4/18.4/17.6 \\
    & 56509.18559 & 56509.21517(2) & 10/3.4/3.4  & 77 & \textsc{wht/ucam} & $u'/g'/r'$ & 18.5/18.5/18.4 \\
    & 56510.12816 & 56510.15480(2) & 10/3.4/3.4  & 56 & \textsc{wht/ucam} & $u'/g'/r'$ & 18.6/18.5/18.4 \\
    & 56872.03795 & 56872.07302(3) & 6/2/2 & 74 & \textsc{wht/ucam} & $u'/g'/r'$ & 18.2/18.2/18.0 \\
    & 56880.16994 & 56880.18780(3) & 10/3.4/3.4  & 41 & \textsc{wht/ucam} & $u'/g'/i'$ & 18.5/18.4/18.2 \\
    & 57198.09643 & 57198.11544(2) & 10/3.4/3.4  & 35 & \textsc{wht/ucam} & $u'/g'/r'$ & 19.2/19.0/19.0 \\
    & 57282.84397 & 57282.85057(2) & 15/5/5  & 20 & \textsc{wht/ucam} & $u'/g'/r'$ & 18.5/18.6/18.4 \\

 \hline  
\end{supertabular}
\twocolumn

\begin{figure}
 \includegraphics[width=2.55cm,angle=270]{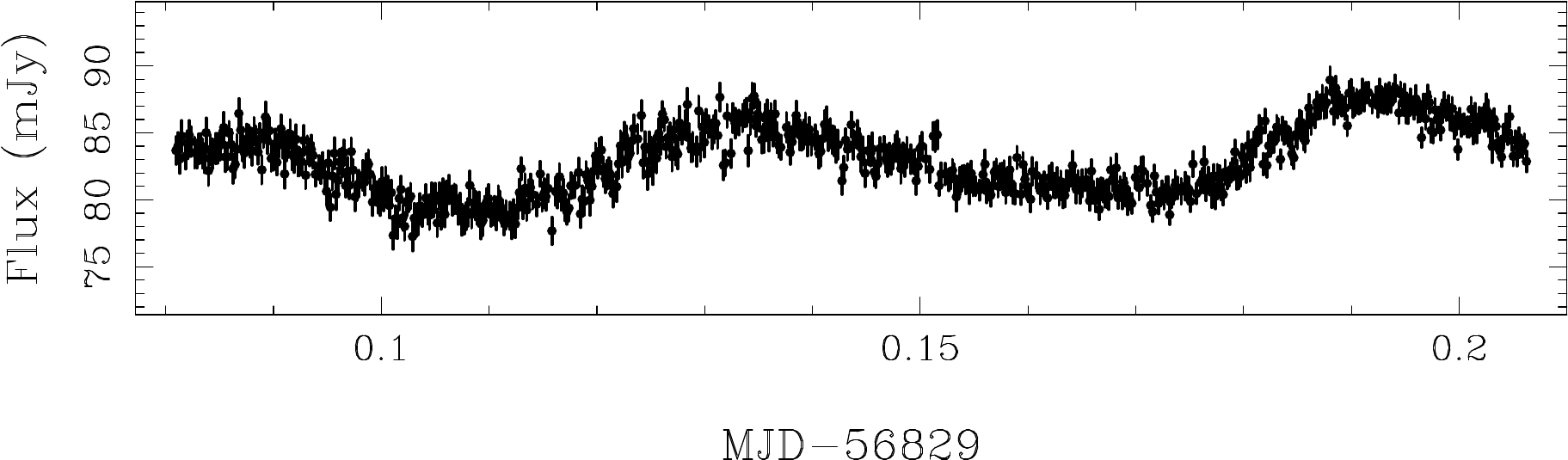}
 \caption{ASASSN-14cl light curve observed on 2014-06-20 with \textit{pt5m}, showing superhumps with a period of approximately 1.4 hours.}
 \label{fig:asassn14cl}
\end{figure}
\begin{figure}
 \includegraphics[width=2.55cm,angle=270]{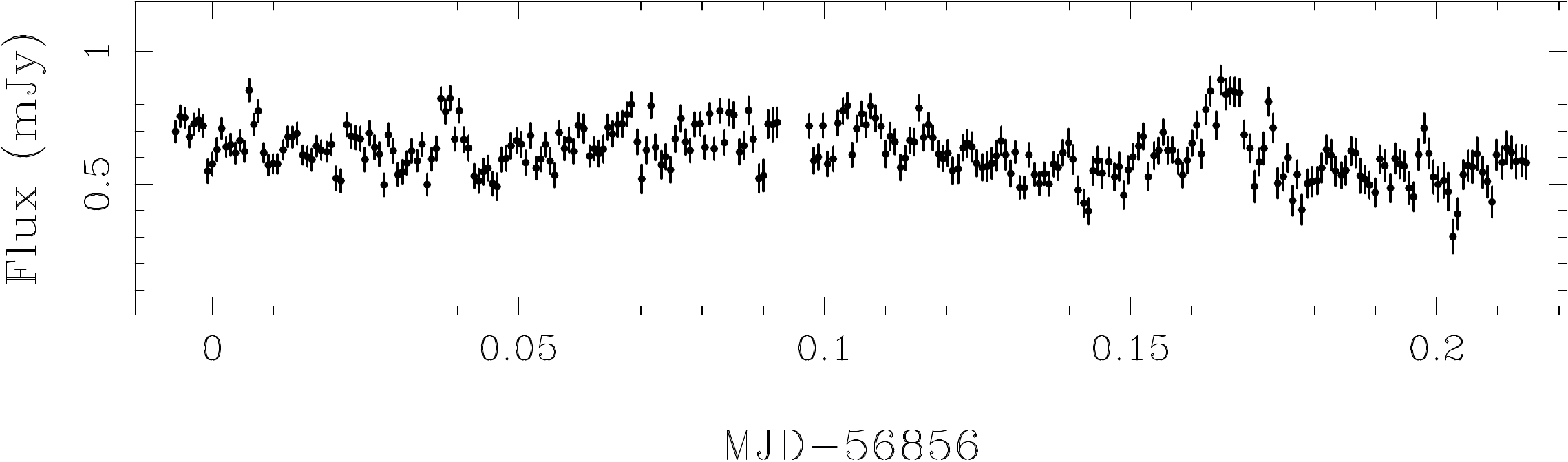}
 \caption{ASASSN-14ds light curve observed during outburst on 2014-07-17 with \textit{pt5m}. There is obvious variability, probably associated with flickering, as well as a possible sinusoidal modulation, but no eclipses.}
 \label{fig:asassn14ds}
\end{figure}
\begin{figure}
\centering
 \includegraphics[width=2.55cm,angle=270]{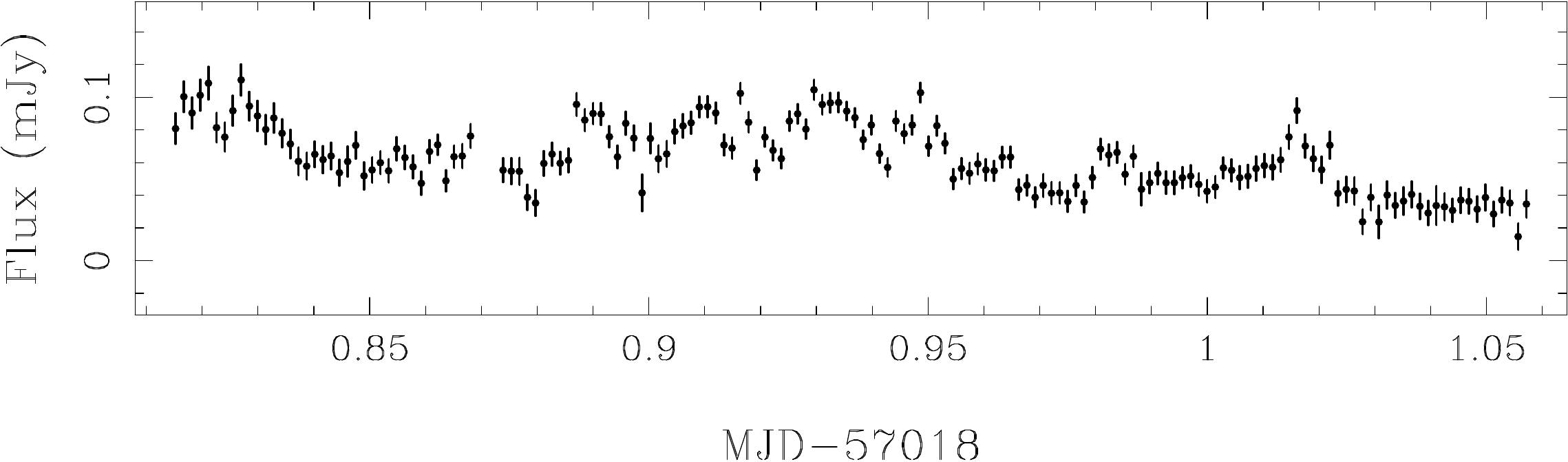}
 \caption{ASASSN-14gl light curve observed during quiescence on 2014-12-27 with \textit{pt5m}.}
 \label{fig:asassn14gl}
\end{figure}
\begin{figure}
 \includegraphics[width=2.55cm,angle=270]{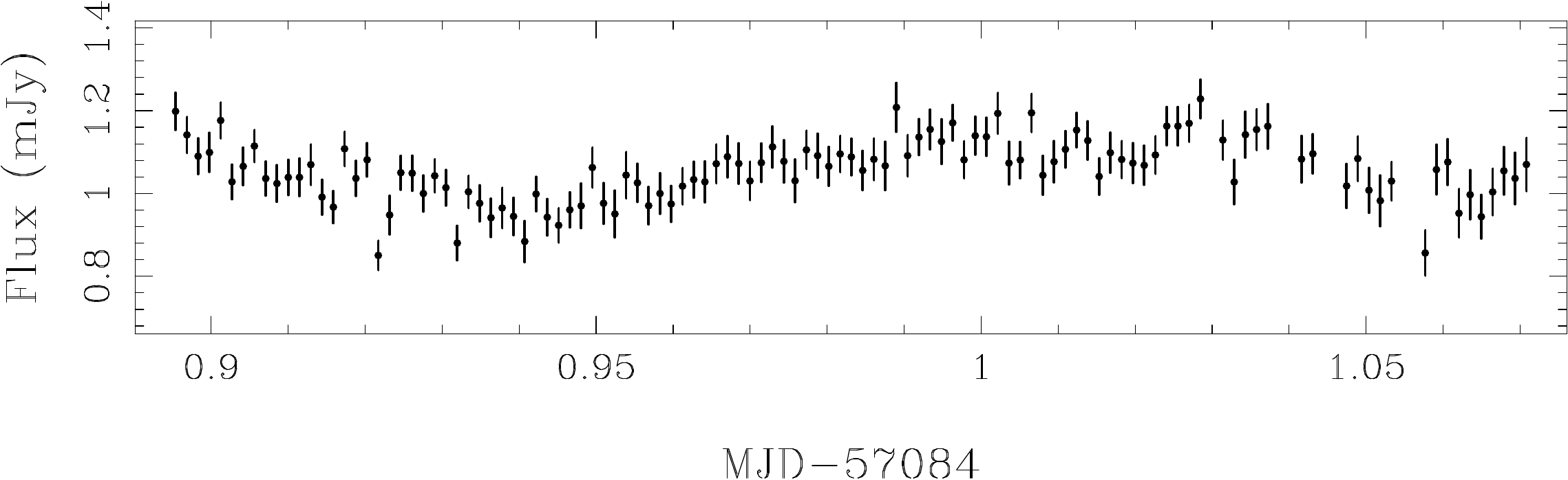}
 \caption{ASASSN-14gu light curve observed on 2015-03-03 with \textit{pt5m}. The periodic modulation is present in outburst and in quiescence, meaning that it cannot be attributed to superhumps.}
 \label{fig:asassn14gu}
\end{figure}
\begin{figure}
 \includegraphics[width=2.55cm,angle=270]{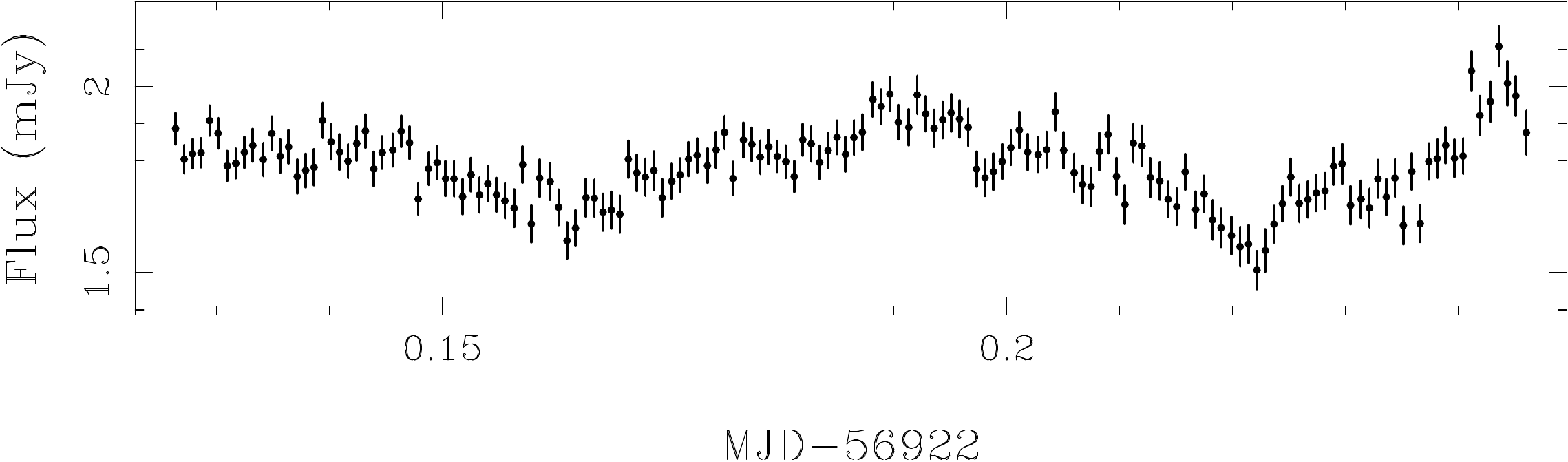}
 \caption{ASASSN-14hk light curve showing superhumps, observed during super-outburst on 2014-09-21 with \textit{pt5m}.}
 \label{fig:asassn14hk}
\end{figure}
\begin{figure}
\centering
 \includegraphics[width=2.55cm,angle=270]{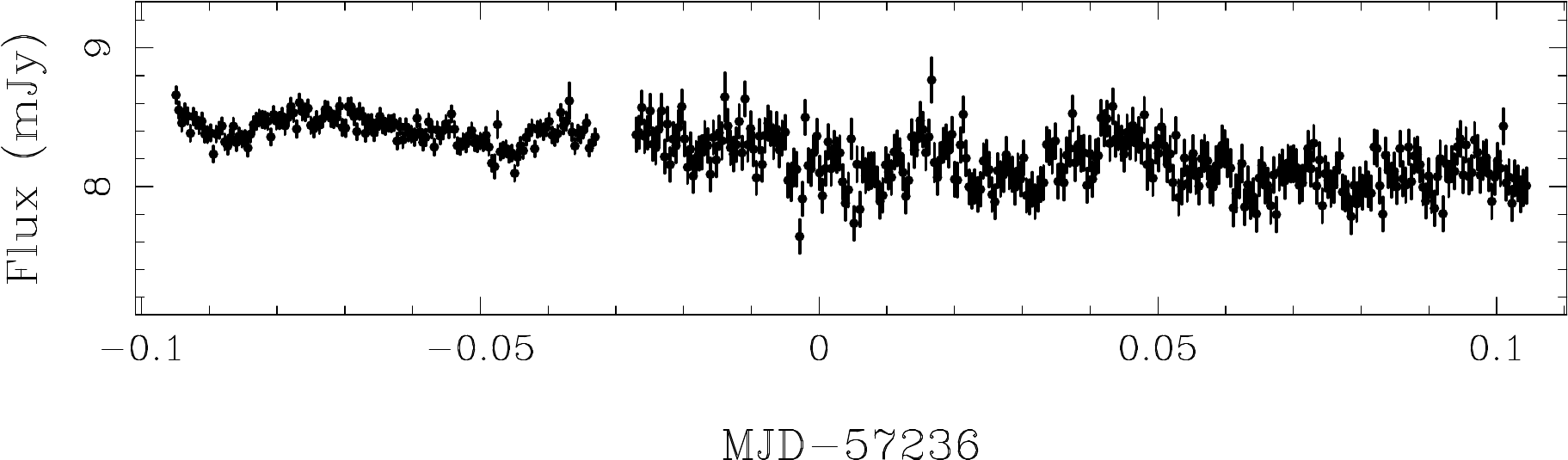}
 \caption{ASASSN-15ni light curve observed on 2015-08-01 during the decline from outburst with \textit{pt5m}. The gap in the light curve is due to the telescope conducting a pier-flip as the target transited the observers meridian. After the pier-flip, a different (fainter) comparison star had to be used. This explains the difference in the size of the error bars before and after the gap.}
 \label{fig:asassn15ni}
\end{figure}
\begin{figure}
 \includegraphics[width=2.55cm,angle=270]{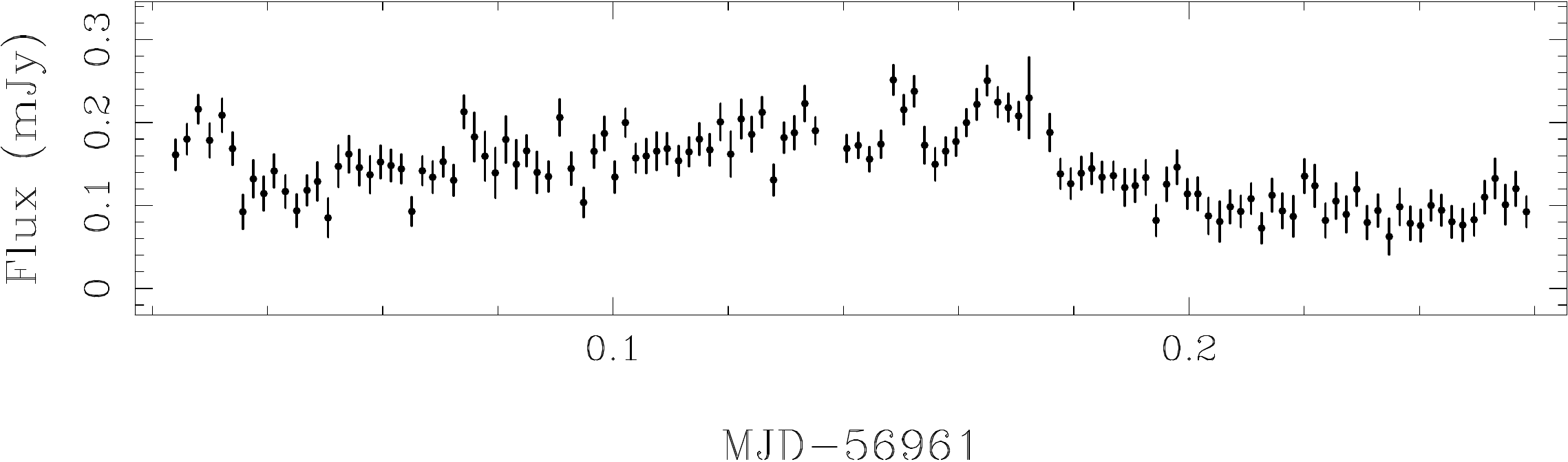}
 \caption{CSS090219:044027+023301 light curve observed during quiescence on 2014-10-30 with \textit{pt5m}.}
 \label{fig:css090219}
\end{figure}
\begin{figure}
\centering
 \includegraphics[width=2.55cm,angle=270]{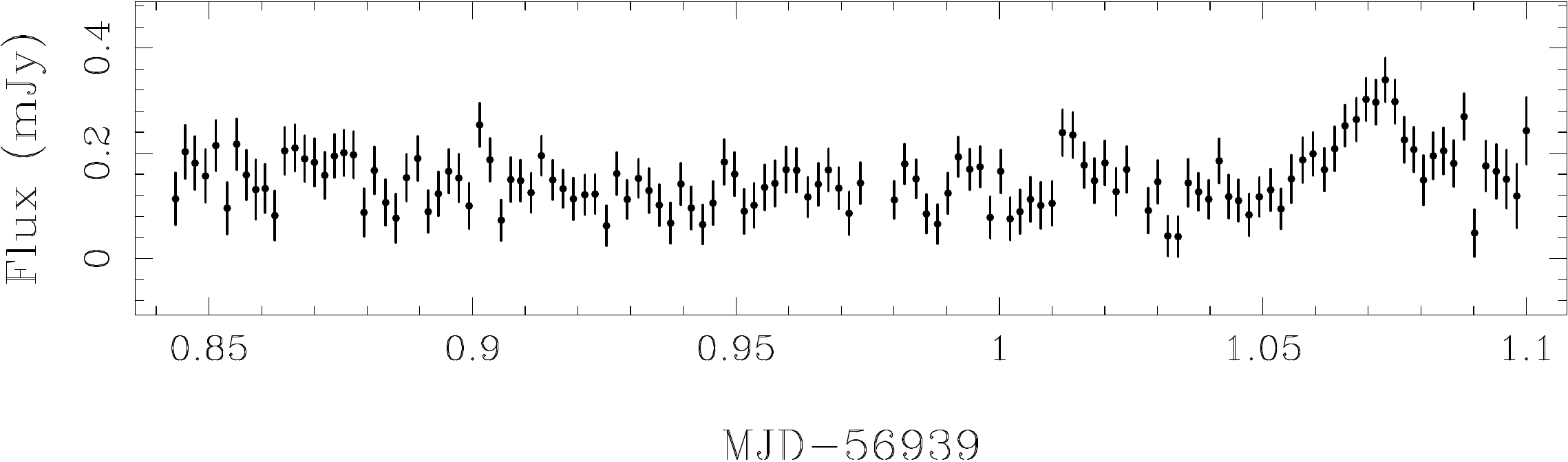}
 \caption{CSS091116:232551-014024 light curve observed during quiescence on 2014-10-09 with \textit{pt5m}.}
 \label{fig:css091116}
\end{figure}
\begin{figure}
\centering
 \includegraphics[width=2.55cm,angle=270]{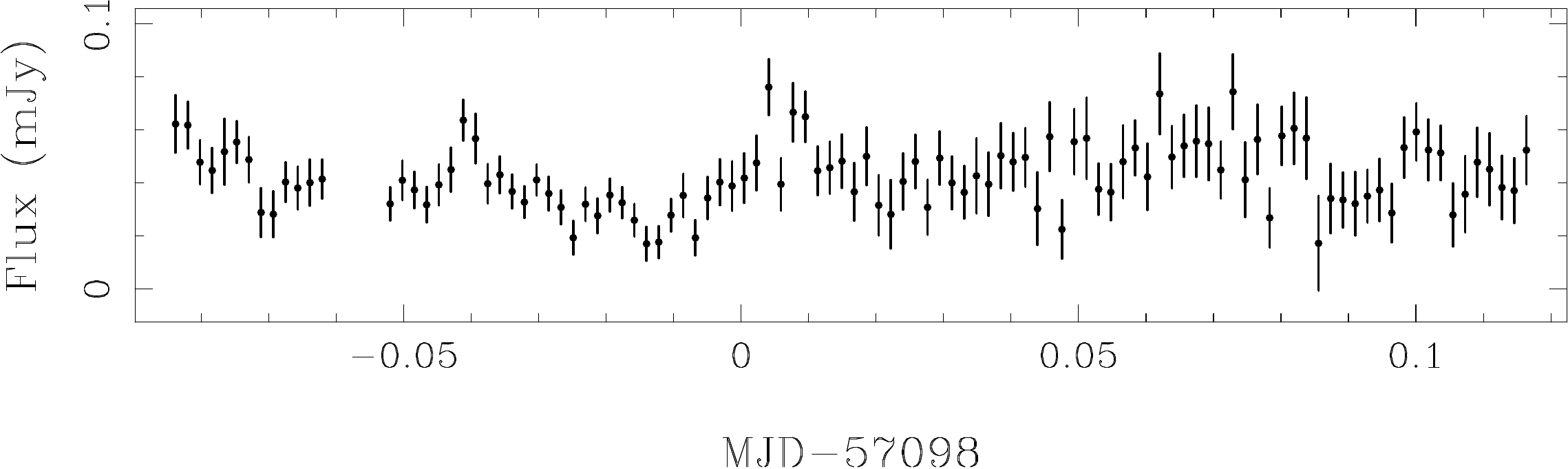}
 \caption{CSS100508:085604+322109 light curve observed in quiescence on 2015-03-16 with \textit{pt5m} in windy conditions. The sporadic variability is attributed to flickering.}
 \label{fig:css1005}
\end{figure}
\begin{figure}
\centering
 \includegraphics[width=2.55cm,angle=270]{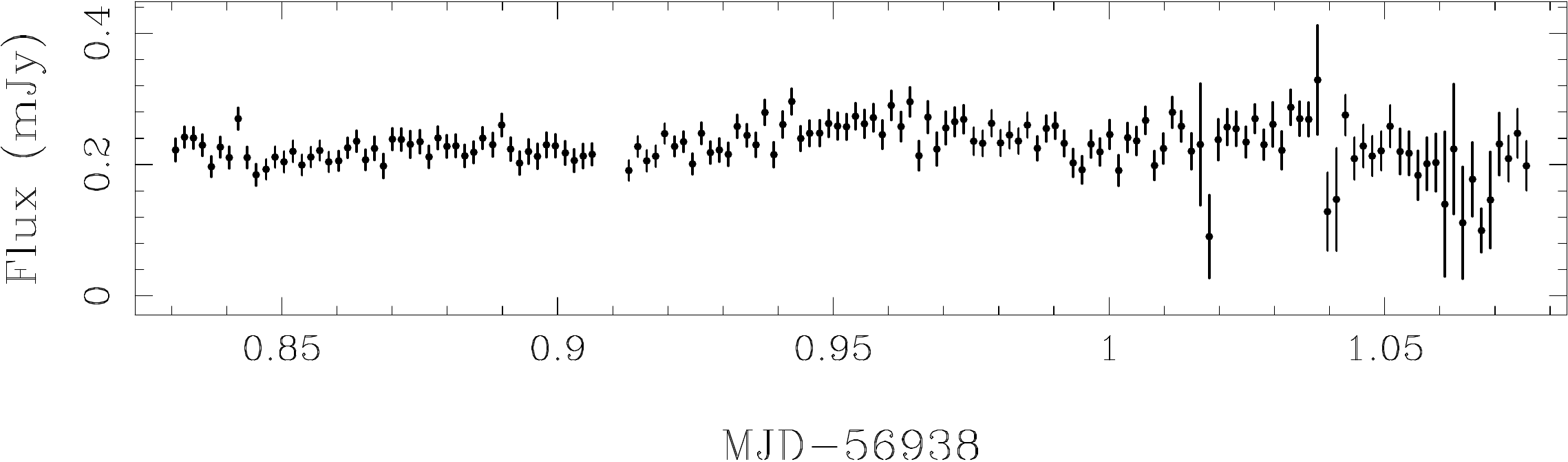}
 \caption{CSS100520:214426+222024 light curve observed during quiescence on 2014-10-08 with \textit{pt5m}. The disruption towards the end is due to patchy cloud cover.}
 \label{fig:css100520}
\end{figure}

\begin{figure}
\centering
 \includegraphics[width=2.55cm,angle=270]{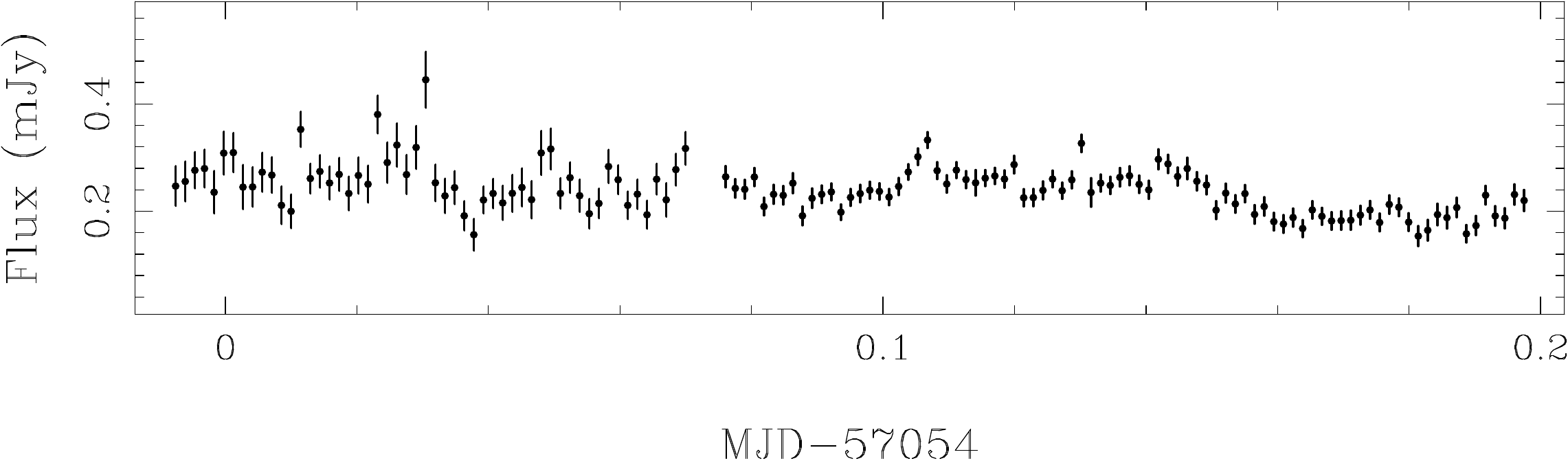}
 \caption{CSS110114:091246-034916 light curve observed during quiescence on 2015-01-31 with \textit{pt5m}.}
 \label{fig:css110114}
\end{figure}
\begin{figure}
\centering
 \includegraphics[width=2.55cm,angle=270]{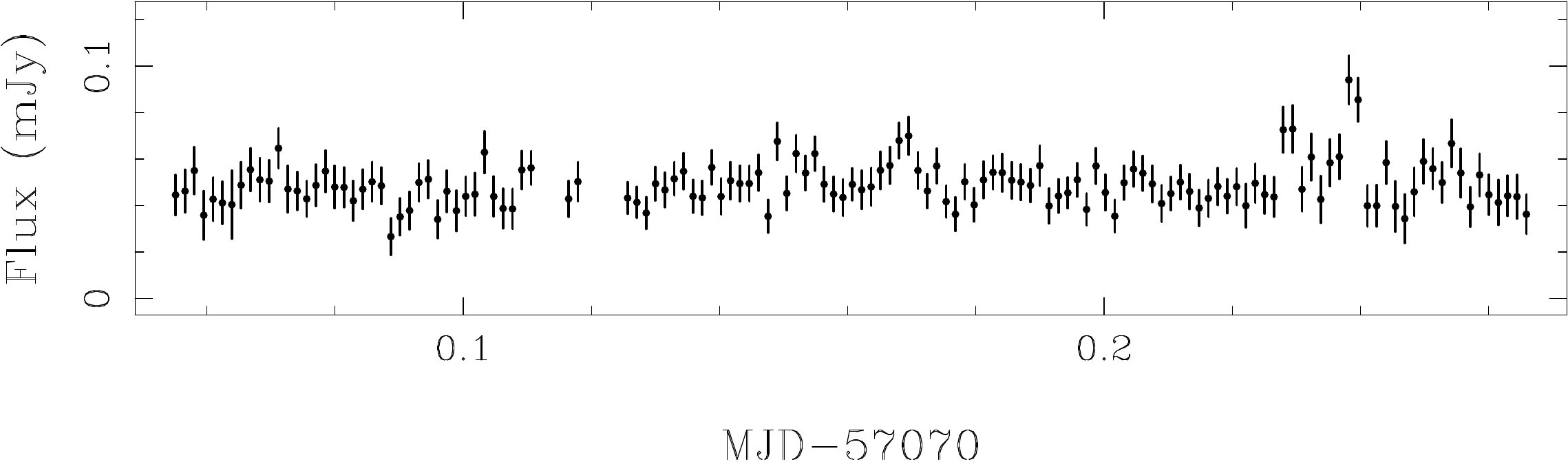}
 \caption{CSS110226:112510+231036 light curve observed during quiescence on 2015-02-16 with \textit{pt5m}.}
 \label{fig:css110226}
\end{figure}
\begin{figure}
 \includegraphics[width=2.55cm,angle=270]{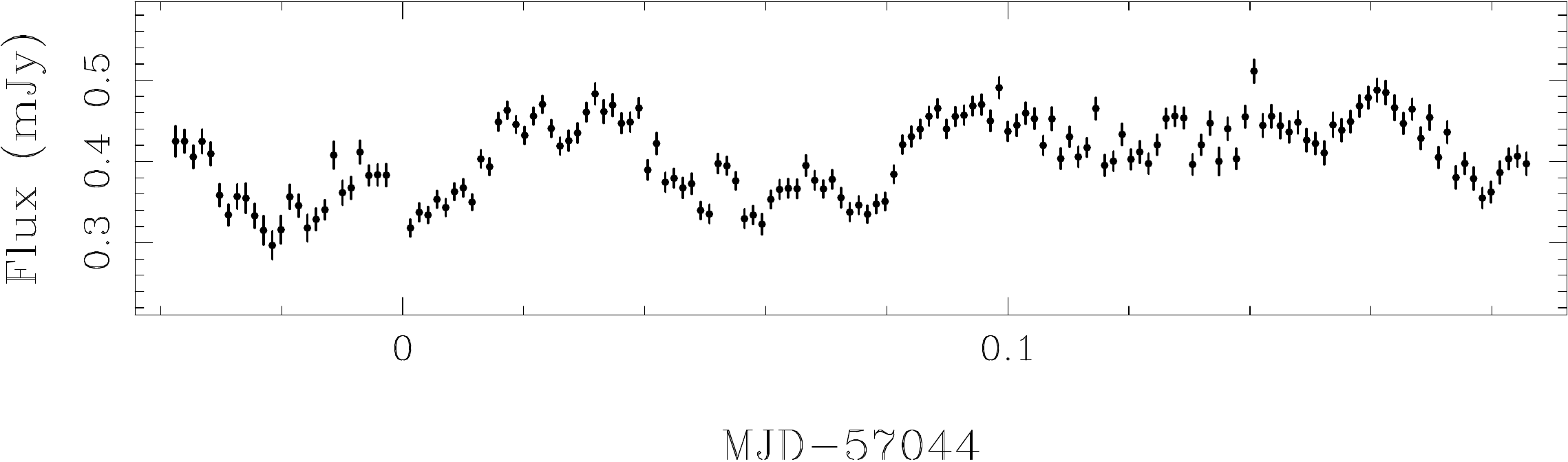}
 \caption{CSS130906:064726+491542 light curve observed during quiescence on 2015-01-21 with \textit{pt5m}. Variability could be attributed to flickering, or movement through hot pixels on the CCD detector (see text for details).}
 \label{fig:css130906}
\end{figure}
\begin{figure}
 \includegraphics[width=2.55cm,angle=270]{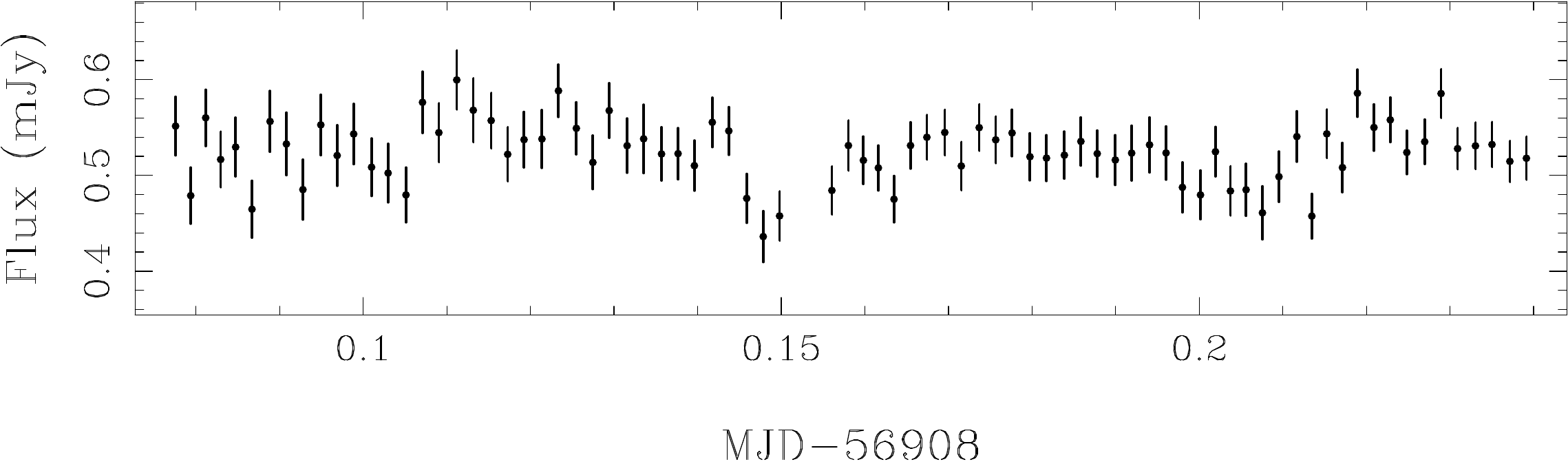}
 \caption{CSS140901:013309+133234 light curve observed on 2014-09-07 with \textit{pt5m}. Here the system is fading from outburst but some superhump signals are still present.}
 \label{fig:css140901}
\end{figure}
\begin{figure}
 \includegraphics[width=2.55cm,angle=270]{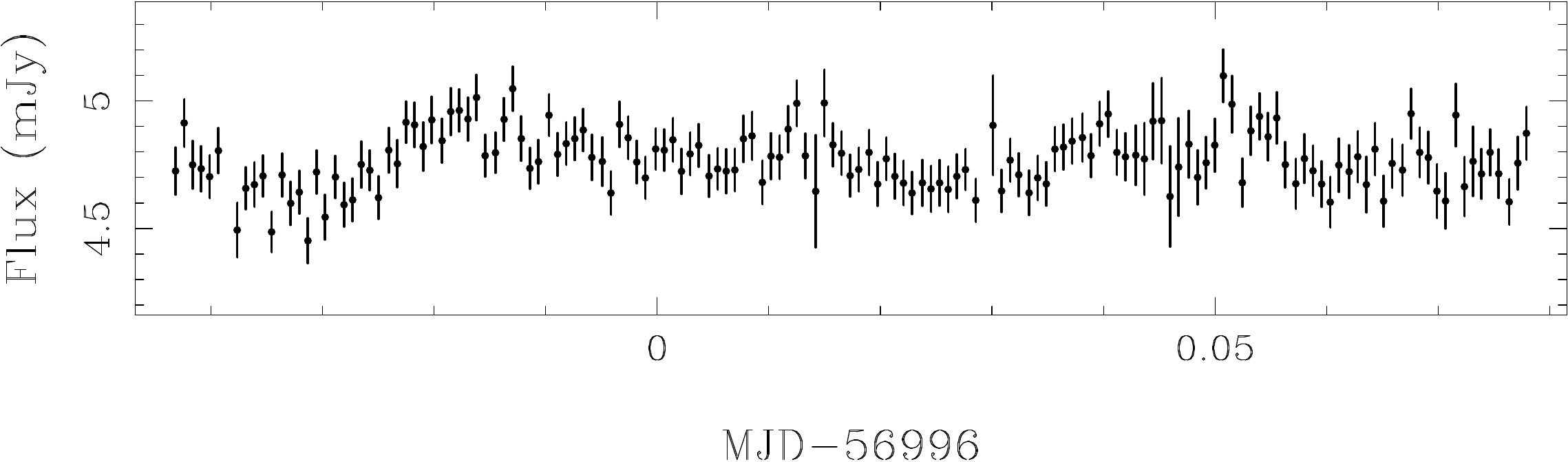}
 \caption{CSS141005:023428-045431 light curve observed whilst fading from outburst on 2014-12-04 with \textit{pt5m}. This light curve is one of several observations on different nights that display periodic features resembling superhumps, although stochastic flickering is not ruled out as the cause for this variability.}
 \label{fig:css141005}
\end{figure}
\begin{figure}
\centering
 \includegraphics[width=2.55cm,angle=270]{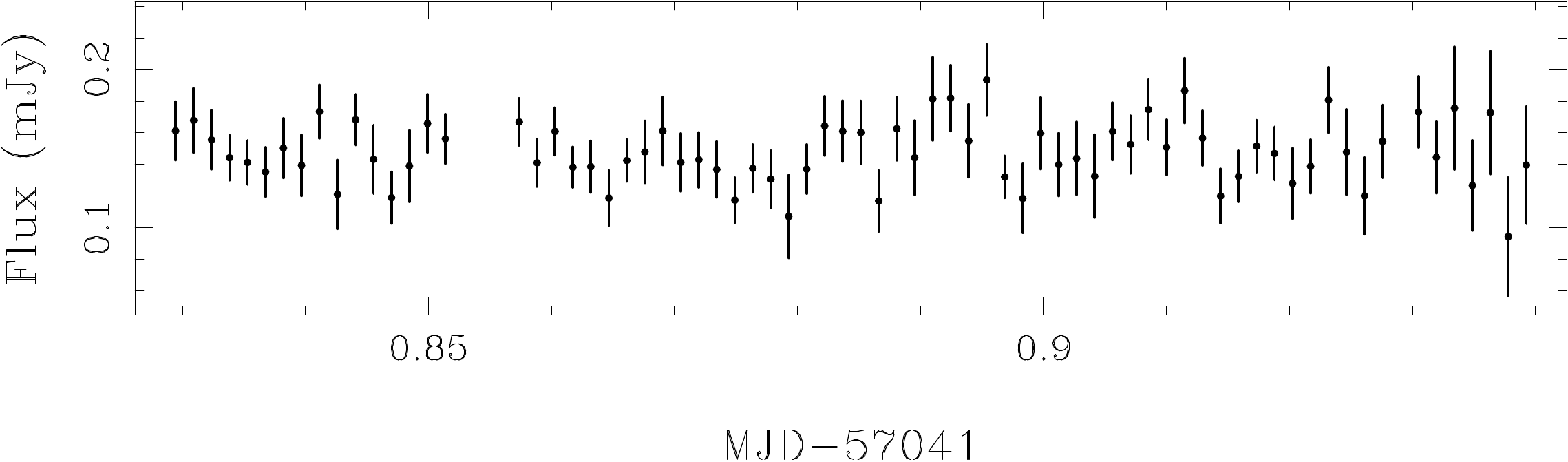}
 \caption{CSS141117:030930+263804 light curve observed in quiescence on 2015-01-19 with \textit{pt5m}. Some thin overhead cirrus affects the quality of the light curve.}
 \label{fig:css141117}
\end{figure}
\begin{figure}
\centering
 \includegraphics[width=2.55cm,angle=270]{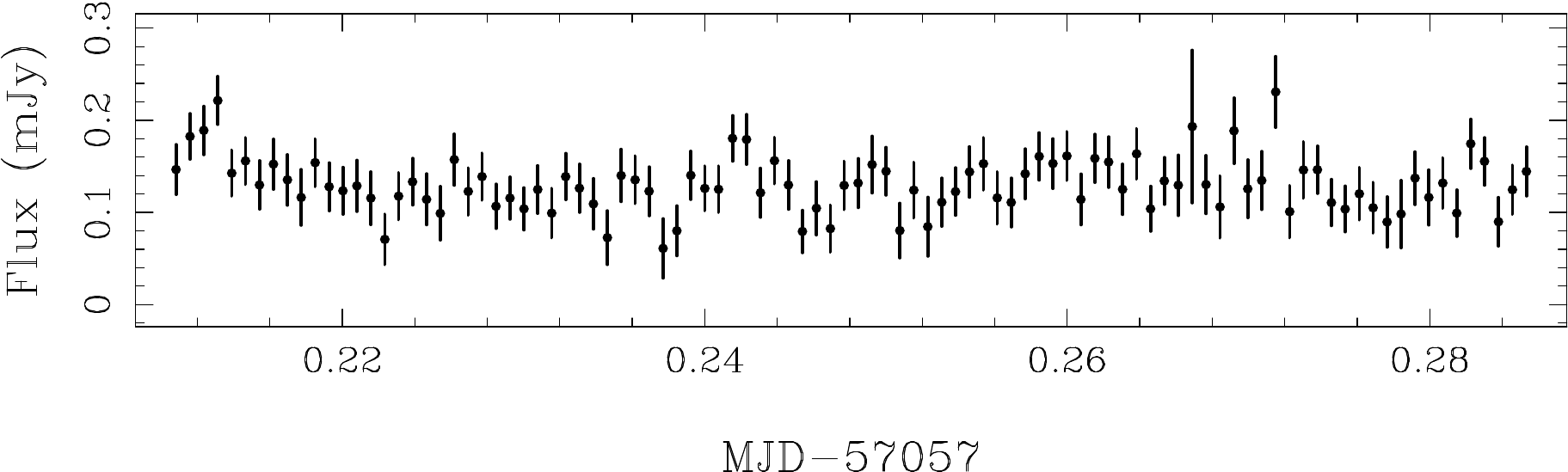}
 \caption{Gaia15aan/ASASSN-14mo light curve observed during the decline from outburst, on 2015-02-03 with \textit{pt5m}.}
 \label{fig:gaia15aan}
\end{figure}
\begin{figure}
\centering
 \includegraphics[width=2.55cm,angle=270]{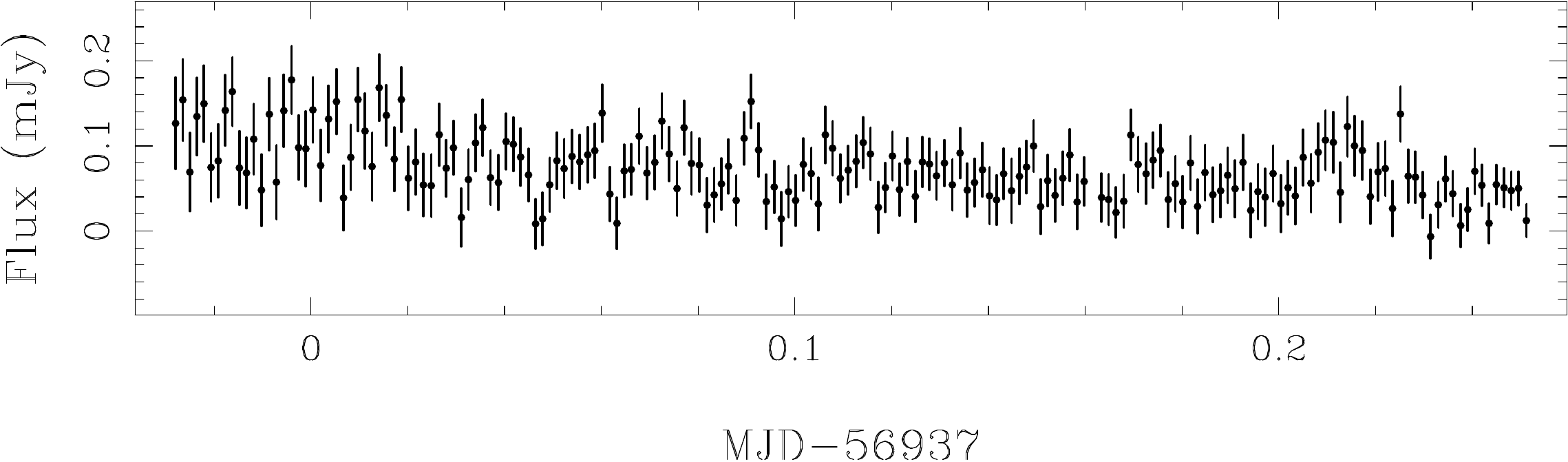}
 \caption{MASTER OT J034045.31+471632.2 light curve observed during quiescence on 2014-10-06 with \textit{pt5m}.}
 \label{fig:mast0340}
\end{figure}

\begin{figure}
 \includegraphics[width=2.55cm,angle=270]{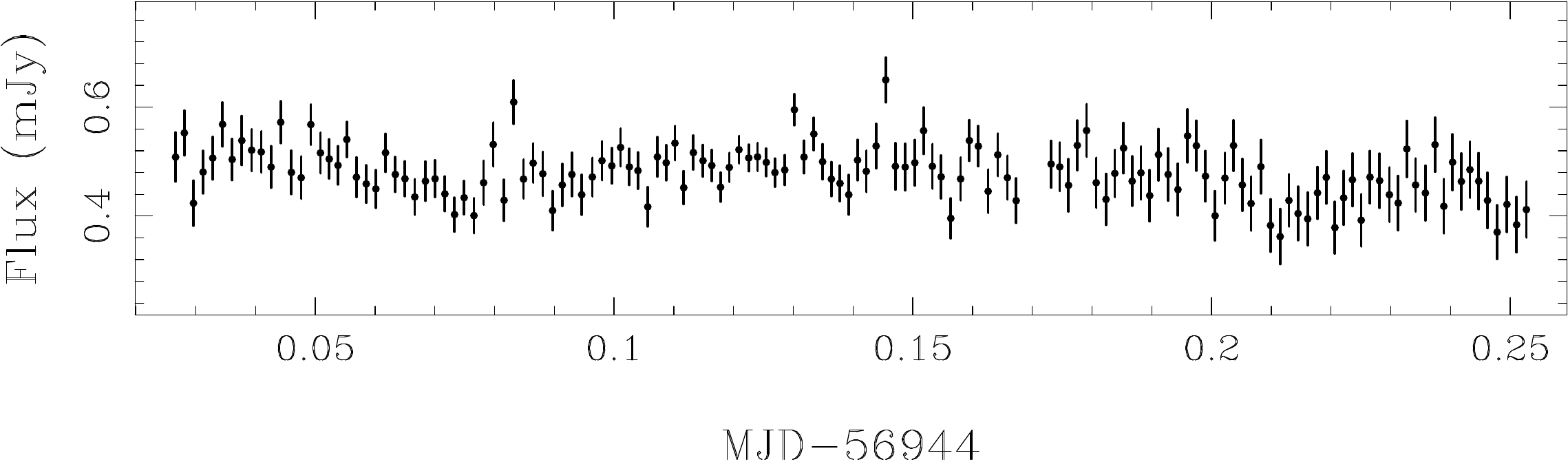}
 \caption{MASTER OT J041923.57+653004.3 light curve observed on 2014-09-21 with \textit{pt5m}. The variability may be periodic, or attributed to flickering.}
 \label{fig:mast0419}
\end{figure}
\begin{figure}
 \includegraphics[width=2.55cm,angle=270]{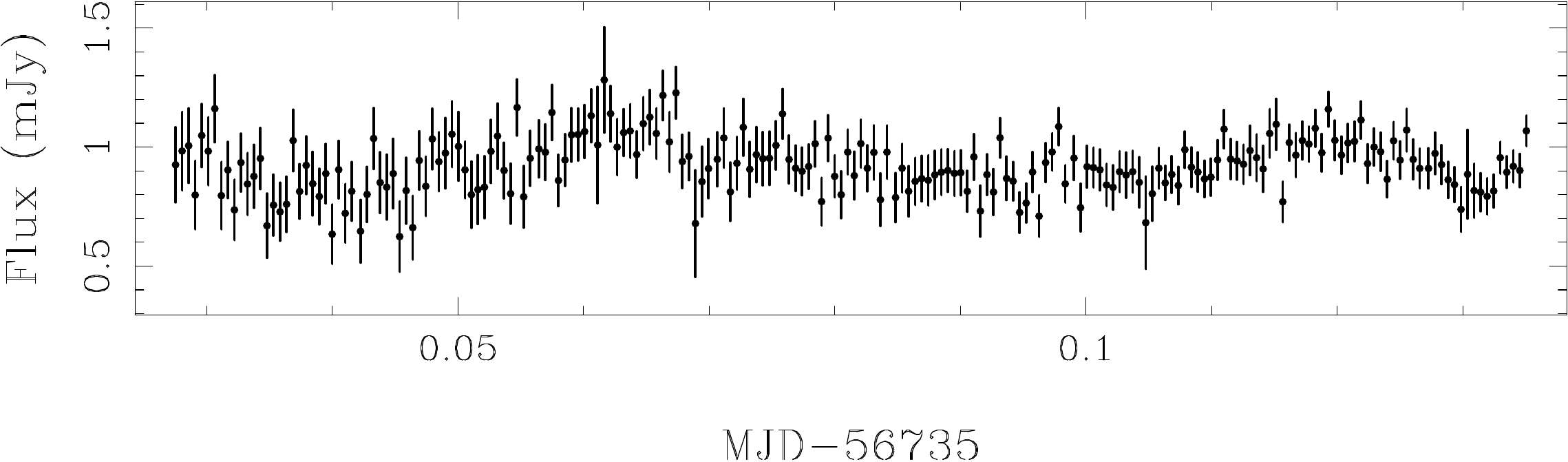}
 \caption{MASTER OT J171921.40+640309.8 light curve showing superhump-like variability, observed during outburst on 2014-03-18 with \textit{pt5m}.}
 \label{fig:mast1719}
\end{figure}
\begin{figure}
\centering
 \includegraphics[width=2.55cm,angle=270]{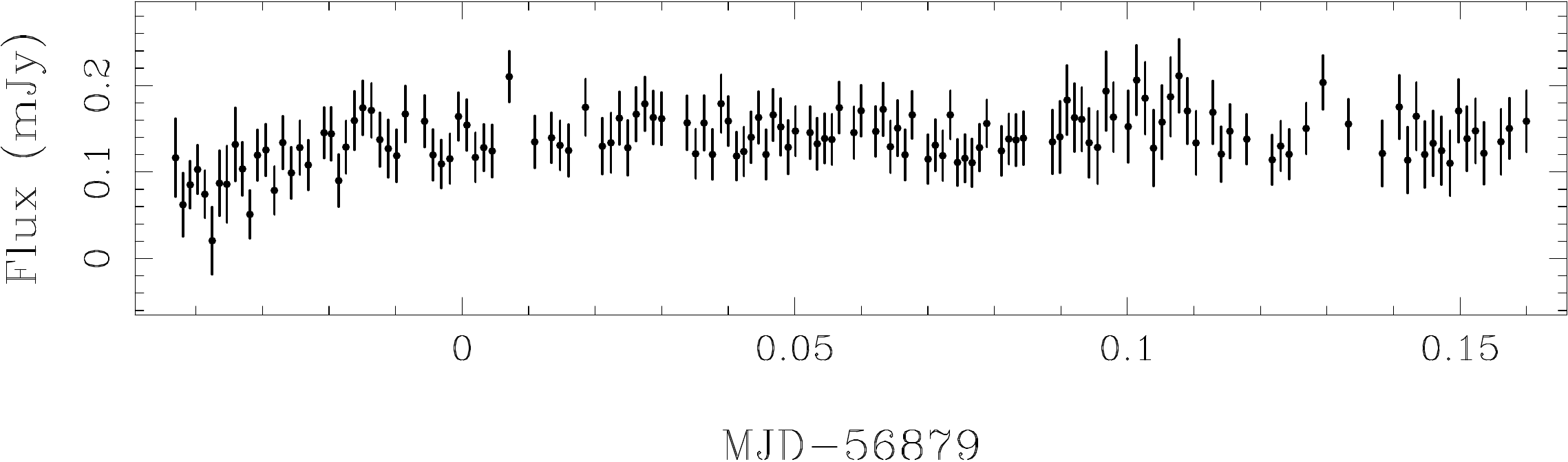}
 \caption{MASTER OT J194955.17+455349.6 light curve observed in quiescence on 2014-08-09 with \textit{pt5m}.}
 \label{fig:mast1949}
\end{figure}
\begin{figure}
\centering
 \includegraphics[width=2.55cm,angle=270]{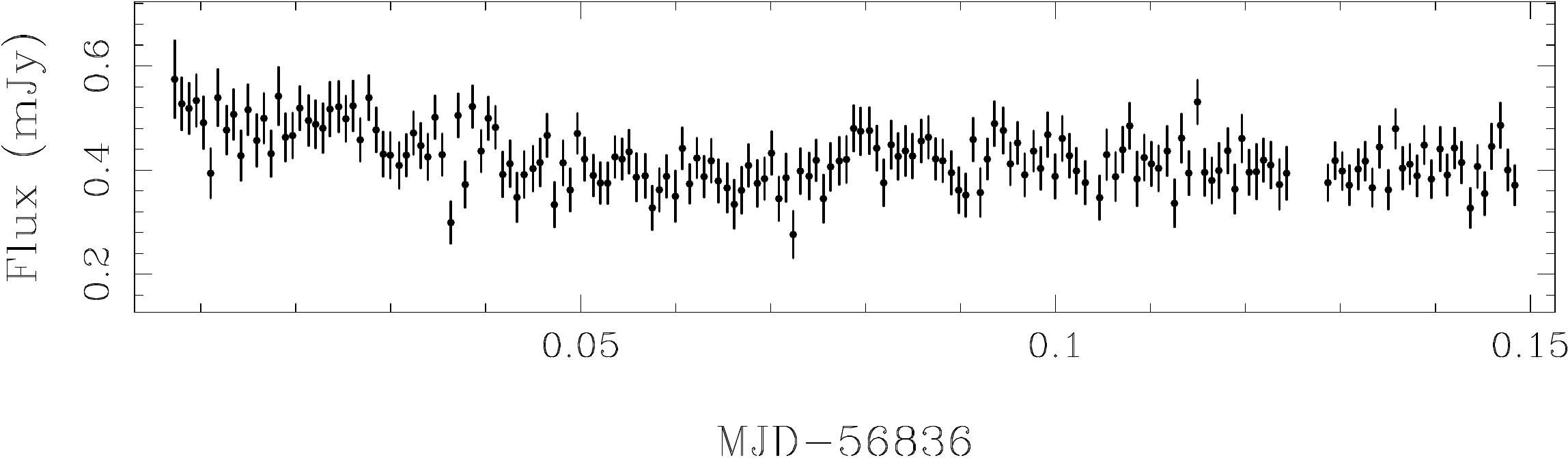}
 \caption{MASTER OT J201121.95+565531.1 light curve observed during the decline from outburst on 2014-06-27 with \textit{pt5m}.}
 \label{fig:mast2011}
\end{figure}
\begin{figure}
\centering
 \includegraphics[width=2.55cm,angle=270]{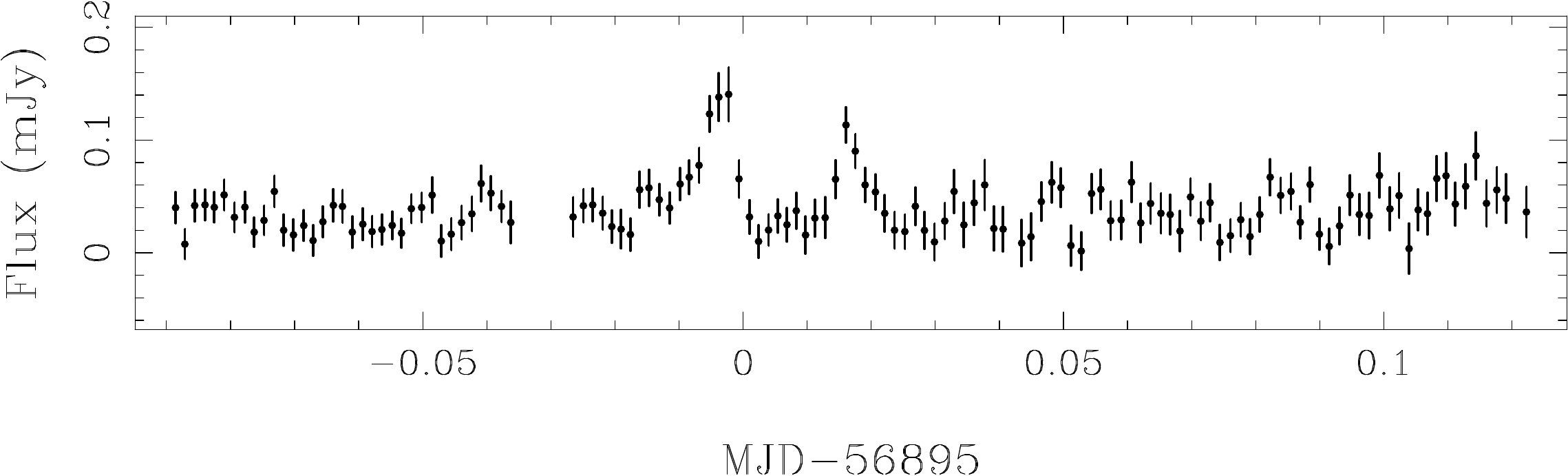}
 \caption{MASTER OT J202157.69+212919.4 light curve observed on 2014-08-25 with \textit{pt5m}. The two peaks in the centre are caused by the apertures tracing across hot pixels on the CCD.}
 \label{fig:mast2021}
\end{figure}
\begin{figure}
\centering
 \includegraphics[width=2.55cm,angle=270]{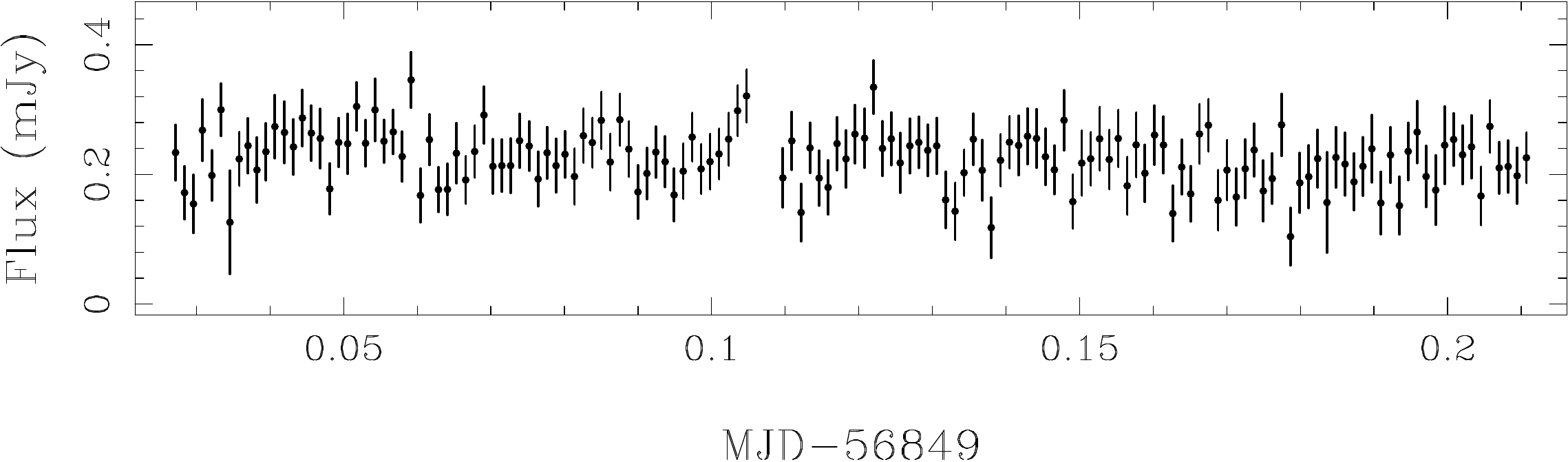}
 \caption{MASTER OT J203421.90+120656.9 light curve observed during outburst on 2014-07-10 with \textit{pt5m}.}
 \label{fig:mast2034}
\end{figure}
\begin{figure}
\centering
 \includegraphics[width=2.55cm,angle=270]{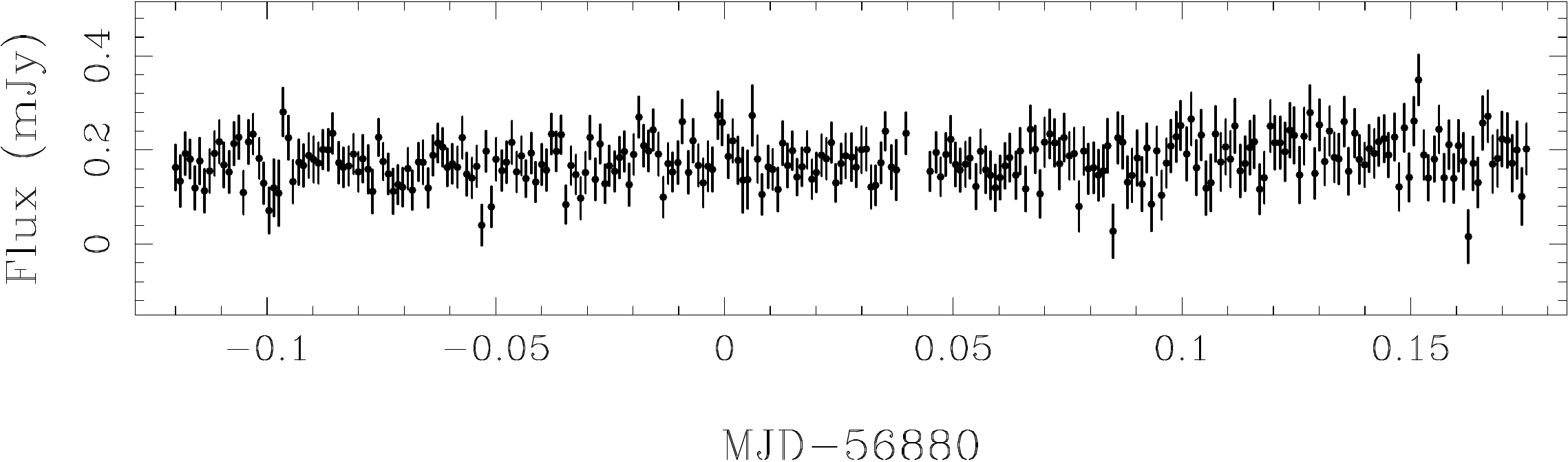}
 \caption{MASTER OT J210316.39+314913.6 light curve observed during quiescence on 2014-08-10 with \textit{pt5m}.}
 \label{fig:master2103}
\end{figure}

\end{document}